\documentclass[11pt,a4paper]{article}
\pdfoutput=1
\usepackage{jcappub}
\usepackage{aasmacros}

\ifx\pdfoutput\undefined
\usepackage[dvips,bookmarks=false]{hyperref}	% This is for arXiv.org
\else
\usepackage{hyperref}	% This is for pdftex
\fi
\hypersetup{colorlinks,bookmarksopen,bookmarksnumbered,citecolor=blue,
linkcolor=black,pdfstartview=FitH,urlcolor=blue}

\usepackage{graphicx}
\usepackage{amsmath}
\usepackage{amssymb}
\usepackage{rotating} 
\usepackage{xspace}
\usepackage{color}

\usepackage{placeins}
\usepackage{rotating}

\newcommand{\be}{\begin{equation}}
\newcommand{\ee}{\end{equation}}
\newcommand{\beq}{\begin{equation}}
\newcommand{\eeq}{\end{equation}}

\makeatletter
\renewcommand{\fnum@table}{\textbf{\tablename~\thetable}}
\renewcommand{\fnum@figure}{\textbf{\figurename~\thefigure}}
\makeatother

\newcommand{\mDM}{m_{\rm{DM}}}
\newcommand{\mmed}{M_{\rm{med}}}
\newcommand{\mMed}{M_{\rm{med}}}
\newcommand{\gDM}{g_{\rm{DM}}}
\newcommand{\gq}{g_q}
\newcommand{\gSM}{g_q}
\newcommand{\sigmaSI}{\sigma_{\rm{SI}}}
\newcommand{\sigmaSD}{\sigma_{\rm{SD}}}
\newcommand{\oh}{\Omega_{\mathrm{DM}} h^2}

\title{Identifying WIMP dark matter from particle and astroparticle data
}
\author[a]{Gianfranco Bertone,}
\author[a,b]{Nassim Bozorgnia,}
\author[c]{Jong Soo Kim,}
\author[a]{Sebastian Liem,}
\author[d]{Christopher McCabe,}
\author[e]{Sydney Otten}
\author[f]{and Roberto Ruiz de Austri}

\affiliation[a]{GRAPPA Institute, Institute for Theoretical Physics Amsterdam\\ 
and Delta Institute for Theoretical Physics, University of Amsterdam, \\
Science Park 904, 1098 XH Amsterdam, The Netherlands
} 

\affiliation[b]{Institute for Particle Physics Phenomenology, Department of Physics,\\
Durham University, Durham, DH1 3LE, United Kingdom
} 

\affiliation[c]{National Institute for Theoretical Physics,\\
School of Physics and Mandelstam Institute for Theoretical Physics,\\
University of the Witwatersrand, Johannesburg, Wits 2050, South Africa} 

\affiliation[d]{Department of Physics, King's College London,\\ Strand, London, WC2R 2LS, United Kingdom
}

\affiliation[e]{
Institute for Theoretical Particle Physics and Cosmology, RWTH Aachen University,\\ 52074 Aachen, Germany
} 

\affiliation[f]{Instituto de F\'{i}sica Corpuscular, IFIC-UV/CSIC,\\ Valencia, Spain
} 

\emailAdd{sebastian.liem@uva.nl}
\emailAdd{g.bertone@uva.nl}
\emailAdd{nassim.bozorgnia@durham.ac.uk}
\emailAdd{jongsoo.kim@tu-dortmund.de}
\emailAdd{christopher.mccabe@kcl.ac.uk}
\emailAdd{sydney.otten@rwth-aachen.de}
\emailAdd{rruiz@ific.uv.es}

\abstract{
One of the most promising strategies to identify the nature of dark matter consists in the search for new particles at accelerators and with so-called direct detection experiments. Working within the framework of simplified models, and making use of machine learning tools to speed up statistical inference, we address the question of what we can learn about dark matter from a detection at the LHC and a forthcoming direct detection experiment. We show that with a combination of accelerator and direct detection data, it is possible to identify newly discovered particles as dark matter, by reconstructing their relic density assuming they are weakly interacting massive particles (WIMPs) thermally produced in the early Universe, and demonstrating that it is consistent with the measured dark matter abundance. An inconsistency between these two quantities would instead point either towards additional physics in the dark sector, or towards a non-standard cosmology, with a thermal history substantially different from that of the standard cosmological model.
}

\keywords{Dark matter detectors, Dark matter experiments, Dark matter theory, Large Hadron Collider, Machine learning, Global analysis}
\arxivnumber{IPPP/17/94, KCL-PH-TH/2017-55, TTK-17-45}

\begin{document}
\maketitle
%%%%%%%%%%%%%%%%%%%%%%%%%%%%%%%%%%%%%%%%%%%%%%%%%%%%%%%%%%%%%%%%%%%%%%%%%%%%%%%%%%%%%
\section{Introduction}
\label{sec:introduction}

An overwhelming body of observational evidence points to the existence of dark matter, an elusive form of matter that permeates the Universe, and is fundamentally different from the particles contained in the Standard Model of particle physics~\cite{Bertone:2010zza}. Understanding the nature of dark matter and the role it plays in the evolution of the early Universe is one of the most compelling problems in cosmology and particle physics~\cite{Bertone:2010zza,Jungman:1995df,Bergstrom00,Bertone05}.

The most widely discussed dark matter candidates are stable, weakly interacting massive particles (WIMPs). A particularly attractive feature of WIMPs is that the production mechanism to explain the abundance of dark matter, \emph{thermal freeze-out}, is simple and entirely understood: WIMPs were in thermal equilibrium in the early Universe, then stopped interacting significantly with any other particles as the Universe expanded, \emph{freezing-out} of equilibrium~\cite{Lee:1977ua,Hut:1977zn,Wolfram:1978gp,Steigman:1979kw,Bernstein:1985th,Scherrer:1985zt,Srednicki:1988ce}. A massive particle with weak interactions freezes-out with the observationally inferred relic density~\cite{Ade:2015xua} if its self-annihilation cross-section is at the weak scale~\cite{Steigman:2012nb}.\footnote{Different definitions of WIMPs are used in the literature. To avoid confusion, here we use the term WIMP to refer to any particle in the $\mathcal{O}(1)$~GeV to $\mathcal{O}(100)~\text{TeV}$ mass range that interacts with Standard Model particles with a strength similar to the weak-interaction, such that the abundance is obtained through the thermal freeze-out mechanism. Our definition is sometimes referred to as a `hidden-sector WIMP'~\cite{Battaglieri:2017aum}.}

WIMPs can be searched for with a variety of detection strategies, for instance by detecting new particles at high-energy colliders; by measuring the recoil energy of nuclei struck by dark matter particles in underground \emph{direct detection} experiments; or by observing the secondary particles produced by the self-annihilation with \emph{indirect detection} experiments. No trace of WIMP dark matter has been found so far but many existing or planned experiments will soon probe a large fraction of the parameter space of the most promising extensions of the Standard Model of particle physics in which WIMPs arise (see e.g.\ refs.~\cite{Bertone:2010at,Arcadi:2017kky}).  

In this context, a particularly interesting question to address is: assuming a positive detection of new particles at colliders or with direct detection experiments, can we identify those particles as WIMP dark matter? A number of studies have attempted to address this question (see e.g.\ refs.~\cite{Baltz:2006fm,Bertone:2007xj,Roszkowski:2017dou,Baum:2017kfa}), typically within the context of a global fitting analysis~\cite{Allanach:2005kz,Hooper:2006wv,Bernal:2008zk,Trotta:2008bp,Bertone:2010rv,Buchmueller:2011ki,Strege:2011pk,Roszkowski:2012uf,Mambrini:2012ue,Arbey:2013iza,Strege:2014ija, Demir:2014jqa, Cerdeno:2015ega, deVries:2015hva, Buchmueller:2015uqa, Bertone:2015tza, Bagnaschi:2015eha, Dutta:2015exw, Ross:2016pml, Liem:2016xpm, Roszkowski:2016bhs, Caron:2016hib, Barr:2016sho, Bagnaschi:2016afc, Bagnaschi:2016xfg, Rogers:2016jrx, Athron:2017ard,Athron:2017yua, Athron:2017kgt, Athron:2017qdc, Bagnaschi:2017tru, Costa:2017gup}. Given a positive detection in one or more experiments, a global fitting analysis enables the parameters of a specified model to be reconstructed. In the case of WIMP dark matter, the reconstructed model parameters also enter into the calculation of the thermal freeze-out abundance, so we can reconstruct the value of the freeze-out abundance that is consistent with the positive detections. We can therefore quantify how compatible the reconstructed freeze-out abundance is with the observationally inferred value; consistent values imply that the positive detections at colliders or direct detection experiments are consistent with WIMP dark matter produced within the standard cosmological model. 

In this study, we seek to address whether we can identify particles as WIMP dark matter within the context of simplified models. In recent years, the dark matter collider community has moved towards simplified models as a useful way to display the results from their searches for dark matter~\cite{Alves:2011wf,Goodman:2011jq, An:2012va, Frandsen:2012rk, Fox:2012ru, Dreiner:2013vla, Cotta:2013jna, Buchmueller:2013dya, Papucci:2014iwa, Buchmueller:2014yoa, Abdallah:2014hon, Malik:2014ggr, Buckley:2014fba, Harris:2014hga, Jacques:2015zha, Haisch:2015ioa, Buchmueller:2015eea, Abdallah:2015ter, Abercrombie:2015wmb, Brennan:2016xjh, Boveia:2016mrp}. The aim of a simplified model is to provide a bottom-up framework to characterise all of the relevant dynamical processes that occur in the production, scattering or self-annihilation of dark matter with a small number of additional particles and parameters. 

We consider two benchmark simplified models of dark matter that are described in section~\ref{sec:models}. The first benchmark contains WIMP dark matter (i.e.\ the abundance is set by thermal freeze-out) while the second benchmark must achieve its abundance through an unspecified mechanism. We assume that both benchmarks will lead to a signal in searches for an energetic jet and large missing transverse momentum (typically referred to as \emph{monojet searches}) at the LHC, together with a signal in a future direct detection experiment. The details of how we generate the signals in the upcoming experiments are described in section~\ref{sec:mocksignals}. We make use of recent developments in machine learning techniques that have increased the potential of global fitting analyses by significantly speeding up much of the expensive calculations~\cite{Caron:2016hib, Bertone:2016mdy,Frate:2017mai}. These techniques are described in section~\ref{sec:ML} and the details of the validation procedure are presented in appendix~\ref{sec:mlvalidation}. The results of our global fitting analysis are given in section~\ref{sec:results}, which show that within the context of our first benchmark model, it is possible to identify the dark matter as WIMP dark matter. Conversely, for the second benchmark model we are able to show that the positive detections are not consistent with WIMP dark matter. Our full conclusions are given in section~\ref{Sec:conclusions}. Additional plots showing the robustness of our conclusions under different assumptions for future monojet searches and under a different statistical interpretation are given in appendices~\ref{sec:addfigs} and~\ref{sec:addfigsprofile}, respectively.

%%%%%%%%%%%%%%%%%%%%%%%%%%%%%%%%%%%%%%%%%%%%%%%%%%%%%%%%%%
%%%%%%%%%%%%%%%%%%%%%%%%%%%%%%%%%%%%%%%%%%%%%%%%%%%%%%%%%%
\section{Simplified models and benchmark points}
\label{sec:models}
%%%%%%%%%%%%%%%%%%%%%%%%%%%%%%%%%%%%%%%%%%%%%%%%%%%%%%%%%%
%%%%%%%%%%%%%%%%%%%%%%%%%%%%%%%%%%%%%%%%%%%%%%%%%%%%%%%%%%

In this section, we describe the two benchmark points for two minimal four-parameter simplified models that we use throughout the paper. We use the minimal simplified models recommended by the LHC Dark Matter Working Group~\cite{Boveia:2016mrp}, which contain only two new particles and four new parameters: the dark matter particle~($\chi$), typically taken as a Dirac fermion; a `mediator' particle ($Z'$ or~$\phi$) that mediates the interaction between dark matter and quarks or gluons; the mass of the dark matter and mediator particles,~$\mDM$ and~$\mMed$ respectively; a coupling of the mediator to dark matter,~$\gDM$; and a coupling of the mediator to quarks,~$\gq$. Couplings to leptons are often set to zero to avoid stringent constraints from di-lepton searches. Finally, the mediator is taken to be either a spin-0 or spin-1 particle.

Benchmark~1 contains a spin-1 mediator with axial-vector couplings (which we'll refer to as an \emph{axial-vector} mediator for simplicity) while benchmark~2 contains a spin-0 mediator with scalar couplings (a \emph{scalar} mediator). As we will explicitly show, benchmark~1 is an example of WIMP dark matter in which the observationally inferred dark matter abundance is obtained through the thermal freeze-out mechanism. In contrast, benchmark~2 predicts an abundance through the freeze-out mechanism that is significantly larger than the observed value. For easy reference, the parameter values and the values of observable quantities are summarised in table~\ref{tab:benchmarks}.

%%%%%%%%%%%%%%%%%%%%%%%%%%%%%%%%%%%%%%%%%%%%%%%%%%%%%%%%%%
\begin{table}[t]
\setlength{\tabcolsep}{10pt}
\renewcommand{\arraystretch}{1.3}
\center
\begin{tabular}{c|cc} 
 & Benchmark 1  & Benchmark 2\\
\hline
Dark matter mass, $\mDM$ [GeV] & 218   & 3 \\
Mediator mass, $\mMed$ [GeV] & 580 & 100 \\
Mediator type & Axial-vector, $Z'$ & Scalar, $\phi$\\
Dark matter-mediator coupling, $\gDM$ & 1 & 1\\
Quark-mediator coupling, $\gq$ & 0.1 & 1 \\
\hline
$\Omega_{\rm{DM}} h^2$ [freeze-out] & $0.12$ & $3\times 10^3$\\
$\langle \sigma^{\chi\bar{\chi}\to q \bar{q}}_{\rm{ann}} \,v \rangle_{\rm{s-wave}}\, [\mathrm{cm}^3 \,\mathrm{s}^{-1}]$  & $9.0 \times 10^{-27}$ & 0\\
$\sigma_{\rm{SI}}\, [\mathrm{cm}^2]$ & 0 & $8.8 \times 10^{-43}$ \\
$\sigma^{p}_{\rm{SD}}\, [\mathrm{cm}^2]$ & $1.9 \times 10^{-42}$ & 0 \\
$\sigma^{n}_{\rm{SD}}\, [\mathrm{cm}^2]$ & $4.2 \times 10^{-42}$ & 0 \\
\end{tabular}
\caption{The two benchmark scenarios considered in our study. Benchmark~1 is an axial-vector mediator model and is an example of WIMP dark matter: $\mDM$ and $\mMed$ have weak-scale masses and the observationally inferred value of~$\Omega_{\rm{DM}} h^2$ is obtained from thermal freeze-out. This model leads to a spin-dependent scattering cross-section within reach of upcoming xenon-based direct detection experiments. Benchmark 2 is the scalar mediator model and predicts a value of $\Omega_{\rm{DM}} h^2$ from thermal freeze-out that is significantly larger than the observed value. To explain the observed dark matter abundance in this model, additional physics not included in the simplified model or a modified cosmological history is required. This model leads to a spin-independent scattering cross-section within reach of the  CRESST-III experiment. Both benchmark models also lead to a signal in a future LHC monojet search.}
\label{tab:benchmarks} 
\end{table}
%%%%%%%%%%%%%%%%%%%%%%%%%%%%%%%%%%%%%%%%%%%%%%%%%%%%%%%%%%

%%%%%%%%%%%%%%%%%%%%%%%%%%%%%%%%%%%%%%%%%%%%%%%%%%%%%%%%%%
\subsection{Benchmark 1: Axial-vector mediator}
%%%%%%%%%%%%%%%%%%%%%%%%%%%%%%%%%%%%%%%%%%%%%%%%%%%%%%%%%%

The first benchmark is an axial-vector mediator model. The interaction Lagrangian for this model is
\begin{equation}
\label{eq:LAV} 
\mathcal{L}^{\rm{int}}_{\text{axial-vector}}=- \gDM \,Z'_{\mu}\, \bar{\chi}\gamma^{\mu}\gamma_5\chi - \gq \sum_{q=u,d,s,c,b,t} Z'_{\mu} \,\bar{q}\gamma^{\mu}\gamma_5q\,.
\end{equation}
To motivate the choice of benchmark parameters, we first consider constraints from dijet and missing transverse energy searches at the LHC, then the relic density calculation, where we choose parameters to obtain the observationally inferred value through the thermal freeze-out mechanism, and finally, constraints on the scattering cross-section from direct detection experiments and IceCube.

%%%%%%%%%%%%%%%%%%%%%%%%%%%%%%%%%%%%%%%%%%%%%%%%%%%%%%%%%%
\subsubsection{LHC searches for dijets and missing transverse energy \label{subsec:dijets}}
%%%%%%%%%%%%%%%%%%%%%%%%%%%%%%%%%%%%%%%%%%%%%%%%%%%%%%%%%%

The production of an axial-vector mediator at the LHC can lead to two distinctive signatures. The first signature occurs if the axial-vector mediator decays back to partons, which shower and hadronise creating collimated jets. This signature is searched for in \emph{dijet} searches. The second signature occurs if the mediator decays to dark matter particles, which leads to signatures with missing transverse energy (MET). As the dark matter escapes the detectors unseen, the dark matter must be produced in association with Standard Model states to trigger an event at the LHC . We will focus on dark matter searches that contain an energetic jet and large MET, generically known as \emph{monojet} searches.

Existing dijet constraints already provide strong constraints on the value of~$\gq$ as a function of~$\mMed$, from the search for a local bump on top of a smooth Standard Model background in the dijet mass spectrum~\cite{An:2012va,Chala:2015ama}. In order to explore models that have an observable direct detection signal, we will limit ourselves to models with $\mMed\ll 1$~TeV. Although this mass range is difficult to constrain at the LHC owing to the large Standard Model background rate, dedicated searches provide stringent constraints. For instance in the range $\mMed\sim550$~-~$600$~GeV, the CMS searches in refs.~\cite{Khachatryan:2016ecr,Sirunyan:2016iap} both place a~95\% CL~limit on $\gq\simeq 0.07$. The ATLAS searches in Refs.~\cite{ATLAS:2016bvn,ATLAS:2016xiv} are also competitive, placing a 95\% CL~limit on $\gq\simeq 0.15$ and $\gq\simeq 0.08$ at $\mMed\sim550$~-~$600$~GeV, respectively.\footnote{There are also constraints from CDF~\cite{Aaltonen:2008dn} but these are weaker than the ATLAS and CMS limits in this mass range. For instance, at $\mMed\sim580$~GeV, CDF excludes $\gq \geq 0.2$.}

These limits assume that the axial-vector mediator cannot decay to dark matter (as occurs for $2\mDM>\mMed$). As discussed in ref.~\cite{Fairbairn:2016iuf}, the dijet signal is proportional to $g_q^2 \times \mathrm{BR}(Z'\to q \bar{q})$. Therefore, when the decay channel to dark matter is open, the dijet limits must be rescaled to take into account the smaller branching ratio~(BR) for the decay of the mediator into quarks. Using the results in ref.~\cite{Buchmueller:2014yoa} for the decay width of the axial-vector mediator, we find that for $\mDM\simeq 218$~GeV, $\mMed\simeq580$~GeV and $\gDM=1$, the CMS limits in this case are $\gq=0.11$. We therefore choose $\gq=0.1$ to maximise future signals while remaining consistent with the dijet constraints.

%%%%%%%%%%%%%%%%%%%%%%%%%%%%%%%%%%%%%%%%%%%%%%%%%%%%%%%%%%
\begin{figure}[t!]
\centering
\includegraphics[width=0.48\columnwidth]{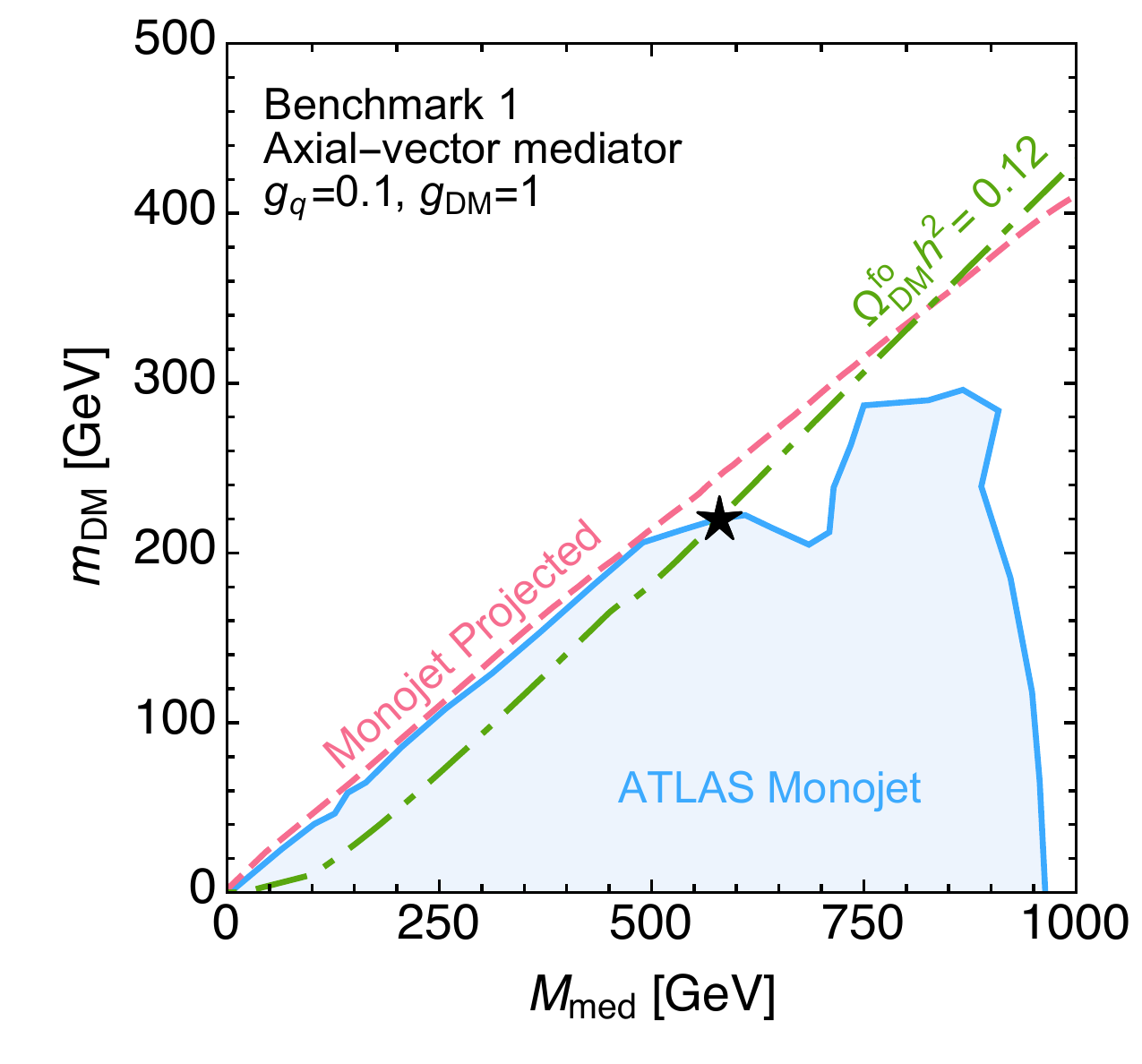} \vspace{3mm}
\includegraphics[width=0.48\columnwidth]{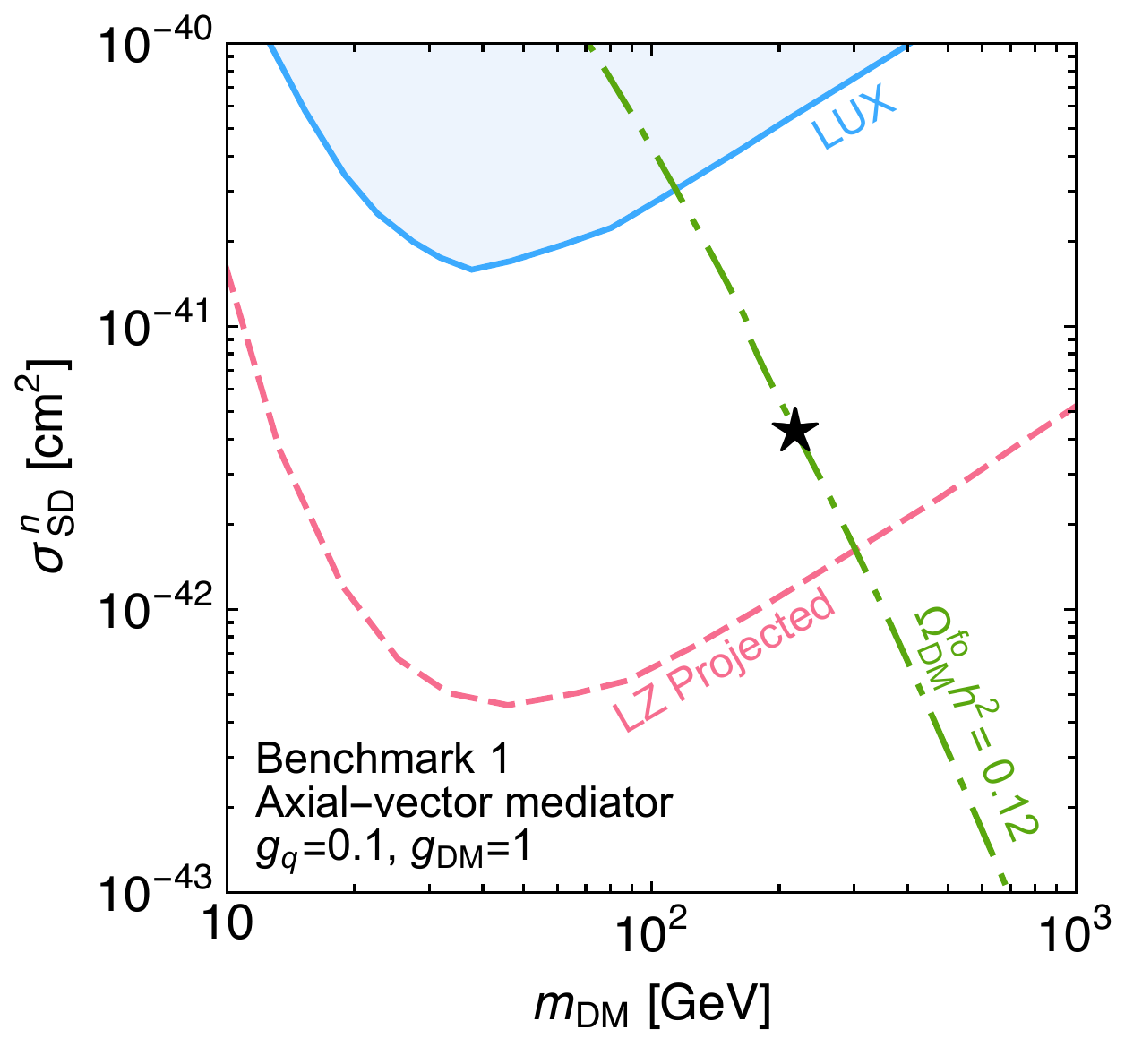}
\caption{The black star shows the benchmark point for the axial-vector mediator model (benchmark~1 in table~\ref{tab:benchmarks}) in the plane relevant for monojet searches (left panel) and direct detection searches (right panel). The blue shaded regions show the parameter space that is currently excluded (extracted from refs.~\cite{ATLASsummary,Akerib:2017kat}) while the red dashed lines show the projected sensitivity for future searches (LZ from ref.~\cite{Akerib:2016lao} and monojet constructed by us). The Panda-X and XENONnT experiments are expected to achieve the same sensitivity as LZ. The dot-dashed green line shows the parameters that produce the observationally inferred relic abundance though thermal freeze-out.
\label{fig:axial_LHC_DD}}
\end{figure}
%%%%%%%%%%%%%%%%%%%%%%%%%%%%%%%%%%%%%%%%%%%%%%%%%%%%%%%%%%

We next discuss constraints from searches involving the pair production of dark matter and either jets or photons in the final state. The ATLAS and CMS monojet searches~\cite{ATLAS:2017dnw,Aaboud:2017phn,CMS:2017tbk} provide the strongest constraint at $\mMed\sim580$~GeV. The blue shaded region in the left panel of figure ~\ref{fig:axial_LHC_DD} shows the current ATLAS monojet limit (extracted from ref.~\cite{ATLASsummary}) on the axial-vector benchmark scenario recommended in ref.~\cite{Albert:2017onk}. This benchmark is similar to our model but includes a non-zero lepton coupling~$g_l$. The 95\%~CL limit is $\mDM\simeq218$~GeV at $\mMed\simeq580$~GeV, $\gDM=1$, $\gq=0.1$ and $g_l=0.1$. As this is a search for jets, the impact of $g_l$ enters only through the mediator width. The inclusion of $g_l=0.1$ changes the width by approximately $10\%$, so this limit on $\mDM$ gives a good indication of the exclusion limit for our benchmark model. A comparable constraint is also obtained from the MET + $\gamma$ ATLAS search~\cite{Aaboud:2017dor} (where the observed limit gained from a $\gtrsim 1\sigma$ upward fluctuation above the expected limit). ATLAS again place a limit on the axial-vector benchmark scenario recommended in ref.~\cite{Albert:2017onk}. The 95\%~CL limit is $\mDM\simeq225\pm10$~GeV at $\mMed=580$~GeV, $\gDM=1$, $\gq=0.1$ and $g_l=0.1$, where the quoted error is the $1\sigma$ theory uncertainty. 

As with the dijet constraints, we choose our benchmark dark matter mass, $\mDM=218$~GeV (indicated by the black star in figure~\ref{fig:axial_LHC_DD}), to lie close to the current exclusion limit to maximise the size of a future signal. This also means that these benchmark parameters are easily within the reach of future searches, as demonstrated by the dashed red line in the left panel of figure~\ref{fig:axial_LHC_DD}, which shows our estimate of the sensitivity for future searches (using the set-up described in sections~\ref{sec:mocksignals} and \ref{sec:ML}).

%%%%%%%%%%%%%%%%%%%%%%%%%%%%%%%%%%%%%%%%%%%%%%%%%%%%%%%%%%
\subsubsection{Thermal freeze-out abundance and indirect detection}
%%%%%%%%%%%%%%%%%%%%%%%%%%%%%%%%%%%%%%%%%%%%%%%%%%%%%%%%%%

We numerically calculate the thermal freeze-out abundance with {\tt micrOMEGAs~4.3}~\cite{Barducci:2016pcb} using the standard model files available in the FeynRules model database~\cite{DMsimp_feynrules}. To motivate our choice of benchmark parameters, it is also instructive to consider the analytic calculation for the process $\chi \bar{\chi}\to Z' \to q \bar{q}$. Over a significant range of the parameter space in this model, both the $s$-wave and $p$-wave terms contribute with a similar strength to the annihilation cross-section at the time of thermal freeze-out. The full expression for the $p$-wave part is rather cumbersome (our expression agrees with the result in ref.~\cite{Busoni:2014gta}) so we only show the $s$-wave part:
\begin{equation}
\label{eq:ax-swave}
\langle \sigma^{\chi\bar{\chi}\to q \bar{q}}_{\rm{ann}} \,v \rangle_{\rm{s-wave}} \approx 9.0 \times 10^{-27}~\mathrm{cm}^3\mathrm{s}^{-1} \, \left( \frac{\gDM\, \gq}{0.1}\right)^2 \left( \frac{580~\mathrm{GeV}}{\mMed}\right)^4  \left(\frac{\beta_q}{0.61} \right)\;,
\end{equation}
where~$v$ is the relative velocity~\cite{Gondolo:1990dk} and $\beta_q = (1- m_q^2/\mDM^2)^{0.5}$. The numerical pre-factor in eq.~\eqref{eq:ax-swave} is valid for dark matter heavier than the top quark mass, which is the case for our benchmark parameters. Our numerical calculation with {\tt micrOMEGAs} finds that the freeze-out density saturates the observed value, $\Omega_{\rm{DM}}h^2 \approx 0.119$~\cite{Ade:2015xua},  when $\mDM\simeq 218$~GeV (with $\mMed=580$~GeV, $\gDM=1$, $\gq=0.1$). This is consistent with the usual approximation for the value of the annihilation cross-section, $\langle \sigma^{\chi\bar{\chi}\to q \bar{q}}_{\rm{ann}} \,v \rangle_{\rm{s-wave}} \approx 10^{-26}~\mathrm{cm}^3\mathrm{s}^{-1}$, that results in the observed abundance~\cite{Jungman:1995df}.

We also comment on indirect detection experiments searching for the product of dark matter self-annihilation since the quantity that these experiments constrain is the $s$-wave annihilation cross-section (i.e.~eq.~\eqref{eq:ax-swave}). The strongest constraint at $\mDM=218$~GeV is from the Fermi-LAT observation of dwarf spheroidal galaxies (dSphs)~\cite{Ackermann:2015zua}, where a cross-section larger than $\langle \sigma^{\chi\bar{\chi}\to q \bar{q}}_{\rm{ann}} \,v \rangle_{\rm{s-wave}}\simeq  5 \times 10^{-26}~\mathrm{cm}^3\mathrm{s}^{-1}$ is excluded at 95\% CL. This is over a factor of five higher than the $s$-wave cross-section predicted by our benchmark. Unfortunately, even the most optimistic future sensitivity studies from both Fermi-LAT (15 years observation of 60 dSphs)~\cite{Charles:2016pgz} and CTA (observing the Galactic Centre)~\cite{Silverwood:2014yza, Lefranc:2015pza} do not quite reach the cross-section predicted by our benchmark point. For this reason we will not consider indirect detection searches further in this study.

%%%%%%%%%%%%%%%%%%%%%%%%%%%%%%%%%%%%%%%%%%%%%%%%%%%%%%%%%%
\subsubsection{Direct detection and IceCube constraints}
%%%%%%%%%%%%%%%%%%%%%%%%%%%%%%%%%%%%%%%%%%%%%%%%%%%%%%%%%%

The axial-vector mediator gives rise to a spin-dependent interaction at direct detection experiments. The naive result for the scattering cross-section is
\begin{align}\label{eq:axialvec:sigmaSD}
\sigma^{\rm{naive}}_{\rm{SD}}&=\frac{3 \, \gq^2\, \gDM^2 \,\mu^2_{n\chi}}{\pi \mmed^4} \left(\Delta^{(p,n)}_u + \Delta^{(p,n)}_d+ \Delta^{(p,n)}_s  \right)^2\\
&\simeq 2.9\times 10^{-42}~\mathrm{cm}^2\cdot\left(\frac{\gq \, \gDM}{0.1}\right)^2\left( \frac{580 \, \mathrm{GeV}}{M_{\rm{med}}}\right)^4\left(\frac{\mu_{n\chi}}{0.93 \, \mathrm{GeV}} \right)^2\,,
\end{align}
where $\mu_{n\chi}= m_n \mDM/(m_n+\mDM)$ is the dark matter-nucleon reduced mass, $m_n \simeq 0.939 \, \mathrm{GeV}$ is the nucleon mass and the $\Delta^{(p,n)}_{u, d, s}$ factors parametrise the quark spin content of the nucleon. We have used the PDG values: $\Delta^{(p)}_u= \Delta^{(n)}_d = 0.84$, $\Delta^{(p)}_d = \Delta^{(n)}_u = -0.43$ and $\Delta_s=-0.09$~\cite{Agashe:2014kda}. 

For this naive result, both the proton-only and neutron-only spin-dependent cross-sections are the same. However, in order to consistently compare the results of LHC and direct detection experiments, we must account for the running of the axial-vector couplings when comparing the results at different scales (see e.g.\ refs.~\cite{Crivellin:2014qxa, DEramo:2014nmf}). While the LHC and annihilation scale is $\mathcal{O}(\mMed)$, typically hundreds of~GeV, the typical direct detection scale is only $\mathcal{O}(100)$~MeV. We use {\tt runDM}~\cite{DEramo:2016gos} to consistently calculate the low-energy proton-only and neutron-only scattering cross-section.\footnote{The {\tt runDM} code captures the leading effects of the running but does not account for the smaller correction from the $q^2$ dependence in the Wilson coefficients of the non-relativistic effective field theory, which is relevant for the axial-vector mediator~\cite{Hoferichter:2015ipa, Bishara:2016hek, Bishara:2017pfq, Bishara:2017nnn}.} Compared to the naive cross-section, we find that for our benchmark points, the neutron-only cross-section increases to $\sigma^{n}_{\rm{SD}}=4.2\times 10^{-42}~\rm{cm}^2$, while the proton-only cross-section decreases to $\sigma^{p}_{\rm{SD}}=1.9\times 10^{-42}~\rm{cm}^2$. 

Both of these cross-sections lie below any current exclusion limits from LUX~\cite{Akerib:2017kat}, PandaX-II~\cite{Fu:2016ega} or PICO-60~\cite{Amole:2017dex} (see right panel of figure~\ref{fig:axial_LHC_DD}). The value of~$\sigma^{n}_{\rm{SD}}$ is easily within reach of the LZ~\cite{Mount:2017qzi}, Panda-X~\cite{Cao:2014jsa} and XENONnT~\cite{Aprile:2015uzo} multi-tonne liquid xenon experiments, as demonstrated by the red dashed line in the right panel of figure~\ref{fig:axial_LHC_DD} (which was extracted from ref.~\cite{Akerib:2016lao}). This cross-section may also be observable with a larger xenon experiment such as DARWIN~\cite{Aalbers:2016jon} in the channel where the $^{129}$Xe nucleus is excited into a low-lying state, with subsequent prompt de-excitation~\cite{Baudis:2013bba,McCabe:2015eia}. The benchmark value of~$\sigma^{p}_{\rm{SD}}$ lies more than an order of magnitude below current IceCube exclusion limits~\cite{Aartsen:2016zhm} and is unlikely to be probed with future IceCube searches. However, it is just within the projected reach of PICO-500~\cite{PICO500}.

%%%%%%%%%%%%%%%%%%%%%%%%%%%%%%%%%%%%%%%%%%%%%%%%%%%%%%%%%%
\subsection{Benchmark 2: Scalar mediator \label{sub:smodels}}
%%%%%%%%%%%%%%%%%%%%%%%%%%%%%%%%%%%%%%%%%%%%%%%%%%%%%%%%%%
The second benchmark is the simplified model with a scalar mediator. The interaction Lagrangian for this model is
\begin{equation}
\label{eq:Scalar}
\mathcal{L}^{\rm{int}}_{\text{scalar}}=- \gDM  \hspace{0.25mm}  \phi \bar{\chi}\chi-  \gq  \hspace{0.5mm} \sum_{q=u,d,s,c,b,t}  \frac{m_q}{v_{\rm{EW}}} \hspace{0.25mm} \phi \bar{q}q \,, 
\end{equation}
where $m_q$ is the quark mass and $v_{\rm{EW}}\simeq246$~GeV is the Higgs vacuum expectation value. In this benchmark, we choose parameters that will result in a monojet signal and direct detection signal at upcoming experiments but which predict a value of the dark matter density from thermal freeze-out that is significantly different from the observationally inferred value. This implies that additional particle physics not included in the simplified model or a modified cosmological history is required to explain the observed dark matter abundance. Here, we won't specify the precise mechanism but see for instance,~refs.~\cite{Gelmini:2006pq,Rehagen:2015zma} for examples of how the relic density could be brought into the observed range.

%%%%%%%%%%%%%%%%%%%%%%%%%%%%%%%%%%%%%%%%%%%%%%%%%%%%%%%%%%
\begin{figure}[t!]
\centering
\includegraphics[width=0.48\columnwidth]{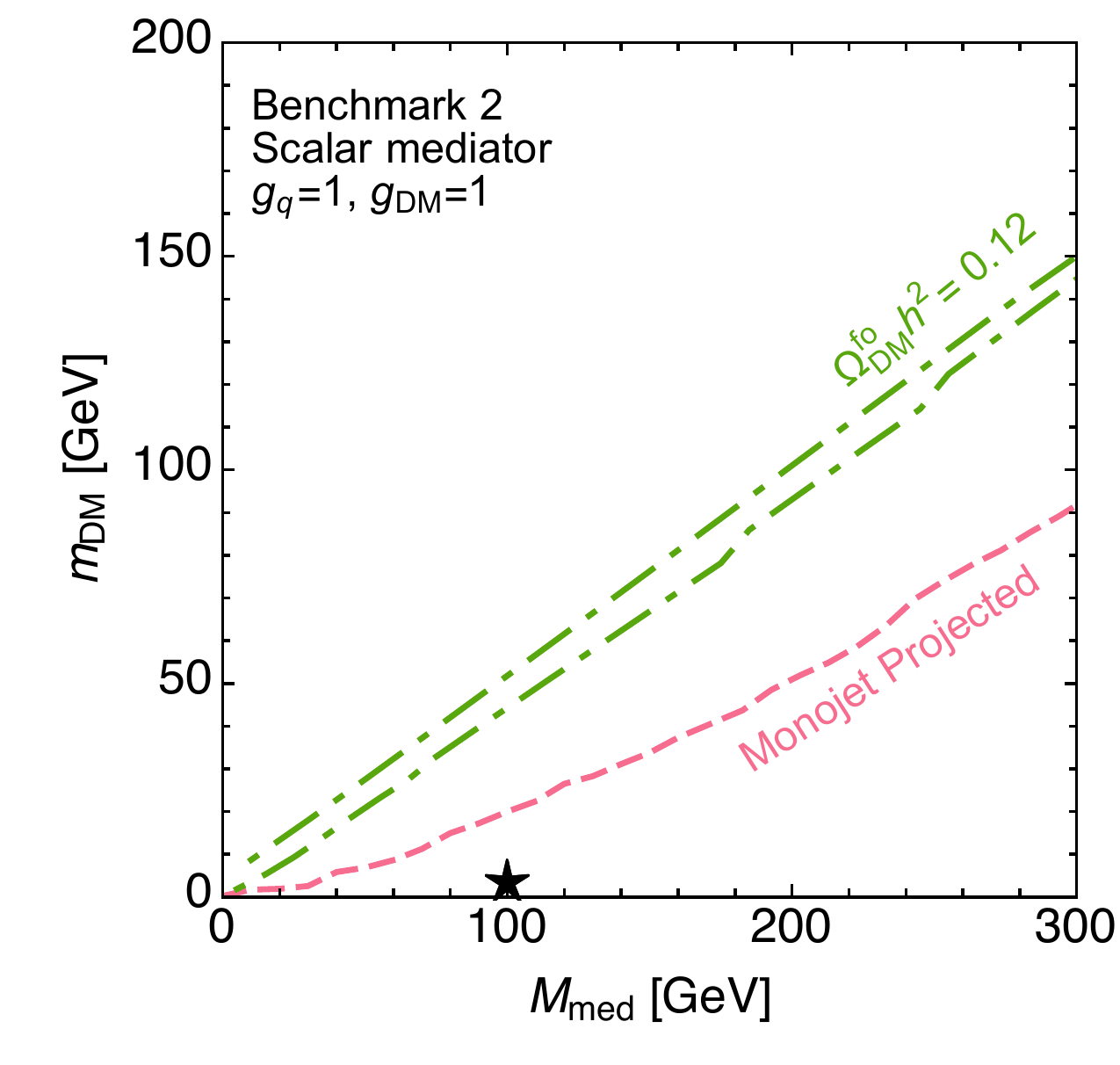} \vspace{3mm}
\includegraphics[width=0.48\columnwidth]{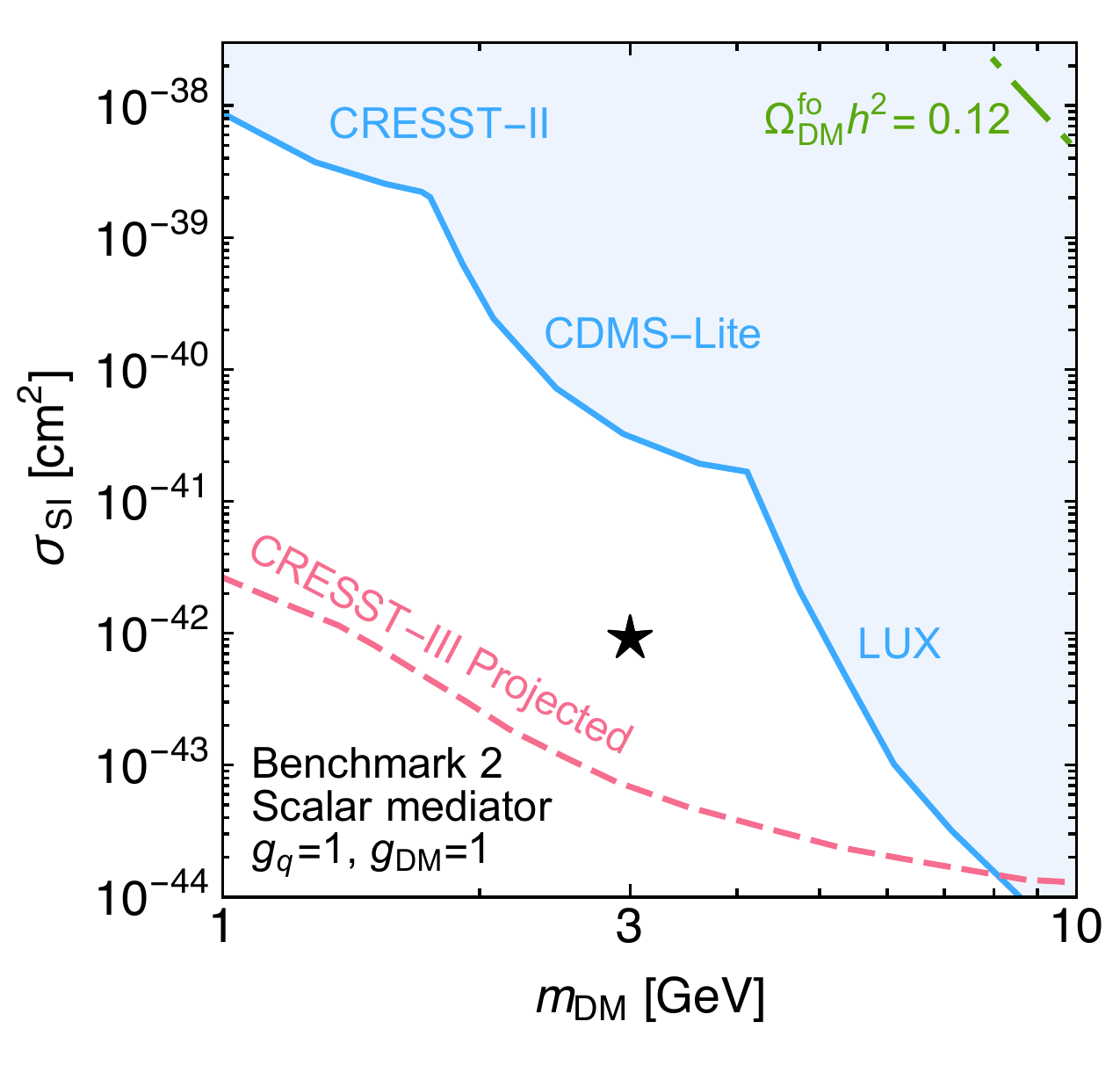}
\caption{
The black star shows the benchmark point for the scalar mediator model (benchmark~2 in table~\ref{tab:benchmarks}) in the plane relevant for monojet searches (left panel) and direct detection searches (right panel). The blue shaded regions show the parameter space that is currently excluded (extracted from refs.~\cite{Akerib:2016vxi,Agnese:2017jvy,Angloher:2015ewa}). There are no constraints on the parameter space in the left panel from current monojet searches. The red dashed lines show the projected sensitivity for future searches (LZ from ref.~\cite{Angloher:2015eza} and monojet constructed by us). Our benchmark point does not lie on the dot-dashed green line, which shows the parameters that produce the observationally inferred relic abundance though thermal freeze-out, implying that the dark matter abundance is obtained through a mechanism other than freeze-out.
\label{fig:scalar_LHC_DD}}
\end{figure}
%%%%%%%%%%%%%%%%%%%%%%%%%%%%%%%%%%%%%%%%%%%%%%%%%%%%%%%%%%

%%%%%%%%%%%%%%%%%%%%%%%%%%%%%%%%%%%%%%%%%%%%%%%%%%%%%%%%%%
\subsubsection{LHC searches for dijets and missing transverse energy}
%%%%%%%%%%%%%%%%%%%%%%%%%%%%%%%%%%%%%%%%%%%%%%%%%%%%%%%%%%

The smallness of the scalar coupling to light quarks $(m_{u,d}/v_{\rm{EW}}\approx 10^{-2})$ means that there are no dijet constraints on this model, nor are there likely to be constraints in the future.

For MET~+~X searches, the quark-mass dependence of the scalar-quark coupling means that the most relevant coupling is to the heaviest quarks~\cite{Haisch:2012kf}. In practice, this means that diagrams that contribute most to the production cross-section have gluons in the initial state. The most sensitive searches are monojet and MET + $\bar{t}t (\bar{b}b)$ searches, which are expected to have a similar sensitivity for small values of~$\mDM$ and~$\mMed$ (see e.g.~refs.~\cite{Aaboud:2017phn, Aaboud:2017aeu}). For $\gq=\gDM=1$, these searches do not currently provide any constraints on~$\mMed$ or~$\mDM$. For instance, the CMS monojet~\cite{Sirunyan:2017hci} and MET~+~$\bar{t}t,(\bar{b}b)$~\cite{Sirunyan:2017xgm} searches are currently sensitive to production cross-sections a factor of two larger than in this model with the parameters $\gq=\gDM=1$, $\mMed \lesssim 100$~GeV and $\mDM\ll \mMed$. For this reason, the left panel of figure~\ref{fig:scalar_LHC_DD} does not contain any current monojet constraints. Our benchmark point is however within reach of future searches (described in further detail in section~\ref{sec:mocksignals}), as demonstrated by the dashed red line in the left panel of figure~\ref{fig:scalar_LHC_DD}

%%%%%%%%%%%%%%%%%%%%%%%%%%%%%%%%%%%%%%%%%%%%%%%%%%%%%%%%%%
\subsubsection{Thermal freeze-out abundance and indirect detection}
%%%%%%%%%%%%%%%%%%%%%%%%%%%%%%%%%%%%%%%%%%%%%%%%%%%%%%%%%%

The smallness of the scalar coupling to light quarks also has important implications for the freeze-out abundance in this model. For this model, the $s$-wave part of the annihilation cross-section for $\chi \bar{\chi} \to \phi \to q \bar{q}$ is zero so no indirect detection signal is predicted. We again performed the freeze-out calculation numerically with {\tt micrOMEGAS~4.3}, and used our own calculation as a check of the {\tt micrOMEGAS} result. The $p$-wave annihilation cross-section is generally very small:
\begin{equation}
\label{eq:scalar-ann}
\begin{split}
\sigma^{\chi\bar{\chi}\to q \bar{q}}_{\rm{ann}} \,v  \sim 2 \times 10^{-31}~\mathrm{cm}^3\mathrm{s}^{-1} \, &\left( \frac{\gDM\, \gq}{1}\right)^2 \left( \frac{100~\mathrm{GeV}}{\mMed}\right)^4\\
&\quad \times \left( \frac{m_c/v_{\rm{EW}}}{0.005}\right)^2 \left( \frac{\mDM}{3~\mathrm{GeV}}\right)^2 \left(\frac{\beta_q}{0.91} \right)^3 \left(\frac{v}{0.3} \right)^2\;,
\end{split}
\end{equation}
so the relic abundance from thermal freeze-out is generally significantly larger than the observed value since, schematically,
\begin{equation}
\label{eq:omDMsch}
\Omega_{\rm{DM}}h^2\,[\text{freeze-out}] \sim 0.1 \times\frac{10^{-26}~\mathrm{cm}^3\mathrm{s}^{-1}}{\sigma^{\chi\bar{\chi}\to q \bar{q}}_{\rm{ann}} \,v}\;.
\end{equation}

The annihilation to gluons also contributes to the total annihilation cross-section and could in principle play a role~\cite{Haisch:2015ioa}. At our benchmark point, however, we find that $\sigma^{\chi\bar{\chi}\to q \bar{q}}_{\rm{ann}} \,v \approx 6.3 \,\sigma^{\chi\bar{\chi}\to g g}_{\rm{ann}} \,v$, so the annihilation to gluons plays a negligible role. In addition, for $\mDM\lesssim 1 $~GeV, the description of the annihilation process in terms of quarks and gluons begins to break down since QCD becomes non-perturbative. For masses above this however, eq.~\eqref{eq:scalar-ann} is still valid.

%%%%%%%%%%%%%%%%%%%%%%%%%%%%%%%%%%%%%%%%%%%%%%%%%%%%%%%%%%
\subsubsection{Direct detection constraints}
%%%%%%%%%%%%%%%%%%%%%%%%%%%%%%%%%%%%%%%%%%%%%%%%%%%%%%%%%%

The scalar mediator gives rise to a spin-independent interaction at direct detection experiments, with an interaction strength with protons and neutrons that is the same (to an excellent approximation). The nucleon scattering cross-section is
\begin{align}
\label{eq:DDSI}
\sigma_{\rm{SI}} &= \frac{\gq^2 \gDM^2}{\pi} \frac{f_N^2 m_n^2}{v_{\rm{EW}}^2} \frac{\mu^2_{n\chi}}{\mmed^4}\,\\
& \simeq 1.5\times 10^{-42}~\mathrm{cm}^2\cdot\left(\frac{\gq \hspace{0.25mm} \gDM}{1}\right)^2\left( \frac{100 \, \mathrm{GeV}}{\mmed}\right)^4\left(\frac{\mu_{n\chi}}{0.93 \, \mathrm{GeV}} \right)^2\,,
\end{align}
where $f_N=0.308$~\cite{Hoferichter:2017olk} parametrises the Higgs coupling to the nucleons in the direct detection target.\footnote{The running of the scalar coupling between the scale probed at the LHC and direct detection experiments is negligible for scalar mediators so can be safely ignored.
} From this result, we see that the dominant constraints on this model arise from direct detection experiments, since the xenon-based experiments LUX~\cite{Akerib:2016vxi}, Panda-X~\cite{Fu:2016ega} and XENON1T~\cite{Aprile:2017iyp} exclude $\sigma_{\rm{SI}}\gtrsim 10^{-46}~\mathrm{cm}^2$ for $\mDM\gtrsim 10~\mathrm{GeV}$. To avoid these constraints, we consider $\mDM=3~\mathrm{GeV}$ as our benchmark dark matter mass. This avoids the constraints from xenon-based experiments, CDMS-Lite~\cite{Agnese:2017jvy} and CRESST-II~\cite{Angloher:2015ewa} while being within reach of upcoming dedicated low-mass direct detection experiments. We focus on the projected sensitivity from CRESST-III~\cite{Angloher:2015eza}, demonstrated in the right panel of figure~\ref{fig:scalar_LHC_DD}, while SuperCDMS is also projected to be sensitive to our benchmark parameters~\cite{Agnese:2016cpb}.

In summary, the scalar mediator benchmark model is an interesting example of a model that predicts observable LHC and direct detection signals but where it is difficult to envisage that the thermal freeze-out mechanism is solely responsible for the dark matter relic abundance.

%%%%%%%%%%%%%%%%%%%%%%%%%%%%%%%%%%%%%%%%%%%%%%%%%%%%%%%%%%
\subsection{Potential limitations of simplified models}
%%%%%%%%%%%%%%%%%%%%%%%%%%%%%%%%%%%%%%%%%%%%%%%%%%%%%%%%%%

Before proceeding, it is worth pausing to consider some of the potential limitations of simplified models (for a more comprehensive discussion, see e.g.\ refs.~\cite{Abdallah:2015ter, Bell:2016ekl, Kahlhoefer:2017dnp, Cui:2017juz}).  As mentioned in the Introduction, simplified models are motivated from a bottom-up perspective so at the very least, they should provide a complete description of the low-energy physics. 

In the context of our study, the models should be valid at the TeV-scale (energies probed by the LHC and in self-annihilation channels). Satisfying this criteria also implies that the models will be valid at direct detection experiments, where the energy scales that are probed are much lower. For a model with an axial-vector mediator to be valid at the TeV-scale, $\mDM$ should satisfy $\mDM \leq \sqrt{\pi/2} \,\mMed/\gDM$ so that the mediator coupling remains perturbative~\cite{Kahlhoefer:2015bea,Duerr:2016tmh}. Furthermore, the couplings should be less than $\mathcal{O}(4\pi)$ so that perturbative unitarity of the scattering amplitude is not violated at the LHC~\cite{Englert:2016joy}. The parameters used in benchmark~1 satisfy both of these criteria (cf.~table~\ref{tab:benchmarks}).

The simplified models must also be consistent with flavour and CP-constraints. We follow the standard procedure for avoiding these constraints by imposing the Minimal Flavour Violation (MFV) structure on the couplings~\cite{DAmbrosio:2002vsn}. For the axial-vector mediator, this meant that the couplings $\gq$ were chosen to be independent of quark-flavour, while for the scalar mediator, the couplings $\gq$ were proportional to the quark Yukawa couplings (equivalent to~$m_q/v_{\rm{EW}})$~\cite{Abdallah:2015ter}.

Further restrictions can also be derived from a top-down perspective. A fully-consistent UV completion of an axial-vector model requires the absence of gauge anomalies. The minimal model with an axial-vector mediator that we consider does not satisfy this requirement~\cite{Liu:2011dh, Duerr:2013dza, Perez:2014qfa, Ekstedt:2016wyi, Jacques:2016dqz, Ismail:2016tod, Ellis:2017tkh, Ismail:2017ulg, Dror:2017nsg} so requires either non-zero couplings to leptons, which results in stringent constraints from di-lepton searches, or the presence of additional particles that cancel the anomaly. As these additional particles can be heavy and beyond the reach of the LHC ($\sim10$~TeV)~\cite{Ismail:2016tod}, their impact on the low-energy signals that we consider are small.  This is the scenario that we envisage in this study. Additionally, UV complete models of the axial-vector simplified model may allow for extra contributions to the total annihilation cross-section and therefore alter the thermal freeze-out calculation~\cite{Bell:2016uhg, Duerr:2017uap}. We do not consider any extra contributions in this work, assuming that they would be sub-dominant to the process $\chi \bar{\chi}\to Z' \to q \bar{q}$. With all of the caveats mentioned above in mind, we now proceed to give the details of our procedure for generating signals at upcoming experiments.

%%%%%%%%%%%%%%%%%%%%%%%%%%%%%%%%%%%%%%%%%%%%%%%%%%%%%%%%%%
%%%%%%%%%%%%%%%%%%%%%%%%%%%%%%%%%%%%%%%%%%%%%%%%%%%%%%%%%%
%%%%%%%%%%%%%%%%%%%%%%%%%%%%%%%%%%%%%%%%%%%%%%%%%%%%%%%%%%
%%%%%%%%%%%%%%%%%%%%%%%%%%%%%%%%%%%%%%%%%%%%%%%%%%%%%%%%%%
\section{Generating mock signals at upcoming experiments \label{sec:mocksignals}}
%%%%%%%%%%%%%%%%%%%%%%%%%%%%%%%%%%%%%%%%%%%%%%%%%%%%%%%%%%

In the previous section, we established that the benchmark parameters are outside current exclusion limits yet are within the reach of near future experiments. In this section we describe our numerical implementation of current and future LHC searches, together with our assumptions for upcoming runs of the LZ and CRESST-III direct detection experiments.

\subsection{Simulating LHC signals}
For both benchmark scenarios we focus on the production of a pair of dark matter particles recoiling against hard initial state radiation, i.e.\
\begin{equation}
pp\to Z^{\prime(*)}+j\;\;\text{or}\;\;\phi^{(*)}+j \to \chi\bar{\chi}+j\,,
\end{equation}
where $^{(*)}$ denotes off-shell $s$-channel resonances. Since the dark matter particles do not interact with the detector, the corresponding signature is a large transverse momentum jet back-to-back to the MET. In general, the monojet signal is accompanied by additional softer jets due to strong initial state QCD bremsstrahlung. As a consequence the experimental collaborations allow for additional jets.

%%%%%%%%%%%%%%%%%%%%%%%%%%%%%%%%%%%%%%%%%%%%%%%%%%%%%%%%%%
\subsubsection{Current monojet searches \label{subsec:currentmono}}
%%%%%%%%%%%%%%%%%%%%%%%%%%%%%%%%%%%%%%%%%%%%%%%%%%%%%%%%%%

We use the results from the 13 TeV ATLAS monojet analysis with an integrated luminosity of~3.2~fb$^{-1}$~\cite{Aaboud:2016tnv} as a constraint in our parameter scan.\footnote{While carrying out this work, ATLAS and CMS published updated results with~36~fb$^{-1}$~\cite{ATLAS:2017dnw,CMS:2017tbk}. The updated ATLAS analysis uses similar signal regions to~\cite{Aaboud:2016tnv}. The new exclusion limits are stronger for light dark matter particles but for our axial-vector benchmark point, the increase in sensitivity is relatively small.} This analysis consists of seven inclusive signal regions (IM1-IM7) as well as six exclusive signal regions (EM1-EM6) with increasing cuts on the MET ranging from 250~GeV to 700~GeV. We make use of the exclusive signal regions as they allow us to take the shape of the MET distribution as input. All signal regions demand a leading jet with a transverse momentum $(p_T)$ of at least 250~GeV. Moreover, a lepton veto is applied and events with more than four jets with $p_T>30$ GeV are rejected. Finally, a cut is imposed on the azimuthal angle between the missing transverse momentum and all jets in order to suppress the QCD background.

We generated parton level signal events for the axial-vector and the scalar mediator scenario with the next--to--leading (NLO) event generator {\tt POWHEG BOX V2} \cite{Nason:2004rx,Frixione:2007vw,Alioli:2010xd,Haisch:2015ioa,Haisch:2013ata}. We applied a parton level cut on the leading jet of 150 GeV while keeping all other settings to their default values. The parton level events are correctly matched with the parton shower of {\tt Pythia~8} \cite{Sjostrand:2007gs} with the NLOPS scheme.  We have simulated 50000 Monte Carlo events for each point of the scalar and axial-vector mediators models and used the {\tt NNPDF3.0~NLO}~\cite{Ball:2014uwa} parton distribution functions. The scale uncertainty is roughly~$10\%$. 

The search has been implemented within the {\tt CheckMATE~2.0.13} framework \cite{Drees:2013wra,Kim:2015wza,Dercks:2016npn}, which is based on the fast detector simulation {\tt Delphes~3.4.0} \cite{deFavereau:2013fsa} with modified detector settings of the ATLAS detector and relies on {\tt FastJet~3.3.0} for jet clustering \cite{Cacciari:2011ma}. The search is fully validated and the validation notes can be found on the official web page~\cite{checkmatewebpage}. With {\tt CheckMATE}, we can determine the number of signal events for all signal regions and we assume a theoretical error of 10\% on the signal. Together with the NLO cross-section from {\tt POWHEG}, we can determine whether the signal can be seen above the Standard Model background. 

%%%%%%%%%%%%%%%%%%%%%%%%%%%%%%%%%%%%%%%%%%%%%%%%%%%%%%%%%%
\subsubsection{Future monojet searches  \label{subsec:futuremono}}
%%%%%%%%%%%%%%%%%%%%%%%%%%%%%%%%%%%%%%%%%%%%%%%%%%%%%%%%%%

We assume that a future monojet search will see an excess from our benchmark models. In our implementation of future LHC searches, we assume an integrated luminosity of 100~fb$^{-1}$ at a centre-of-mass energy of 13~TeV. This scenario corresponds to the expected integrated luminosity collected by the end of LHC Run~2.

We estimate the sensitivity of future monojet searches by extrapolating the current ATLAS search to higher integrated luminosities while keeping the same signal region definitions. An advantage of this approach is that we can rescale the number of predicted Standard Model background events given by the ATLAS collaboration. When rescaling the integrated luminosity from 3.2~fb$^{-1}$ to~100~fb$^{-1}$, the number of background events are scaled with the ratio~100/3.2. This ansatz assumes that no significant detector upgrades will be made or that there will be a notable change in pileup.

The projected sensitivity depends on the uncertainty of the predicted Standard Model rates in the respective signal regions. Recent ATLAS monojet searches show that good improvements in reducing systematic uncertainties have been achieved (e.g.\ from around~5\% to~3\% from ref.~\cite{Aaboud:2016tnv} to ref.~\cite{ATLAS:2017dnw}). It is therefore reasonable to assume that this error will continue to decrease with additional data. Unfortunately, the overall systematic error and its relative contribution to the statistical uncertainty is difficult to predict for future searches. Owing to this uncertainty, we therefore consider two scenarios; in the first scenario, we assume that the systematic error improves to 1\%, while in the second scenario, we assume 3\%. As the current uncertainty is around~3\%, this second scenario corresponds to the pessimistic assumption that the systematic error undergoes no further reduction.

%%%%%%%%%%%%%%%%%%%%%%%%%%%%%%%%%%%%%%%%%%%%%%%%%%%%%%%%%%
\subsubsection{Implementation of other LHC constraints  \label{subsec:LHCothers}}
%%%%%%%%%%%%%%%%%%%%%%%%%%%%%%%%%%%%%%%%%%%%%%%%%%%%%%%%%%

So far, we have only considered the monojet signature. However, as discussed in Section~\ref{subsec:dijets}, the mediator can also decay back to partons and thus resonant dijet searches can provide strong constraints. We implemented a trigger based analysis which is sensitive to resonance masses as low as 425~GeV~\cite{ATLAS:2016xiv}. The selection criteria are similar to common dijet searches with the exception that the leading trigger jets must have $p_T>185$~GeV.

We also implemented ref.~\cite{ATLAS:2016bvn} that considered very light resonances. This analysis focuses on dijet resonances which are produced in association with a photon or a jet. Apart from the dijet selection cuts they demand $p_T (\gamma)>150$~GeV and $p_T (j)>430$~GeV. If the mediator mass is too small, then the extra jet cannot be separated from the dijet configuration and a lower limit of 350~GeV on the resonance mass must be imposed.

We have implemented both ATLAS searches within the {\tt CheckMATE} framework. The MC events are generated with {\tt MadGraph5aMC@NLO} \cite{Alwall:2014hca} with the UFO model files obtained by the {\tt FeynRules} implementation \cite{Alloul:2013bka}. The parton level events are further passed to {\tt Pythia~8}. We impose an upper limit on the total width of $\Gamma_{\rm{med}}\le0.1 \mMed$, i.e.\ we only test a model point against dijet constraints if the width-to-mass ratio is less than~10\%. We use these searches only to constrain our model i.e.~we do not model an excess in these searches with an integrated luminosity of 100~fb$^{-1}$.

Finally, we have not implemented MET~+~$\bar{t}t\,(\bar{b}b)$ searches for the scalar model, or the MET + $\gamma$ search for the axial-vector model since the expected sensitivity of these searches is comparable to or lower than from the monojet search~\cite{Haisch:2015ioa}.

%%%%%%%%%%%%%%%%%%%%%%%%%%%%%%%%%%%%%%%%%%%%%%%%%%%%%%%%%%
%%%%%%%%%%%%%%%%%%%%%%%%%%%%%%%%%%%%%%%%%%%%%%%%%%%%%%%%%%
%%%%%%%%%%%%%%%%%%%%%%%%%%%%%%%%%%%%%%%%%%%%%%%%%%%%%%%%%%

%%%%%%%%%%%%%%%%%%%%%%%%%%%%%%%%%%%%%%%%%%%%%%%%%%%%%%%%%%
\subsection{Simulating direct detection signals \label{sec:simDD}}
%%%%%%%%%%%%%%%%%%%%%%%%%%%%%%%%%%%%%%%%%%%%%%%%%%%%%%%%%%

The values of the parameters for benchmarks 1 and 2 are chosen such that they are within the reach of the next generation of direct detection experiments: LZ and CRESST-III for benchmark~1 and~2, respectively. Below we describe how we simulate the dark matter signal in each of these experiments. For the dark matter distribution in the Galaxy, we assume the so-called Standard Halo Model with canonical parameter choices. Namely, a Maxwell-Boltzmann velocity distribution truncated at the escape speed of 544~km/s, a local dark matter density of 0.3~GeV$/$cm$^3$, and a local circular speed of 220~km/s. We do not take into account the effect of astrophysical uncertainties on the parameter reconstruction in this work, which could play a role for the scalar mediator benchmark where $\mDM=3$~GeV (see e.g.~refs.~\cite{McCabe:2010zh, Bozorgnia:2016ogo}).

%%%%%%%%%%%%%%%%%%%%%%%%%%%%%%%%%%%%%%%%%%%%%%%%%%%%%%%%%%
\subsubsection{LZ} \label{sec:lz}
%%%%%%%%%%%%%%%%%%%%%%%%%%%%%%%%%%%%%%%%%%%%%%%%%%%%%%%%%%

The LZ (LUX-ZEPLIN)~\cite{Akerib:2015cja,Mount:2017qzi} direct detection experiment is a next generation liquid xenon experiment. It is the successor to the LUX experiment~\cite{Akerib:2016vxi} and is expected to start operation during~2020. It will search for the scintillation and ionization signals produced by a recoiling nucleus due to the scattering of a dark matter particle in the detector. The XENONnT and upgraded PandaX experiments are expected to have a similar sensitivity to the LZ experiment. Owing to the more detailed LZ design proposals~\cite{Akerib:2015cja,Mount:2017qzi}, we focus on~LZ but our results are expected to also apply to XENONnT and PandaX.

We assume an exposure of 1000~days and 5.6~tonnes of~Xe, and a nuclear recoil energy range of $[6,30]$ keV~\cite{Akerib:2015cja}. We adopt an energy resolution of $\sigma_{\rm det} (E)=0.065~{\rm keV} (E/{\rm keV}) +0.24~{\rm keV} \sqrt{E/{\rm keV}}$, obtained from a fit to the black points (the most LZ-like scenario) in the right panel of figure~1 in ref.~\cite{Schumann:2015cpa}. We assume 2.37~background events for a $5.6 \times 10^6$~kg~day exposure in the energy range $[6,30]$~keV, as given in table 3.8.1.1 of ref.~\cite{Akerib:2015cja}. This corresponds to a rate of $1.76 \times 10^{-8}$ counts$/$(keV kg day), assuming that the background rate is constant in energy. This assumption is reasonable since the LZ background rate is dominated by $pp$ solar-neutrinos and~Rn and~Kr radio-nuclides, all of which are approximately constant in energy for the low-energy range used in our study (see e.g.\ ref.~\cite{McCabe:2015eia}). The backgrounds from $^{136}$Xe decay and neutron scattering are energy dependent but are sub-dominant. To simplify our analysis, We assume a 100\% nuclear recoil efficiency above 6~keV, which is consistent with LZ simulations~\cite{McCabe:2017rln}. Furthermore, we only consider events in the lower half of the nuclear recoil band so include a~50\% acceptance of the nuclear recoils. We use the Xe nuclear structure functions for spin-dependent dark matter-nucleus scattering from ref.~\cite{Klos:2013rwa}, and assume an isotope abundance for $^{129}$Xe and $^{131}$Xe of 26.4\% and 21.2\%, respectively.

In our analysis we consider six energy bins of width 4~keV in the range of $[6,30]$ keV. For our benchmark parameters, $\mDM=218$~GeV and $\sigma_{\rm{SD}}^n=4.2 \times 10^{-42}$~cm$^2$ (cf.\ table~\ref{tab:benchmarks}) we obtain 3.94, 2.96, 2.22, 1.68, 1.28 and 0.98 events in each 4~keV energy bin, respectively. This compares to a constant background count rate of 0.394 in each energy bin.

%%%%%%%%%%%%%%%%%%%%%%%%%%%%%%%%%%%%%%%%%%%%%%%%%%%%%%%%%%
\subsubsection{CRESST-III} \label{sec:cresst}
%%%%%%%%%%%%%%%%%%%%%%%%%%%%%%%%%%%%%%%%%%%%%%%%%%%%%%%%%%

The CRESST experiment uses CaWO$_4$ crystals to search for the scintillation and phonon signals due to  the scattering of dark matter particles with the target nuclei in the detector. The CRESST-II Phase 2 experiment released two major results during the last few years: one with the TUM40 detector that reached an overall background of 3.5 counts$/$(keV kg day)~\cite{Angloher:2014myn}, and the other with the Lise detector which reached the low energy threshold of 0.3 keV~\cite{Angloher:2015ewa}  More recently, the first results with CRESST-III detectors have been presented~\cite{Petricca:2017zdp}. These detectors achieved the design low energy threshold of 0.1~keV. Taken together, the CRESST results set the strongest exclusion limit on the spin-independent dark matter-nucleon cross-section in the mass range between approximately 0.4 GeV and 1.7 GeV (ref.~\cite{Agnese:2017jvy} and~\cite{Dolan:2017xbu} set exclusion limits above and below this mass range respectively).

In this study, we adopt the detector configuration of CRESST-III Phase~3, which is expected to deliver data in 2019--2020~\cite{Angloher:2015eza}. CRESST-III Phase~3 is projected to consist of~100 24~\!g detectors that have a nuclear recoil energy threshold of 100~eV. The projected background rate is 0.035~counts$/$(keV\,kg\,day), which is approximately 100 times smaller than the rate in Phase~2~\cite{Angloher:2014myn}. The expected sensitivity of this next generation CRESST experiment is shown in figure~6 of ref.~\cite{Angloher:2015eza}: it is expected to be sensitive to $\sigma_{\rm{SI}} \approx 7\times10^{-44}~\mathrm{cm}^2$ at~$\mDM=3$~GeV, approximately an order of magnitude lower than our benchmark cross-section (cf.\ table~\ref{tab:benchmarks}). In our simulation, we assume an exposure of 250~kg~day, which is sufficient to discover our benchmark point with approximately six months of data taking. This exposure is also well within the CRESST collaboration's total projection of a 1000~kg~day exposure from two years of data taking. 

The energy resolution obtained with the Lise detector depended linearly on the deposited energy with a slope of 6.6~eV$/$keV~\cite{Angloher:2015ewa}. We assume a similar behaviour for CRESST-III and adopt a detector resolution of $\sigma_{\rm{det}}(E)=20~{\rm eV} + 6.6~{\rm eV} (E/{\rm keV})$~\cite{cresst-pv}. We adopt a detection efficiency of~56\% down to the energy threshold~\cite{cresst-pv}, which is the asymptotic value of the TUM40 detector efficiency and a reasonable expectation for the new detectors~\cite{Angloher:2014myn}. To obtain the final dark matter signal rate, we multiply this detection efficiency by~50\% to take into account the nuclear recoil acceptance region (see figure~5 of ref.~\cite{Angloher:2015ewa}). 

In our simulation we consider the following configuration of energy bins in the range of $[0.1,10]$~keV, such that we obtain an observable signal over background for our benchmark dark matter model given the exposure, energy resolution, and detection efficiency discussed above: $[0.1,0.2]$~keV, $[0.2,0.3]$~keV, $[0.3,1]$~keV, $[1,5]$~keV, and $[5,10]$~keV. For our benchmark parameters,  $\mDM=3$~GeV and  $\sigma_{\rm{SI}}=8.8 \times 10^{-43}$~cm$^2$, the number of signal events in each energy bin is 5.1, 1.7, 1.1, 0.17, and $2.3 \times 10^{-5}$, respectively. This compares to a background count rate of 0.875, 0.875, 6.13, 35.0, and 43.8, respectively. We have checked that increasing the number of energy bins beyond the five~bins considered here, or increasing the energy range beyond 10~keV does not change our results.

%%%%%%%%%%%%%%%%%%%%%%%%%%%%%%%%%%%%%%%%%%%%%%%%%%%%%%%%%%
\section{Parameter reconstruction \label{sec:ML}}
%%%%%%%%%%%%%%%%%%%%%%%%%%%%%%%%%%%%%%%%%%%%%%%%%%%%%%%%%%

In this work we adopt a Bayesian framework for the statistical interpretation of our results. The cornerstone of Bayesian inference is provided by Bayes' theorem,
\begin{equation}
p(\boldsymbol{\theta}|\mathbf{D}) = \frac{\mathcal{L}(\mathbf{D}|\boldsymbol{\theta})p(\boldsymbol{\theta})}{p(\mathbf{D})}\;,
\label{eq:bayes}
\end{equation}
where~$\mathbf{D}$ are the data and~$\boldsymbol{\theta}$ are the model parameters. The posterior probability distribution function (pdf), $p(\boldsymbol{\theta}|\mathbf{D})$, is the product of the likelihood function, $\mathcal{L}(\mathbf{D}|\boldsymbol{\theta})$, and the prior pdf, $p(\boldsymbol{\theta})$. Finally, $p(\mathbf{D})$ is the Bayesian evidence, which for parameter inference acts 
only as an overall normalisation constant. It is therefore not considered in this analysis.

In order to study the constraints on a single parameter of interest $\theta_i$, we consider the one-dimensional marginal posterior pdf. The marginal~pdf is obtained from the full posterior distribution by integrating (marginalising) over the unwanted parameters in the $n$-dimensional parameter space:
\begin{equation}
p(\theta_i|\mathbf{D}) = \int p( \boldsymbol{\theta} |\mathbf{D}) \,d\theta_1\dots d\theta_{i-1}\, d\theta_{i+1}\dots d\theta_n.
\label{eq:marg}
\end{equation}
This result can be easily generalised to obtain the marginal pdf for more than one parameter of interest. 

%%%%%%%%%%%%%%%%%%%%%%%%%%%%%%%%%%%%%%%%%%%%%%%%%%%%%%%%%%
\begin{table}[t]
\setlength{\tabcolsep}{10pt}
\renewcommand{\arraystretch}{1.3}
\center
\begin{tabular}{r|cc} 
  & Axial-vector mediator  & Scalar mediator\\
\hline
$\mDM$ & 1 GeV -- 1 TeV & 0.1 GeV -- 1 TeV \\
$\mMed$ & 1 GeV -- 1 TeV &  1 GeV -- 1 TeV \\
$\gDM$ & 0.1 -- 10 & 0.1 -- 10\\
$\gq$ & 0.01 -- 1 & 0.1 -- 10 \\
\end{tabular}
\caption{The prior ranges for the axial-vector and scalar mediator models. The prior distributions are uniform over the common logarithm of the model parameters.}
\label{tab:priors} 
\end{table}
%%%%%%%%%%%%%%%%%%%%%%%%%%%%%%%%%%%%%%%%%%%%%%%%%%%%%%%%%%

We choose uniform priors on the common logarithm of the model parameters that enter into eq.~\eqref{eq:bayes} because we want to give \emph{a priori} equal probability to all the scales considered for the couplings and the masses.
In table \ref{tab:priors} we tally our prior ranges for the model parameters of the axial-vector and scalar cases.

In order to sample the posterior we employ the MultiNest algorithm~\cite{Feroz:2008xx}, which is based on the method of nested sampling~\cite{Skilling04}.
MultiNest is an efficient sampler for likelihood functions with a very complex structure. 

%%%%%%%%%%%%%%%%%%%%%%%%%%%%%%%%%%%%%%%%%%%%%%%%%%%%%%%%%%
\subsection{Incorporating experimental signals and constraints}
%%%%%%%%%%%%%%%%%%%%%%%%%%%%%%%%%%%%%%%%%%%%%%%%%%%%%%%%%%

The likelihood function that enters eq.~\eqref{eq:bayes} is composed of several pieces, each corresponding to a different independent type of experimental constraint that we apply in this analysis. The total likelihood can be expressed as
\begin{equation}
\mathcal{L} =  \mathcal{L}_{\mathrm{LHC}} \times \mathcal{L}_{\mathrm{DD}} \;,
\end{equation}
where we next discuss $\mathcal{L}_{\mathrm{LHC}}$ and $\mathcal{L}_{\mathrm{DD}}$ in turn.

%%%%%%%%%%%%%%%%%%%%%%%%%%%%%%%%%%%%%%%%%%%%%%%%%%%%%%%%%%
\subsubsection{LHC likelihood}\label{sec:lhclikelihood}
%%%%%%%%%%%%%%%%%%%%%%%%%%%%%%%%%%%%%%%%%%%%%%%%%%%%%%%%%%

The LHC likelihood, $\mathcal{L}_{\mathrm{LHC}}$, is composed of two parts. The first corresponds to constraints from dijet searches, for which we have implemented a step function on the signal following ref.~\cite{ATLAS:2016xiv}. The second corresponds to a future signal observed in the monojet channel. The monojet signal regions (EM1-EM6, cf.~section~\ref{subsec:currentmono}) considered in our analysis are exclusive so the likelihood corresponding to the monojet signal is simply the product of the likelihoods from each of the six exclusive signal regions: $\mathcal{L} = { \prod_{i=1}^{6} \mathcal{L}_i}$. 

The likelihood for each signal region~$i$ is given by
\begin{equation}
\mathcal{L}_i(n_i|s,b,\boldsymbol{\nu}) = \mathrm{Poisson}\left(n_i|\lambda_{\rm{LHC}}[s,b,\boldsymbol{\nu}]\right) \times \mathcal{L}_G(\boldsymbol{\nu})\;.
\end{equation}
The first factor reflects the Poisson probability of observing~$n_i$ events in the $i$-th signal region given the mean value~$\lambda_s$. The mean value is given by
\begin{equation}
\lambda_{\rm{LHC}} [s,b,\boldsymbol{\nu}]= s\,(1+\Delta_s \nu_s) + b\,(1+\Delta_b \nu_b)\;,
\end{equation}
where~$s$ and~$b$ are the mean values of the signal and background,~$\Delta_s$ and~$\Delta_b$ are systematic uncertainties of the signal and background, and~$\boldsymbol{\nu} = \{\nu_s, \nu_b\}$ are nuisance parameters that parametrise systematic uncertainties. These uncertainties are constrained by the likelihood term~$\mathcal{L}_G(\boldsymbol{\nu})$, which is taken as a product of standard normal distributions for each component of~$\boldsymbol{\nu}$. 

For the systematic uncertainty on the signal,~$\Delta_s$, we assume two 
sources: one from the cross-section determination of~$10 \%$ (cf.~section~\ref{subsec:currentmono}) and the other from the uncertainty in the predictions of the efficiencies by the distributed Gaussian processes machine learning predictors (described further in section~\ref{sec:nn}). These two sources of uncertainty are uncorrelated so we add them in quadrature. For the systematic uncertainty on the background,~$\Delta_b$, we assume two scenarios as described in section~\ref{subsec:futuremono}, namely an uncertainty of 1\% and 3\%.

Finally, the four parameters from our simplified models enter through the mean signal value i.e.~$s=s(\mDM,\mMed,\gSM,\gDM)$. We obtain the marginal likelihood for a signal region by integrating over the nuisance parameters,
\begin{equation}
\mathcal{L}_i(n_i|s,b) = \int \mathcal{L}_i(n_i|s,b,\boldsymbol{\nu})p(\boldsymbol{\nu})d\boldsymbol{\nu}\; , 
\end{equation}
where the prior over~$\boldsymbol{\nu}$ is uniform around~$\boldsymbol{\nu} = 0$ and of length~six standard deviations on either side.

%%%%%%%%%%%%%%%%%%%%%%%%%%%%%%%%%%%%%%%%%%%%%%%%%%%%%%%%%%
\subsubsection{Direct detection likelihood}
%%%%%%%%%%%%%%%%%%%%%%%%%%%%%%%%%%%%%%%%%%%%%%%%%%%%%%%%%%

The dark matter direct detection likelihood, $\mathcal{L}_{\mathrm{DD}}$, is split in two parts. One part includes current results from the LUX and PICO experiments, as implemented in the DDCalc package~\cite{Workgroup:2017lvb}, while the other part includes the likelihood for the upcoming~LZ and~CRESST-III experiments for the axial-vector and scalar mediator benchmarks, respectively. The likelihood function for~LZ and~CRESST-III is given by the product of Poisson likelihoods over~$m$ energy bins, $\mathcal{L} = { \prod_{i=1}^m \mathcal{L}_i}$, where
\begin{equation}
\mathcal{L}_i(n_i | \mDM, \sigma)= \mathrm{Poisson}\left(n_i | \lambda_{\rm{DD}}[ s(\mDM, \sigma), b]\right)\;,
\end{equation}
is the Poisson probability of observing~$n_i$ events in the $i$-th energy bin given the mean number of events $\lambda_{\rm{DD}} [(\mDM, \sigma), b] = s (\mDM, \sigma) + b$. Here,~$s(\mDM, \sigma)$ and~$b$ are the number of signal and background events in the $i$-th energy bin respectively, while~$\sigma$ refers to $\sigma^{n}_{\rm{SD}}$ when the likelihood is computed for~LZ, and to $\sigma_{\rm{SI}}$ when it is computed for CRESST-III. Specific values of~$s$ and~$b$ for our benchmark scenarios are given in~section~\ref{sec:simDD}.

%%%%%%%%%%%%%%%%%%%%%%%%%%%%%%%%%%%%%%%%%%%%%%%%%%%%%%%%%%
\subsection{Fast likelihood evaluation with machine learning tools \label{sec:nn}}
%%%%%%%%%%%%%%%%%%%%%%%%%%%%%%%%%%%%%%%%%%%%%%%%%%%%%%%%%%

Performing a global analysis of theories with even moderately large parameter spaces, such as the simplified models we study in this paper, involves evaluating the likelihood tens to hundreds of thousands of times. 
This makes incorporating LHC results computationally expensive as the LHC likelihood evaluation requires simulating the collision and detection on an event-by-event basis.
To accelerate this evaluation of the LHC likelihood, we make use of machine learning to bypass all of the simulation steps.
We construct machine learning models that learn the mapping from model parameters~$\boldsymbol{\theta}$ to the signal quantities, such as the number of events in a signal region~$n$.
These machine learning models can then predict the signal quantities from the model parameters without performing any of the simulation steps, which drastically reduces the overall computing time.

In this subsection, we describe our implementation of the various machine learning tools that we use to accelerate the likelihood evaluation.
For clarity, we have collected the configuration details of our machine learning models and all of the figures detailing the performance of these models in appendix~\ref{sec:mlvalidation}, and focus on the high-level description of our procedure.

We use three different machine learning methods: random forest classifier~\cite{Ho:1995:RDF:844379.844681, Breiman2001} to implement the dijet constraints, and a combination of deep neural networks and distributed Gaussian processes~\cite{Bertone:2016mdy, 2015arXiv150202843D} to compute the signal events for the monojet likelihood.
All of these algorithms fall under the umbrella of supervised learning, in which the algorithm learns the rules, or mapping, between the inputs and the output from examples.
The typical way to assess the quality of the taught algorithm is to randomly split the dataset of examples into a \emph{training} dataset and a \emph{testing} dataset.
The training dataset is used by the algorithm to infer an approximate representation of the mapping. The quality of the inferred representation is then assessed using the testing set.
For both the scalar and the axial-vector models, we generated, using the set-up described in section~\ref{sec:mocksignals}, datasets of example inputs with associated outputs. 
In the scalar case we generated $\sim 20000$ model points by sampling uniformly in the common logarithm of each model parameter over the ranges given in table~\ref{tab:priors}.
For the axial-vector case we similarly generated $\sim 20000$ model points with again the same prior as in table~\ref{tab:priors} but with the exception that $\log_{10} \gq$ ranged from~$-1$ to~$1$.
We extended this dataset by an additional $\sim 15000$ model points sampling from $\log_{10} \gq \in [-2,-0.8]$ in order to cover the whole region of our interest.

The random forest classifier was trained on the axial-vector dataset in order to be able to mark points as excluded or allowed by the dijet constraint during the global analysis. Random forests are an ensemble method where the predictions of many decision trees trained on subsets of the data are averaged together. 
This increases classification accuracy while protecting against over-fitting.

In order to evaluate the monojet likelihood we need to compute the number of signal events~$n$ in each signal region.
The number can be written as $n = L \cdot \sigma_{\rm{prod}} \cdot \epsilon_{\rm{eff}}$ where~$L$ is the time-integrated luminosity, $\sigma_{\rm{prod}}$ the production cross-section for the relevant processes and $\epsilon_{\rm{eff}}$ encodes the detector efficiencies and the cut acceptances.
We use a combination of two methods: deep neural networks to compute~$\sigma_{\rm{prod}}$, and distributed Gaussian processes to compute~$\epsilon_{\rm{eff}}$.
Gaussian processes use the training data and a calibrated measure of similarity between points, i.e. a \emph{kernel} function, to predict the functional value at a previously unseen point.
The prediction is not only a mean estimate but a full (Gaussian) posterior distribution and thus the prediction comes with an uncertainty estimate.
Deep neural networks are collections of (fully) connected layers consisting of artificial neurons which acts as an approximate representation of a mapping from some input to a target, which here, is the mapping from simplified model parameters to monojet production cross-section.
The propagation of the input through the neural network proceeds via multiplication with the weights, i.e. the connection factor between neurons, and a possible shift called bias for each neuron. To the sum of both, each neuron applies a non-linear function, the activation function. A gradient-descent based optimisation algorithm is then used to search for the global minimum of a function measuring the error of the predictions by updating the weights and biases.

%%%%%%%%%%%%%%%%%%%%%%%%%%%%%%%%%%%%%%%%%%%%%%%%%%%%%%%%%%
%%%%%%%%%%%%%%%%%%%%%%%%%%%%%%%%%%%%%%%%%%%%%%%%%%%%%%%%%%
%%%%%%%%%%%%%%%%%%%%%%%%%%%%%%%%%%%%%%%%%%%%%%%%%%%%%%%%%%
%%%%%%%%%%%%%%%%%%%%%%%%%%%%%%%%%%%%%%%%%%%%%%%%%%%%%%%%%%
%%%%%%%%%%%%%%%%%%%%%%%%%%%%%%%%%%%%%%%%%%%%%%%%%%%%%%%%%%
\section{Results and interpretation of our global fitting analysis \label{sec:results}}
%%%%%%%%%%%%%%%%%%%%%%%%%%%%%%%%%%%%%%%%%%%%%%%%%%%%%%%%%%
%%%%%%%%%%%%%%%%%%%%%%%%%%%%%%%%%%%%%%%%%%%%%%%%%%%%%%%%%%

We now discuss the results and implications of our global analysis of the two benchmark models introduced in section~\ref{sec:models}.
We first discuss the results for the axial-vector model, which contains WIMP dark matter (i.e.\ the observationally inferred dark matter density is obtained from thermal freeze-out), before turning to the scalar model, where the density from thermal freeze-out is significantly larger than the observed value.
In this section, we focus on the marginalised posterior distributions that assume a~1\% systematic error on the Standard Model background rates in future LHC monojet searches (as detailed in Section~\ref{subsec:futuremono}).
For clarity, we have relegated the results with a~3\% systematic error on the background rates to appendix~\ref{sec:addfigs} and the profile likelihood distributions (with a~1\% systematic error) to appendix~\ref{sec:addfigsprofile} since the conclusions drawn are similar in all cases.

%%%%%%%%%%%%%%%%%%%%%%%%%%%%%%%%%%%%%%%%%%%%%%%%%%%%%%%%%%
\subsection{Benchmark 1: Axial-vector mediator}
%%%%%%%%%%%%%%%%%%%%%%%%%%%%%%%%%%%%%%%%%%%%%%%%%%%%%%%%%%

\begin{figure}[t!]
\centering
\includegraphics{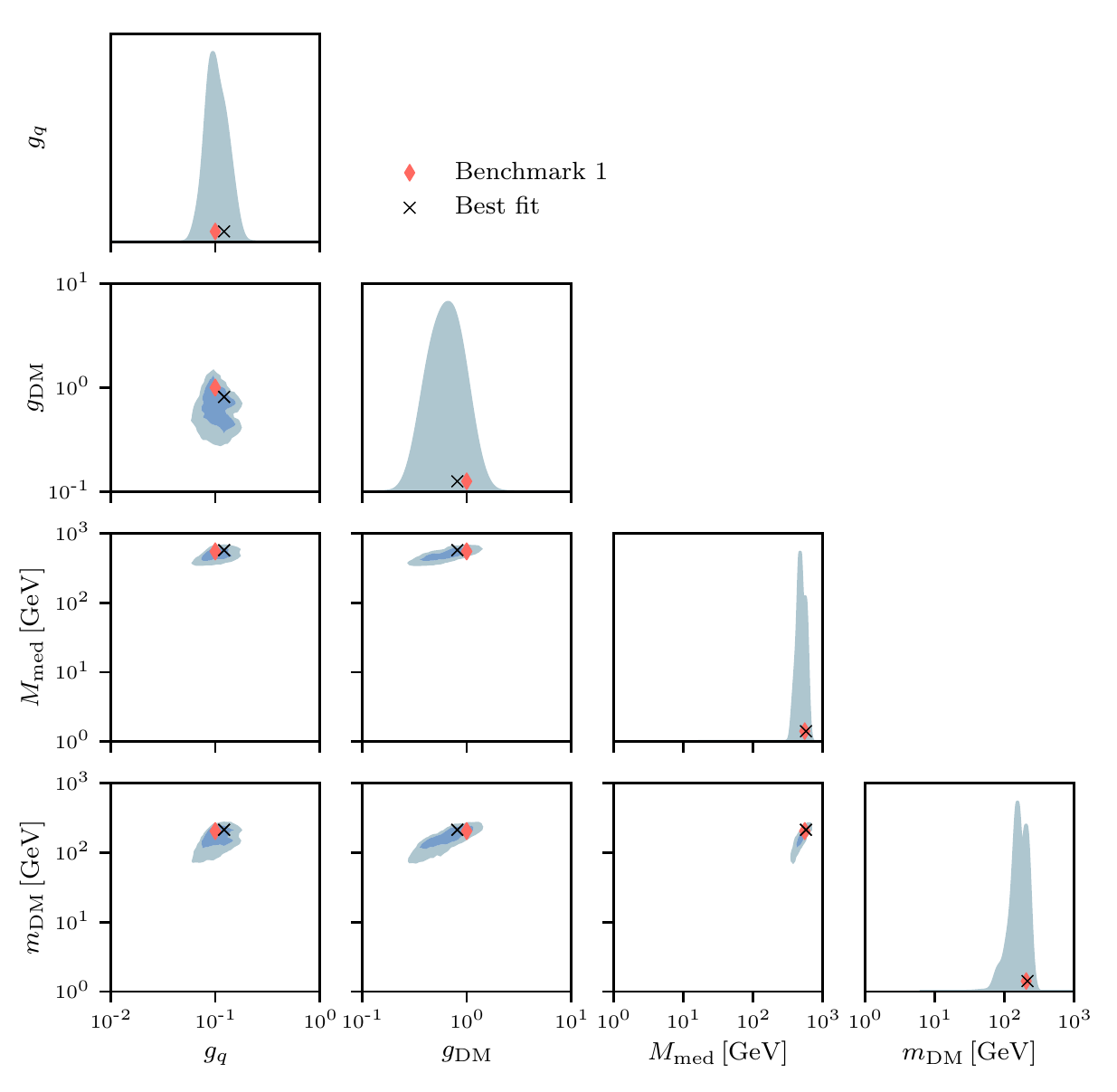}
\caption{
\textbf{Axial-vector mediator:}
Posteriors over the model parameters assuming future excesses in an ATLAS monojet search (with 1\% syst.) and LZ (as described in section~\ref{sec:mocksignals}).
The full one-dimensional posteriors are shown. In the two-dimensional case the~68\% (dark blue) and~95\% (light blue) credibility regions are shown. Also marked are the best fit point (black cross), and the benchmark model (red diamond). All of the reconstructed parameters are consistent with the benchmark parameters.}
\label{fig:axialvector_1pc_params}
\end{figure}

\begin{figure}[t!]
\centering
\includegraphics{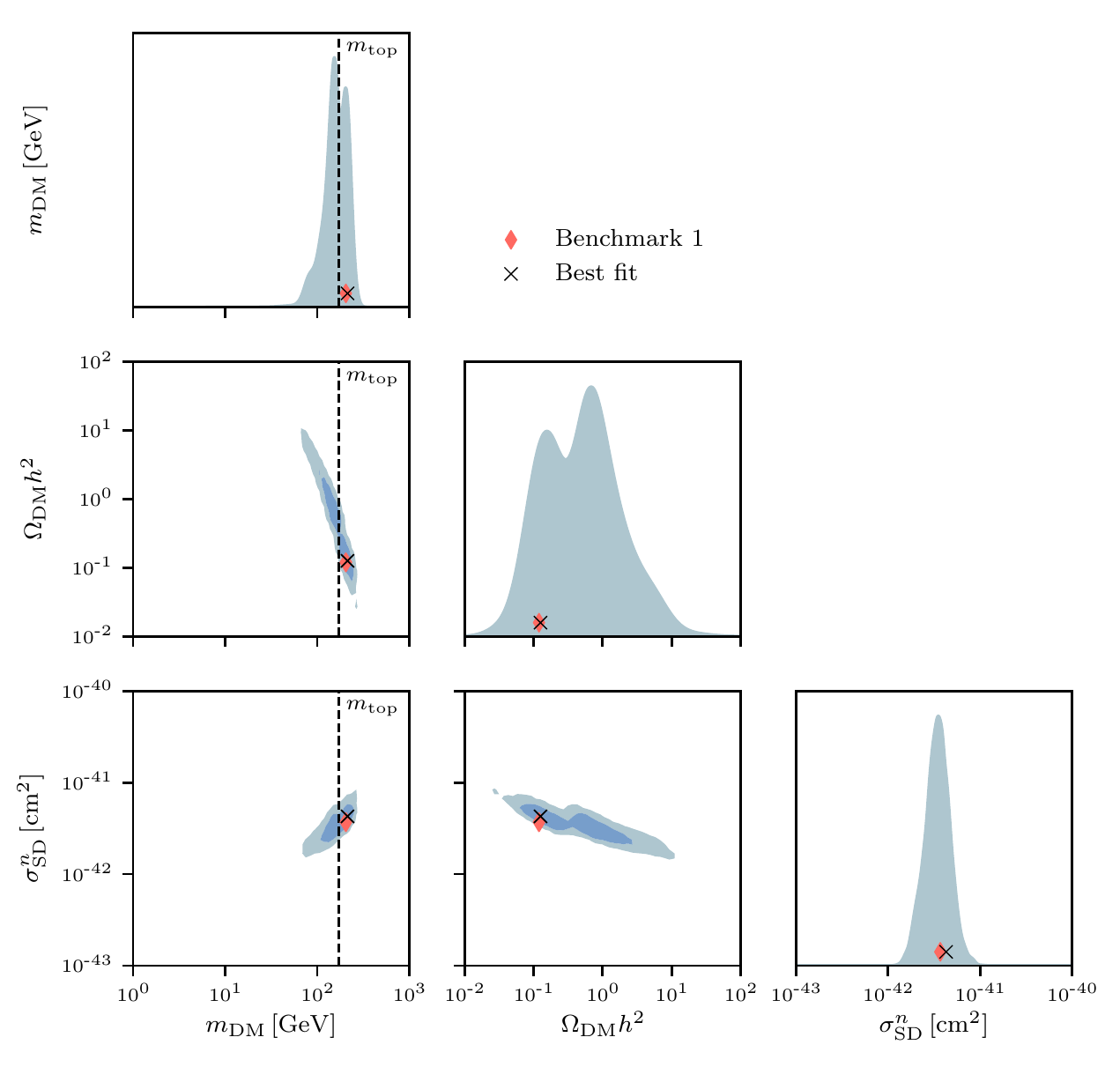}
\caption{
\textbf{Axial-vector mediator:}
Posteriors over the phenomenological dark matter quantities assuming future excesses in an ATLAS monojet search (with 1\% syst.) and LZ (as described in section~\ref{sec:mocksignals}).
The full one-dimensional posteriors are shown. In the two-dimensional case the 68\% (dark blue) and 95\% (light blue) credibility regions are shown. Black dashed line indicate where $\mDM = m_{\mathrm{top}}$, the top quark mass.
Also marked are the best fit point (black cross), and the benchmark model (red diamond). The reconstructed value of the relic density~$\oh$ is consistent with the observationally inferred dark matter relic density ($\oh\approx0.12$), as can be seen in the marginalised posterior.
\label{fig:axialvector_1pc_pheno}}
\end{figure}

Figures~\ref{fig:axialvector_1pc_params} and~\ref{fig:axialvector_1pc_pheno} show the marginalised posteriors from our global analysis of the (synthetic) excesses in LZ and the ATLAS monojet search. The red diamond and the black cross in figures~\ref{fig:axialvector_1pc_params} to~\ref{fig:scalar_1pc_pheno} show the benchmark parameters in table~\ref{tab:benchmarks} and the best fit point, respectively.
Figure~\ref{fig:axialvector_1pc_params} focuses on the posteriors for the four parameters that define the axial-vector simplified model, while figure~\ref{fig:axialvector_1pc_pheno} focuses on the more phenomenological parameters. 

%%%%%%%%%%%%%%%%%%%%%%%%%%%%%%%%%%%%%%%%%%%%%%%%%%%%%%%%%%
%%%%%%%%%%%%%%%%%%%%%%%%%%%%%%%%%%%%%%%%%%%%%%%%%%%%%%%%%%
% lg m_MED : (2.624397264774494, 2.7580517532956845) (two-sided)
% lg m_DM : (2.075586047538792, 2.328737507977979) (two-sided)
% lg g_DM : (-0.38241769871718845, -0.01621316755765076) (two-sided)
% lg naive_sigmaSD : (-41.73405244499853, -41.482235267767) (two-sided)
% lg sigmaSD : (-41.60071682176755, -41.33342379213232) (two-sided)
% lg g_MED : (-1.088072951399026, -0.8785289549132516) (two-sided)
% lg oh2 : (-0.9197466635866507, 0.24733849118239676) (two-sided)

Figure~\ref{fig:axialvector_1pc_params} shows that we are able to constrain at the~68\% credibility level~(CL): the dark matter mass within the range $\mDM \in [119, 213]$ GeV; the mediator mass within $\mMed \in [421, 573]$ GeV; the mediator--dark matter coupling within $\gDM \in [0.42, 0.96]$; and the mediator--quark coupling within $\gSM \in [0.08, 0.13]$. These values for the reconstructed parameters are consistent with the benchmark parameters (cf.~table~\ref{tab:benchmarks}). For~$\gSM$, the benchmark parameter lies squarely in the middle of the 68\%~CL range, while the $\mDM$, $\mMed$ and~$\gDM$ benchmark parameters lie just outside the 68\%~CL range.

%%%%%%%%%%%%%%%%%%%%%%%%%%%%%%%%%%%%%%%%%%%%%%%%%%%%%%%%%%
%%%%%%%%%%%%%%%%%%%%%%%%%%%%%%%%%%%%%%%%%%%%%%%%%%%%%%%%%%

In figure~\ref{fig:axialvector_1pc_pheno}, we additionally show the posteriors for the reconstructed dark matter relic density~$\oh$ and the spin-dependent scattering cross-section~$\sigmaSD^n$. Mainly from the mock signal in LZ, we are able to constrain at 68\% CL, $\sigmaSD^{n} \in [2.51, 4.64] \times 10^{-42}$ cm$^2$, while the dark matter relic density is constrained to $\oh \in [0.12, 1.77]$. The reconstructed value of the relic density~$\oh$ is consistent with the observationally inferred dark matter relic density, as can be clearly seen in the marginalised posterior.

Figures~\ref{fig:axialvector_1pc_params} and~\ref{fig:axialvector_1pc_pheno} contain a number of interesting features. The effect of the top quark mass~$m_t$ threshold can be seen in the two-dimensional posteriors of $\oh$--$\mDM$ in figure~\ref{fig:axialvector_1pc_pheno}, where two distinct peaks in the 68\%~CL regions can clearly be seen. One region lies above~$m_t$ and the other below~$m_t$. This is a reflection of the dependence of the annihilation cross-section on the quark mass~$m_q$, where specifically, the $s$-wave contribution is proportional to~$m_q^2$~\cite{Busoni:2014gta}. The reconstructed value of~$\oh$ is also strongly dependent on~$\mDM$ since the $p$-wave contribution to the annihilation cross-section, which dominates for~$\mDM<m_t$, is proportional to~$\mDM^2$~\cite{Busoni:2014gta}. As~$\mDM$ decreases, the annihilation cross-section decreases, so the value of~$\oh$ must increase (cf.~eq.~\eqref{eq:omDMsch}). The dependence of~$\oh$ on~$\mDM$ is responsible for the behaviour of the~$\oh$ one-dimensional~posterior in figure~\ref{fig:axialvector_1pc_pheno}, which has a clear double peak structure. In the axial-vector mediator model, the precision of the reconstructed value of~$\oh$ is therefore controlled by our accuracy in reconstructing the dark matter mass~$\mDM$.

In this context, the LZ (mock) excess is useful because it places a lower bound on~$\mDM$. For small values of~$\mDM$, the recoil energy spectrum of the xenon nucleus drops off too quickly in energy such that the fit to the observed (synthetic) signal is poor (i.e.\ small values of~$\mDM$ are not able to explain the presence of signal events at higher energy). At higher values of~$\mDM$, the xenon nucleus' energy recoil spectrum flattens and the sensitivity to $\mDM$ is reduced~\cite{Green:2008rd}. However, here the monojet signal provides a complementary upper bound on~$\mDM$. The monojet signal is only sensitive to on-shell production of the axial-vector mediator, which corresponds to $\mDM< \mMed/2$ (this is immediately apparent from the monojet constraints shown in the left panel of figure~\ref{fig:axial_LHC_DD}, where only the parameter space~$\mDM< \mMed/2$ is constrained). The mock monojet signal is largely responsible for the reconstruction of~$\mMed$, so the maximum value of $\mMed$ (around~750~GeV in the~one-dimensional posterior shown in figure~\ref{fig:axialvector_1pc_params}) corresponds to a maximum value of $\mDM$ of around 325~GeV, which is indeed what we observe in figure~\ref{fig:axialvector_1pc_params}.

%%%%%%%%%%%%%%%%%%%%%%%%%%%%%%%%%%%%%%%%%%%%%%%%%%%%%%%%%%
\subsection{Benchmark 2: Scalar mediator}
%%%%%%%%%%%%%%%%%%%%%%%%%%%%%%%%%%%%%%%%%%%%%%%%%%%%%%%%%%

\begin{figure}[t!]
\centering
\includegraphics{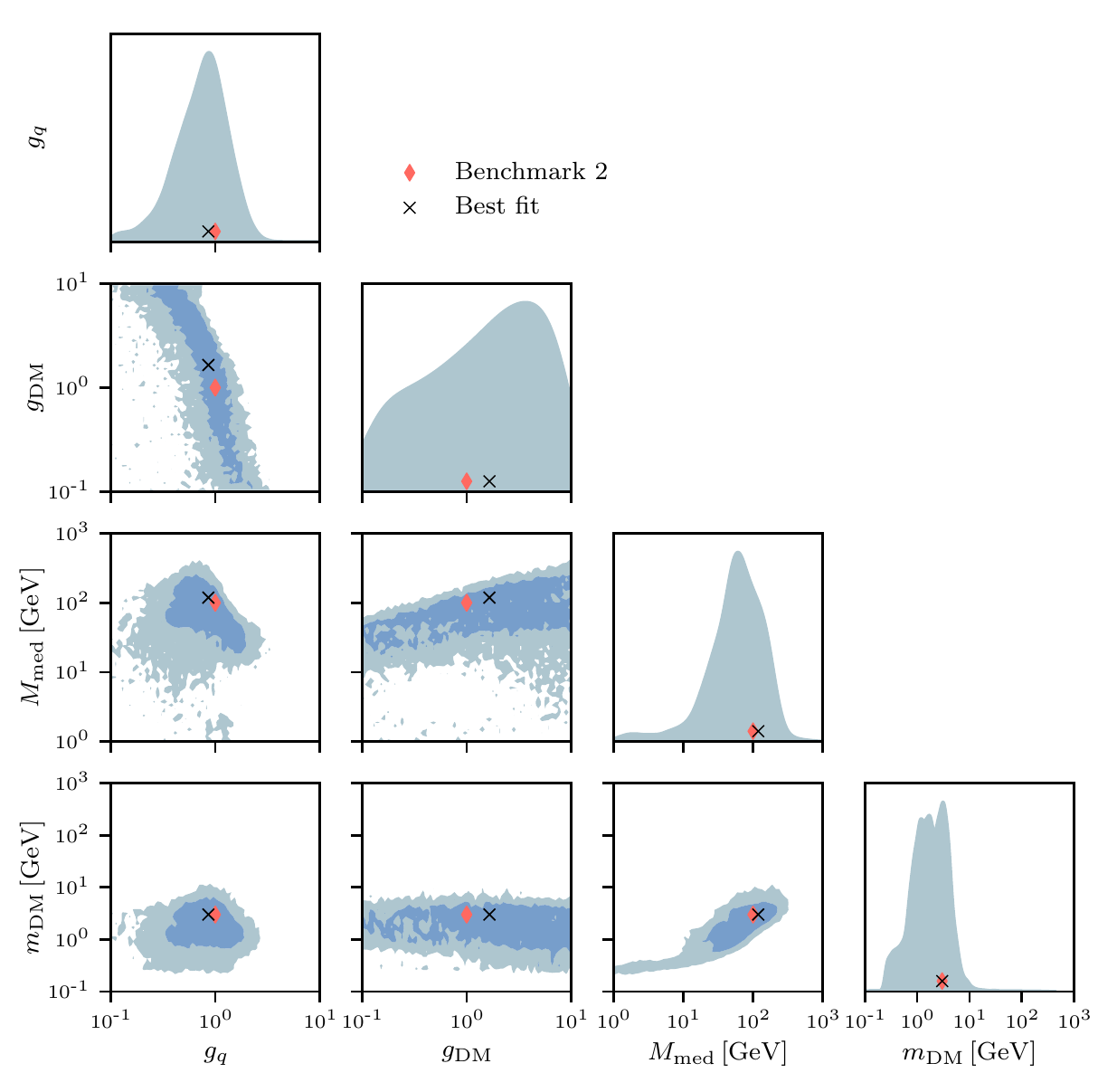}
\caption{
\textbf{Scalar mediator:}
Posteriors over the model parameters assuming future excesses in an ATLAS monojet search (with 1\% syst.) and CRESST-III (as described in section~\ref{sec:mocksignals}).
The full one-dimensional posteriors are shown, in the two-dimensional case the 68\% (dark blue) and 95\% (light blue) credibility regions are shown.
Also marked are the best fit point (black cross), and the benchmark model (red diamond). All of the reconstructed parameters are consistent with the benchmark parameters. The weak constraint on~$\gDM$ is explained in the text.
\label{fig:scalar_1pc_params}}
\end{figure}

\begin{figure}[t!]
\centering
\includegraphics{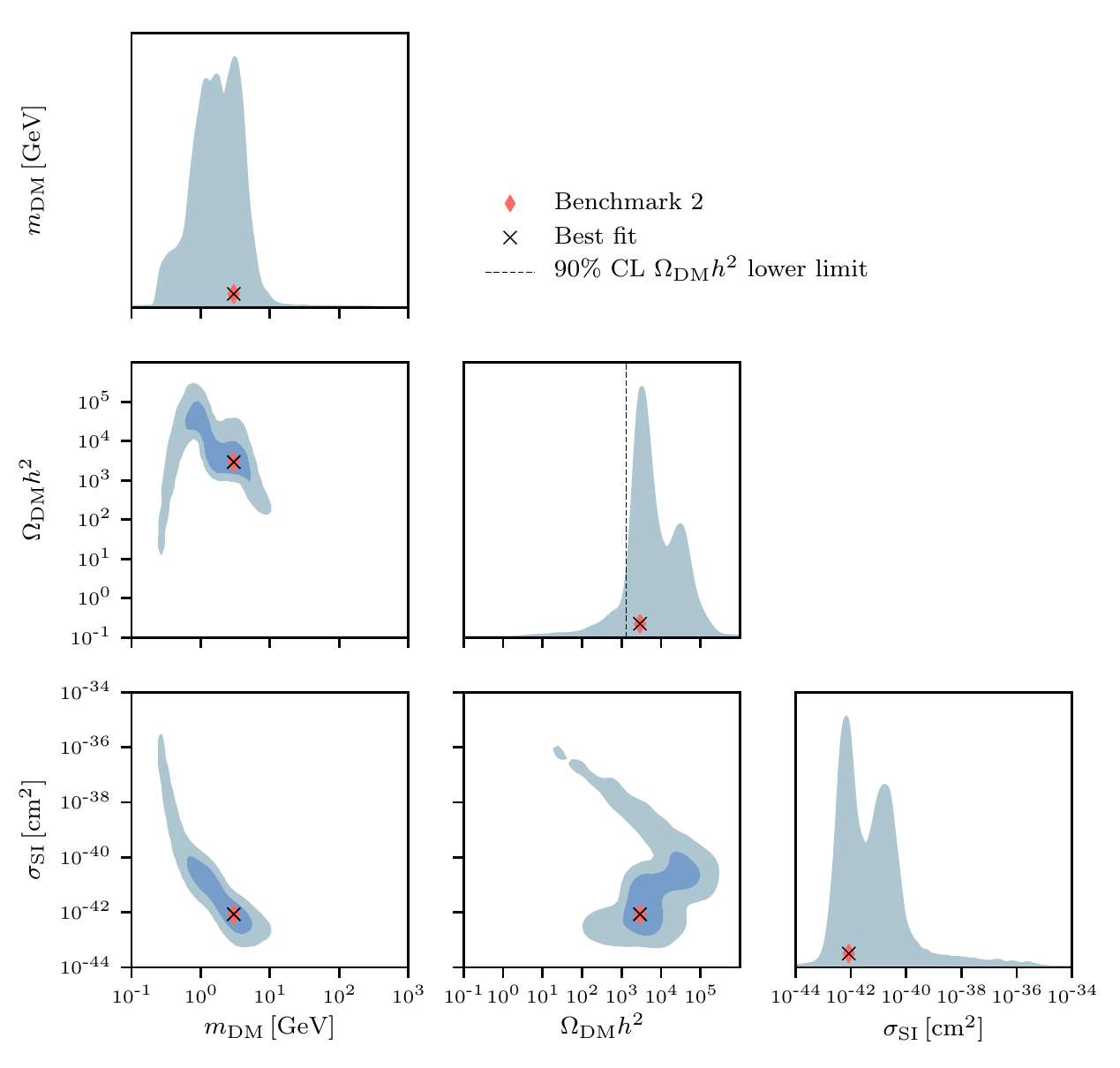}
\caption{
\textbf{Scalar mediator:}
Posteriors over the phenomenologically interesting quantities assuming future excesses in an ATLAS monojet search (with 1\% syst.) and CRESST-III (as described in section~\ref{sec:mocksignals}).
The full one-dimensional posteriors are shown, in the two-dimensional case the 68\% (dark blue) and 95\% (light blue) credibility regions are shown.
Also marked are the best fit point (black cross), and the benchmark model (red diamond).
The reconstructed value of the relic density~$\oh$ is more than four orders of magnitude larger than the observationally inferred value, implying the presence of additional particle physics not included in the simplified model or that a modified cosmological history is required to explain the observed dark matter abundance within this model.
\label{fig:scalar_1pc_pheno}}
\end{figure}

In figures~\ref{fig:scalar_1pc_params} and~\ref{fig:scalar_1pc_pheno}, we show the marginalised posteriors of the scalar mediator model parameters that follow from (synthetic) excesses in CRESST-III and the ATLAS monojet analysis.
Figure~\ref{fig:scalar_1pc_params} show the posteriors for the four parameters that define the scalar simplified model, while figure~\ref{fig:scalar_1pc_pheno} display on the phenomenological observables. 

%%%%%%%%%%%%%%%%%%%%%%%%%%%%%%%%%%%%%%%%%%%%%%%%%%%%%%%%%%
%%%%%%%%%%%%%%%%%%%%%%%%%%%%%%%%%%%%%%%%%%%%%%%%%%%%%%%%%%
% scalar 1%
% lg oh2 : 3.114819400725677 (ll)
% lg g_DM : (-0.4902343637026698, 0.7011444116871935) (two-sided)
% lg m_DM : (-0.10989148070434197, 0.569386347398565) (two-sided)
% lg g_MED : (-0.3853491258338513, 0.10514206575899822) (two-sided)
% lg m_MED : (1.394732196954521, 2.1282250554047732) (two-sided)
% lg sigmaSI : (-42.30280985599578, -40.35457099610666) (two-sided)

Figure~\ref{fig:scalar_1pc_params} shows that we are able to constrain the dark matter mass $\mDM \in [0.32,  5.03]$~GeV (68\% CL), reconstruct the mediator mass $\mMed \in [24.8, 134.3]$~GeV (68\% CL), and the quark--mediator coupling~$\gSM$ to within $[0.41, 1.27]$ (68\% CL). There is only weak sensitivity to the dark matter--mediator coupling~$\gDM$ leaving it unconstrained, with a weak preference for larger values.
These reconstructed values are compatible with the benchmark parameters given in table~\ref{tab:benchmarks}, with all the benchmark parameters lying within the~68\%~CL range.

%%%%%%%%%%%%%%%%%%%%%%%%%%%%%%%%%%%%%%%%%%%%%%%%%%%%%%%%%%
%%%%%%%%%%%%%%%%%%%%%%%%%%%%%%%%%%%%%%%%%%%%%%%%%%%%%%%%%%

In figure~\ref{fig:scalar_1pc_pheno} we show the posteriors for the reconstructed dark matter relic density~$\oh$ and the spin-independent scattering cross-section~$\sigmaSI$. The scattering cross-section is constrained to lie in the range $\sigmaSI \in [0.49, 44] \times 10^{-42}$ cm$^2$ (68\% CL), consistent with the benchmark value. The relic density is bounded from below $\oh > 1.3 \times 10^3$ (90\% CL), approximately four orders of magnitude larger than the observationally inferred value. For this model to be a reasonable dark matter theory, either new particle physics contributions to the annihilation cross-section or modified thermal history must be introduced in order to reduce the relic density to its observed value.

As with the axial-vector mediator results, there are a number of interesting features in the scalar mediator'€™s posterior distributions shown in figures~\ref{fig:scalar_1pc_params} and~\ref{fig:scalar_1pc_pheno}. We begin by discussing the insensitivity of the monojet search to the dark matter--mediator coupling~$\gDM$ observed in figure~\ref{fig:scalar_1pc_params}. The monojet search is sensitive to on-shell production of the mediator, which subsequently decays invisibly to a pair of dark matter particles. In the narrow width approximation, this means that the signal is proportional to $g_q^2 \times \mathrm{BR}(\phi \to \bar{\chi} \chi)$, where the $g_q^2$ factor comes from the production of the mediator and $\mathrm{BR}(\phi \to \bar{\chi} \chi)$ from its subsequent decay. The branching ratio is approximately
\begin{equation}
\mathrm{BR}(\phi \to \bar{\chi} \chi) \approx \frac{ \gDM^2} { \sum_q 3 g_q^2 \cdot(m_q/v_{\rm{EW}})^2+ \gDM^2}\;,
\end{equation}
where the sum extends over all quarks that satisfy $\mDM<\mMed/2$. Considering the benchmark value $\mMed \simeq 100$~GeV, the bottom-quark dominates in the sum. Now, since $m_b/v_{\rm{EW}} =4.2/246 \sim 10^{-2}$, we see that in-fact, $\mathrm{BR}(\phi \to \bar{\chi} \chi) \approx 1$. This implies that the monojet search is essentially independent of~$\gDM$ for the scalar mediator model.

The next feature that deserves comment is the tail feature that is observed in the $\mDM$--$\sigma_{\rm{SI}}$, $\mDM$--$\oh$, $\oh$--$\sigma_{\rm{SI}}$ plots in figure~\ref{fig:scalar_1pc_pheno}, and the $\mMed$--$\mDM$ plot in figure~\ref{fig:scalar_1pc_params}. The feature in all plots has a common origin, arising from fitting the (synthetic) CRESST-III excess with sub-threshold events that have fluctuated above the energy threshold into the signal region. As the event rate for events that must fluctuate up into the signal region is small, a very large scattering cross-section~$\sigma_{\rm{SI}}$ is required to explain the signal (cf.~the $\mDM$--$\sigma_{\rm{SI}}$ plot in figure~\ref{fig:scalar_1pc_pheno} where we see the largest cross-section is six orders of magnitude above the best fit point). Such a large cross-section can only be explained with a smaller value of $\mMed$ (cf.~eq.~\eqref{eq:DDSI} where $\sigma_{\rm{SI}}\propto \mMed^{-4}$). This is the reason for the tail at small values of $\mMed$ in the $\mMed$--$\mDM$ plot in figure~\ref{fig:scalar_1pc_params}. A small value of~$\mMed$ at small~$\mDM$ and large~$\sigma_{\rm{SI}}$ is also responsible for the tails seen in the $\mDM$--$\oh$, $\oh$--$\sigma_{\rm{SI}}$ plots in figure~\ref{fig:scalar_1pc_pheno}. As~$\mMed$ becomes closer to~$2\mDM$, an $s$-channel resonance begins to develop in the dark matter self-annihilation cross-section, meaning that it increases significantly. A large increase in the annihilation cross-section leads to a smaller value of~$\oh$ (cf.~eq.~\eqref{eq:omDMsch}). As discussed in ref.~\cite{Angloher:2017zkf}, the treatment of sub-threshold events is somewhat subtle, so to some extent, these tails may be considered as an artefact of our treatment of sub-threshold events in CRESST-III.

In any case, however, the tail doesn't affect our conclusion, which for clarity, we again restate: for the scalar mediator, the reconstructed value of the relic density is bounded from below $\oh > 1.3 \times 10^3$ (90\% CL), approximately four orders of magnitude larger than the observationally inferred value. We have therefore been able to show that for this model, additional particle physics not included in the simplified model or a modified cosmological history is required to explain the observed dark matter abundance.

%%%%%%%%%%%%%%%%%%%%%%%%%%%%%%%%%%%%%%%%%%%%%%%%%%%%%%%%%%
%%%%%%%%%%%%%%%%%%%%%%%%%%%%%%%%%%%%%%%%%%%%%%%%%%%%%%%%%%
%%%%%%%%%%%%%%%%%%%%%%%%%%%%%%%%%%%%%%%%%%%%%%%%%%%%%%%%%%

%%%%%%%%%%%%%%%%%%%%%%%%%%%%%%%%%%%%%%%%%%%%%%%%%%%%%%%%%%
\section{Summary and conclusions \label{Sec:conclusions}}
%%%%%%%%%%%%%%%%%%%%%%%%%%%%%%%%%%%%%%%%%%%%%%%%%%%%%%%%%%

In this article we have tackled the question of whether it is possible to reconstruct the relic density of new particles discovered at colliders and with direct detection experiments, and to identify them with WIMP dark matter. Since the interpretation of experimental data is possible only in the framework of a well-defined particle physics model, we focused here on simplified models, which emerged over the last few years as a useful tool to interpret and display LHC searches.   

We considered two benchmarks: a model with an axial-vector mediator, in which the dark matter particle achieves the correct abundance by thermal freeze-out; and a model with a scalar mediator, in which the dark matter relic density is set through an unspecified mechanism. Both benchmarks are compatible with current searches, and would lead by 2019-2020 to detectable signals in monojet searches at the LHC, as well as at future direct detection experiments like LZ, XENONnT, PandaX, or CRESST-III. 

For both benchmarks, we generated mock LHC and direct detection data, and we then assessed how well one can reconstruct the particle physics parameters, and corresponding phenomenological quantities -- in particular the relic density, in order to investigate the identification of the new particles with dark matter -- from a combined analysis of those data. The reconstruction procedure was made possible by the introduction of state-of-the-art machine learning techniques that significantly sped up much of the expensive calculations.

In the case of the {\it axial-vector benchmark model}, we have demonstrated (see figure~\ref{fig:axialvector_1pc_pheno}) that upcoming observations would allow us to reconstruct within a factor of 2, at 68\% CL, the dark matter mass ($\mDM \in [119, 213]$ GeV) and the spin-dependent cross-section ($\sigmaSD^{n} \in [2.51, 4.64] \times 10^{-42}$ cm$^2$)
. The compatibility of the reconstructed relic density ($\oh \in [0.12, 1.77]$) with the dark matter abundance inferred from cosmological observations would provide a powerful validation of the particle physics model adopted, as well as of the standard cosmological model. {\it Our analysis thus demonstrates that a combination of accelerator and direct detection data may allow us to identify newly discovered particles as the dark matter in the universe}.

In the case of the {\it scalar benchmark model}, we could reconstruct (at 68\% CL) within an order of magnitude the dark matter mass ($\mDM \in [0.32,  5.03]$ GeV) and the spin-independent cross-section ($\sigmaSI \in [0.49, 44] \times 10^{-42}$ cm$^2$). 
Interestingly, the reconstruction procedure leads to a (90\% CR) lower bound on the relic density of $\oh > 1303$, a result clearly incompatible with cosmological data. The data would point in this case  to additional physics in the dark sector, or to non-standard cosmologies, with a thermal history substantially different from that of the standard model of cosmology. 

%%%%%%%%%%%%%%%%%%%%%%%%%%%%%%%%%%%%%%%%%%%%%%%%%%%%%%%%%%
\subsection*{Acknowledgements}
%%%%%%%%%%%%%%%%%%%%%%%%%%%%%%%%%%%%%%%%%%%%%%%%%%%%%%%%%%

We are grateful to Fady Bishara, Oliver Buchmueller, Tristan du Pree and David Salek for discussions. G.B. (PI), N.B., and S.L.~acknowledge support from the European Research Council
through the ERC starting grant WIMPs Kairos. C.M.~acknowledges support from the Science and Technology Facilities Council (STFC) Grant ST/N004663/1. R. RdA is supported by the Ram\'on y Cajal program of the Spanish MICINN, the Elusives European ITN project (H2020-MSCA-ITN-2015//674896-ELUSIVES), the  ``SOM Sabor y origen de la Materia" (PROMETEOII/2014/050) and Centro de excelencia Severo Ochoa Program under grant SEV-2014-0398.

\appendix
%%%%%%%%%%%%%%%%%%%%%%%%%%%%%%%%%%%%%%%%%%%%%%%%%%%%%%%%%%%%%%%%%%%%%%%%%%%%
\section{Details, validation and performance of the machine learning tools}
\label{sec:mlvalidation}
%%%%%%%%%%%%%%%%%%%%%%%%%%%%%%%%%%%%%%%%%%%%%%%%%%%%%%%%%%%%%%%%%%%%%%%%%%%%

In this appendix we have collected the performance figures of the various machine learning tools used to compute the LHC likelihood, and their configurations and training procedures.

\begin{figure}[!t]
\centering
\includegraphics[width=0.95\columnwidth]{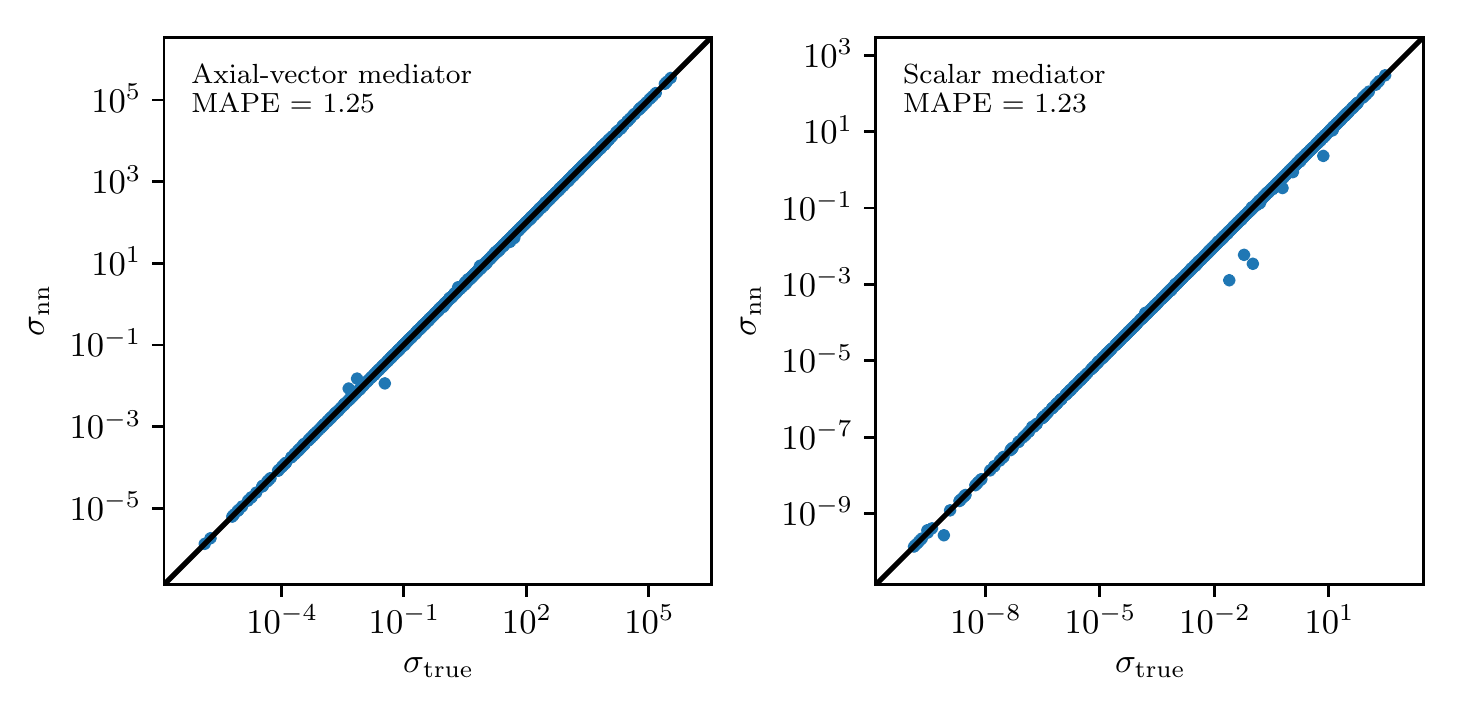}
\caption{The performance of the deep neural networks described in section~\ref{sec:nn} and appendix~\ref{sec:mlvalidation} in predicting the monojet production cross-section for the axial-vector (left) and the scalar mediator (right) models.
The blue dots are the results for testing datasets ($N=947$ and $N=999$ respectively), previously unseen by the network. The horizontal axis (labelled $\sigma_{\rm{true}}$) is the cross-section as calculated by the traditional MC set-up (described in section~\ref{sec:mocksignals}), while the vertical axis (labelled $\sigma_{\rm{nn}}$) is the deep neural network prediction.
The black line indicates perfect agreement i.e.\ $\sigma_{\mathrm{nn}} = \sigma_{\mathrm{true}}$.
Also given is the MAPE for each deep neural network on their respective validation dataset.
\label{fig:xsec_nn_performance}}
\end{figure}

The performance on the testing datasets for the deep neural networks in predicting the monojet production cross-section is shown in figure~\ref{fig:xsec_nn_performance}. Good agreement is found in both cases, with a mean absolute percentage error (MAPE) of $1.2$ -- $1.3 \%$, which is sub-dominant to the theoretical error. 
We used the Keras deep learning library~\cite{chollet2015keras} with a Tensorflow backend to implement the neural networks.
To train the networks the data was preprocessed; all inputs were scaled to the interval $[0,1]$ via min-max scaling, and a natural logarithm has been applied to the output, the production cross-section, since the values range across many orders of magnitude.
The feed-forward neural networks in both cases consist of three fully connected hidden layers with~200 neurons each with s-shaped rectified linear activation units (SReLU)~\cite{Jin:2015aa}.
To guard against over-fitting we used early stopping regularisation with a patience of~900 epochs.
To find a reasonable local minimum of the loss function, we used the Adam optimiser~\cite{Kingma:2014aa} with default values beside the learning rate. Learning rate scheduling has been implemented with a starting value of~0.001, decreasing by a factor of~3 every~7000 epochs or when exceeding the patience until~$10^{-7}$ is reached.
A batch size of~3200 was found to be suitable.
In order to minimise a meaningful objective, we customised the loss function such that the MAPE of the true, non-logarithmic cross-sections is minimised.
As mentioned in section 4.2, the training data for the axial-vector case comes in two batches which are sampled from a disjunct volume of the parameter space. Since the huge difference in point density of the batches prevent our neural networks from performing well when training a single neural network on both batches, we used one neural network for each batch and stacked both nets. The performance shown in figure~\ref{fig:xsec_nn_performance} is for the batch containing~20000 samples and because the performance for the other batch was very similar, we omit the presentation of its performance plot. 

\begin{figure}[!t]
\centering
\includegraphics[width=0.84\columnwidth]{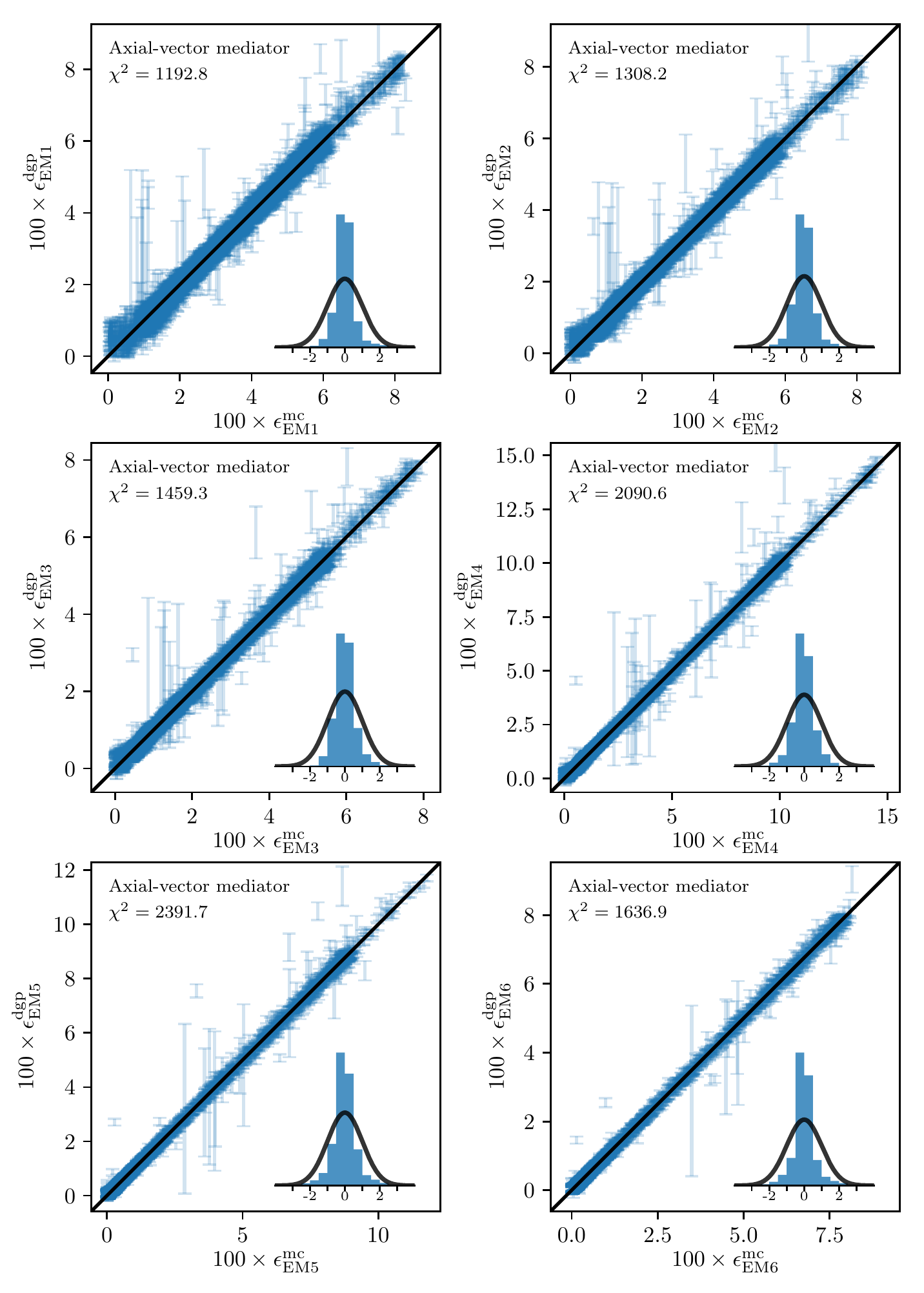}
\caption{
\textbf{Axial-vector mediator:}
The performance of the DGPs described in section~\ref{sec:nn} and appendix~\ref{sec:mlvalidation} in predicting the signal region efficiencies for the ATLAS monojet signal regions.
The blue error bars are the results for the testing dataset ($N = 3000$), previously unseen by the DGPs. The horizontal axes show the rescaled efficiencies as calculated by the traditional MC set-up for each of the six exclusive signal regions (see section~\ref{sec:mocksignals}), while the vertical axes show the rescaled DGP prediction.
The black line indicates perfect agreement i.e.\ $\epsilon^{\mathrm{dgp}} = \epsilon^{\mathrm{mc}}$.
Also indicated is the MAPE, and the $\chi^2$ of the validation dataset.
The inset in each panel shows how the distribution of the $(\epsilon^{\mathrm{dgp}} - \epsilon^{\mathrm{mc}})/\sigma_{\mathrm{dgp}}$ compares with the standard normal distribution.
\label{fig:axialvector_eff_dgp_performance}}
\end{figure}

\begin{figure}[!t]
\centering
\includegraphics[width=0.84\columnwidth]{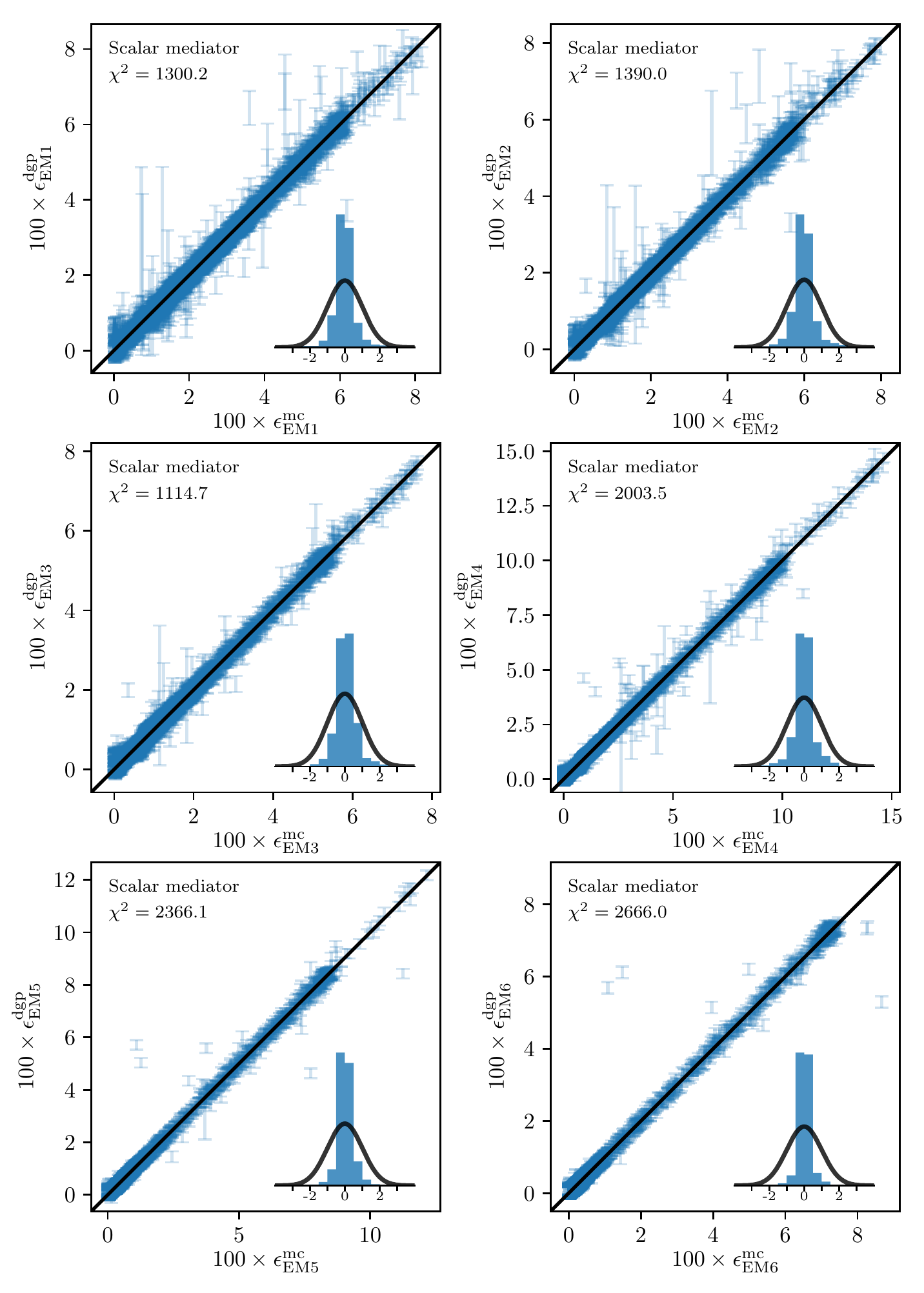}
\caption{
\textbf{Scalar mediator:}
The performance of the DGPs described in section~\ref{sec:nn} and appendix~\ref{sec:mlvalidation} in predicting the signal region efficiencies for the ATLAS monojet signal regions.
The blue error bars are the results for the testing dataset ($N = 3000$), previously unseen by the DGPs. The horizontal axes show the rescaled efficiencies as calculated by the traditional MC set-up for each of the six exclusive signal regions (see section~\ref{sec:mocksignals}), while the vertical axes show the rescaled DGP prediction.
The black line indicates perfect agreement i.e.\ $\epsilon^{\mathrm{dgp}} = \epsilon^{\mathrm{mc}}$.
Also indicated is the MAPE, and the $\chi^2$ of the validation dataset.
The inset in each panel shows how the distribution of the $(\epsilon^{\mathrm{dgp}} - \epsilon^{\mathrm{mc}})/\sigma_{\mathrm{dgp}}$ compares with the standard normal distribution.
\label{fig:scalar_eff_dgp_performance}}
\end{figure}

Figures~\ref{fig:axialvector_eff_dgp_performance} and~\ref{fig:scalar_eff_dgp_performance} show the distributed Gaussian processes (DGPs) performance in predicting the signal region efficiency for the six ATLAS monojet exclusive signal regions (EM1 - EM6).
The $\chi^2$ are less than expected, indicating that the Gaussian processes are overly conservative when estimating the errors. 
As seen in the inset panels, the distributions of $(\epsilon^{\mathrm{dgp}} - \epsilon^{\mathrm{mc}})/\sigma_{\mathrm{dgp}}$ for the test points are more peaked than the expected standard normal.
However, the discrepancies are conservative in nature, and are not significant enough to affect the final results.
The procedure to train the DGP predictors was as follows: 
An additional input feature was constructed, the mass gap $\mMed - 2\mDM$, which captures the reduced signal region efficiency when the eponymous monojet becomes soft as less energy becomes available.
The inputs were preprocessed by rotating them to eliminate correlations, and
both the input and the output were rescaled so that each was distributed as a standard normal.
The DGP architecture we used was a single layer of~32 experts (each a GP), with a standard squared exponential kernel with independent length scale for each dimension, often referred to as automatic relevance detection (ARD). For more context on GPs we refer the reader to, e.g., ref.~\cite{Rasmussen:2005:GPM:1162254}.
The training data was split among the experts using the k-dimensional (KD) tree algorithm, the dataset was split into~32 subsets and each expert was assigned two random subsets, ensuring some overlap between the experts.

\begin{figure}[!t]
\centering
\includegraphics{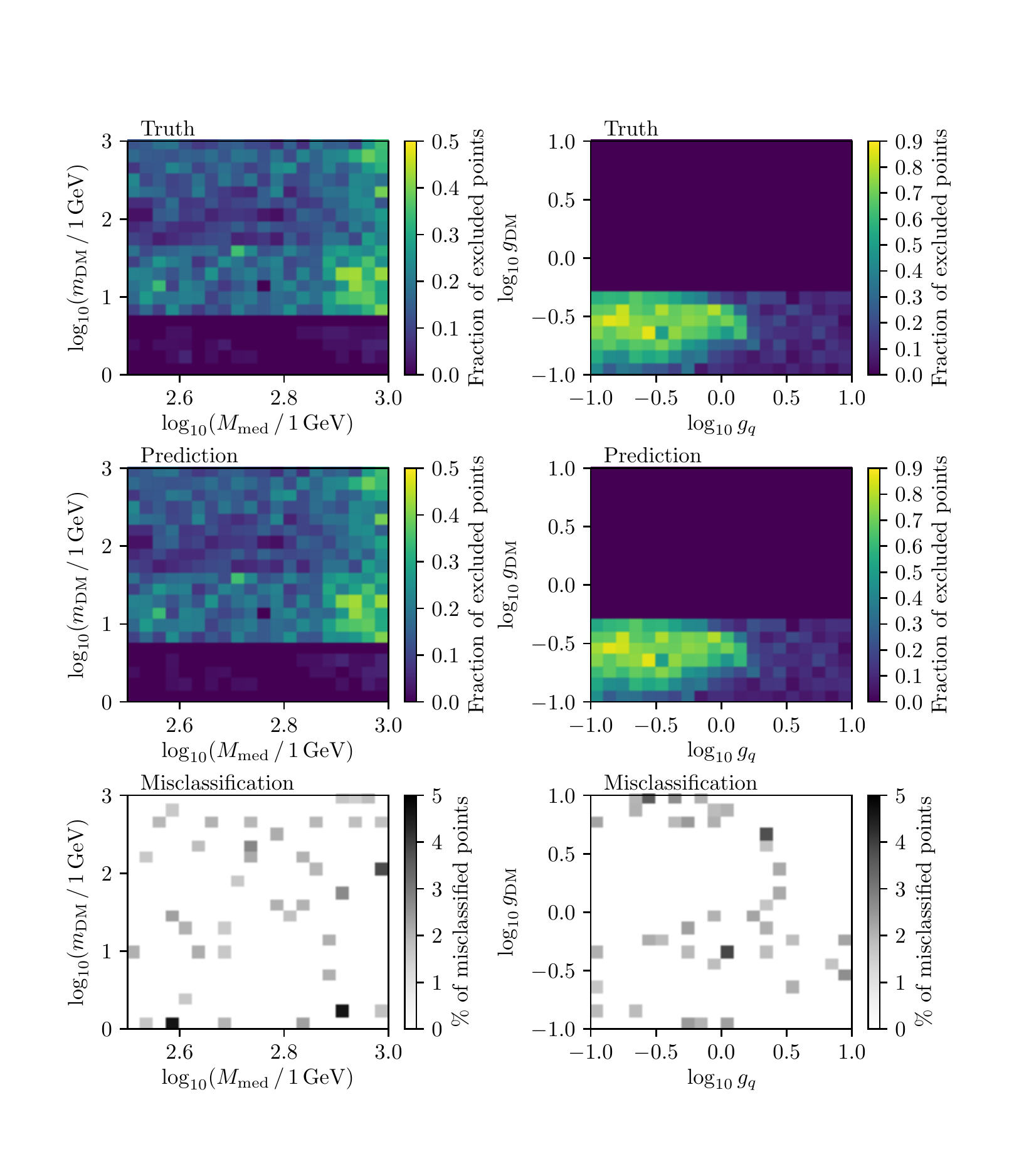}
\caption{Comparison of points excluded by the dijet search using the traditional MC set-up (described in section~\ref{subsec:futuremono}) and the predictions by the random forest classifier (described in section~\ref{sec:nn} and appendix~\ref{sec:mlvalidation}).
The fraction of excluded points are shown projected on two planes: $(\mMed, \mDM)$ and $(\gSM, \gDM)$.
The panels labelled ``Truth'' show the results from the MC set-up, while the panels labelled ``Prediction'' shows the predictions from the random forest classifier.
The last row, with panels labelled ``Misclassification'' show the misclassification, i.e.\ how the truth and prediction columns differ.
\label{fig:dijet_randomforest}}
\end{figure}

Finally, figure~\ref{fig:dijet_randomforest} shows the performance of the dijet exclusion classifier for all of the example dataset.
We used the scikit-learn implementation~\cite{scikit-learn} with~250 decision trees, max~3 features, and `entropy' as the splitting criterion.
No specific preprocessing was necessary, as decision trees are quite robust to input variance.
Using 4000 model points as the test dataset we found that the classifier misclassified~43 points, or~$1.1\%$. 

%%%%%%%%%%%%%%%%%%%%%%%%%%%%%%%%%%%%%%%%%%%%%%%%%%%%%%%%%%%%%%%%%%%%%%%%%%%%

%%%%%%%%%%%%%%%%%%%%%%%%%%%%%%%%%%%%%%%%%%%%%%%%%%%%%%%%%%%%%%%%%%%%%%%%%%%%
%%%%%%%%%%%%%%%%%%%%%%%%%%%%%%%%%%%%%%%%%%%%%%%%%%%%%%%%%%%%%%%%%%%%%%%%%%%%

%%%%%%%%%%%%%%%%%%%%%%%%%%%%%%%%%%%%%%%%%%%%%%%%%%%%%%%%%%%%%%%%%%%%%%%%%%%%
%%%%%%%%%%%%%%%%%%%%%%%%%%%%%%%%%%%%%%%%%%%%%%%%%%%%%%%%%%%%%%%%%%%%%%%%%%%%

%%%%%%%%%%%%%%%%%%%%%%%%%%%%%%%%%%%%%%%%%%%%%%%%%%%%%%%%%%%%%%%%%%%%%%%%%%%%
%%%%%%%%%%%%%%%%%%%%%%%%%%%%%%%%%%%%%%%%%%%%%%%%%%%%%%%%%%%%%%%%%%%%%%%%%%%%

%%%%%%%%%%%%%%%%%%%%%%%%%%%%%%%%%%%%%%%%%%%%%%%%%%%%%%%%%%%%%%%%%%%%%%%%%%%%
%%%%%%%%%%%%%%%%%%%%%%%%%%%%%%%%%%%%%%%%%%%%%%%%%%%%%%%%%%%%%%%%%%%%%%%%%%%%
\section{Posterior distributions with a larger systematic uncertainty  \label{sec:addfigs}}
%%%%%%%%%%%%%%%%%%%%%%%%%%%%%%%%%%%%%%%%%%%%%%%%%%%%%%%%%%%%%%%%%%%%%%%%%%%%
%%%%%%%%%%%%%%%%%%%%%%%%%%%%%%%%%%%%%%%%%%%%%%%%%%%%%%%%%%%%%%%%%%%%%%%%%%%%

\begin{figure}[t!]
\centering
\includegraphics[width=0.95\columnwidth]{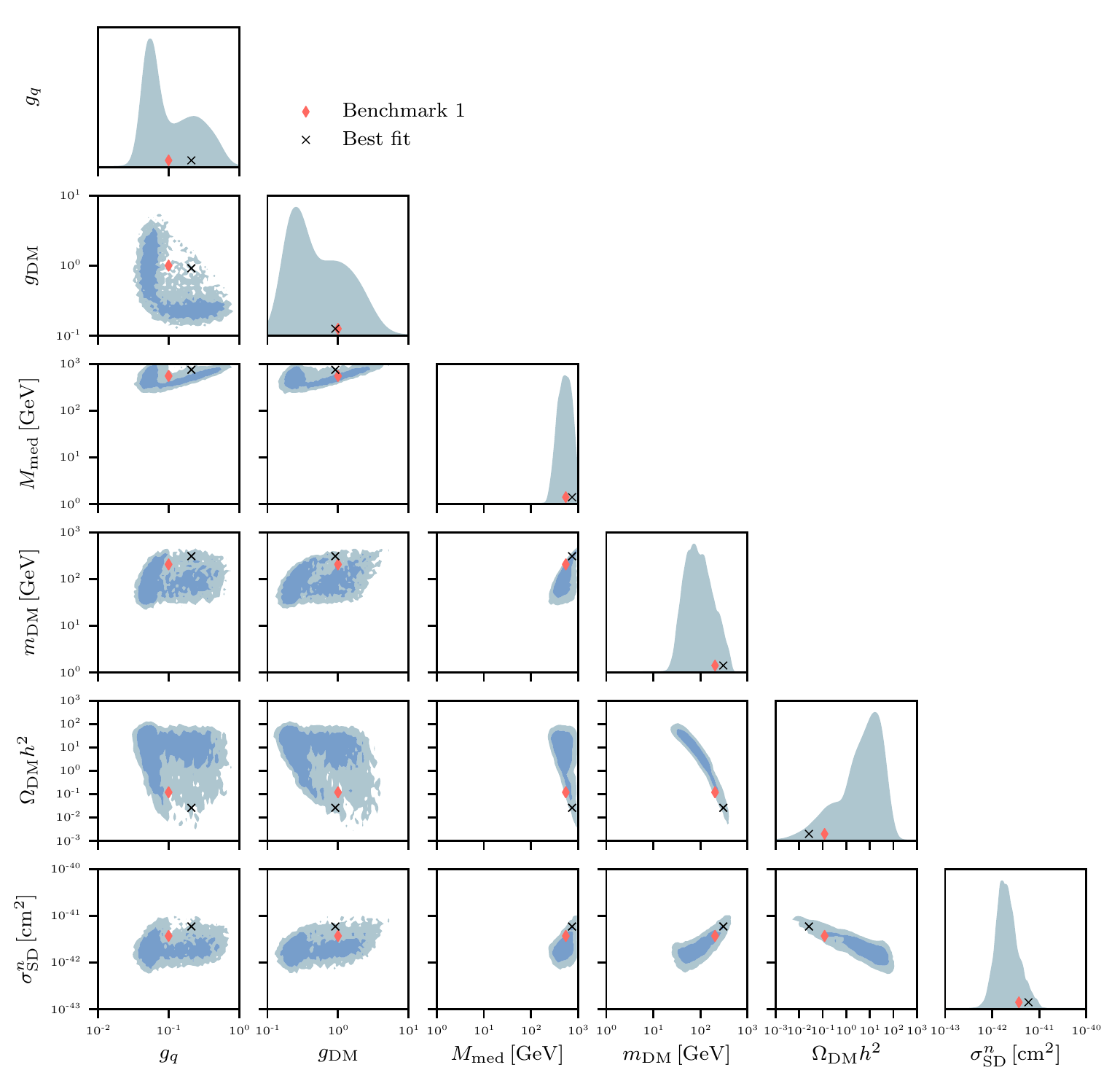}
\caption{
\textbf{Axial-vector mediator:}
Posteriors over the model parameters and the phenomenologically interesting quantities assuming future excesses in LZ and the ATLAS monojet search, assuming a~3\% systematic error on the Standard Model rates.
The full one-dimensional posteriors are shown. In the two-dimensional case, the~68\% (dark blue) and~95\% (light blue) credibility regions are shown.
Also marked are the best fit point (black cross), and the benchmark model (red diamond). 
These results are similar to the posterior distributions in figures~\ref{fig:axialvector_1pc_params} and~\ref{fig:axialvector_1pc_pheno} so we are able to draw the same conclusions.
\label{fig:axialvector_3pc}}
\end{figure}

\begin{figure}[t!]
\centering
\includegraphics[width=0.95\columnwidth]{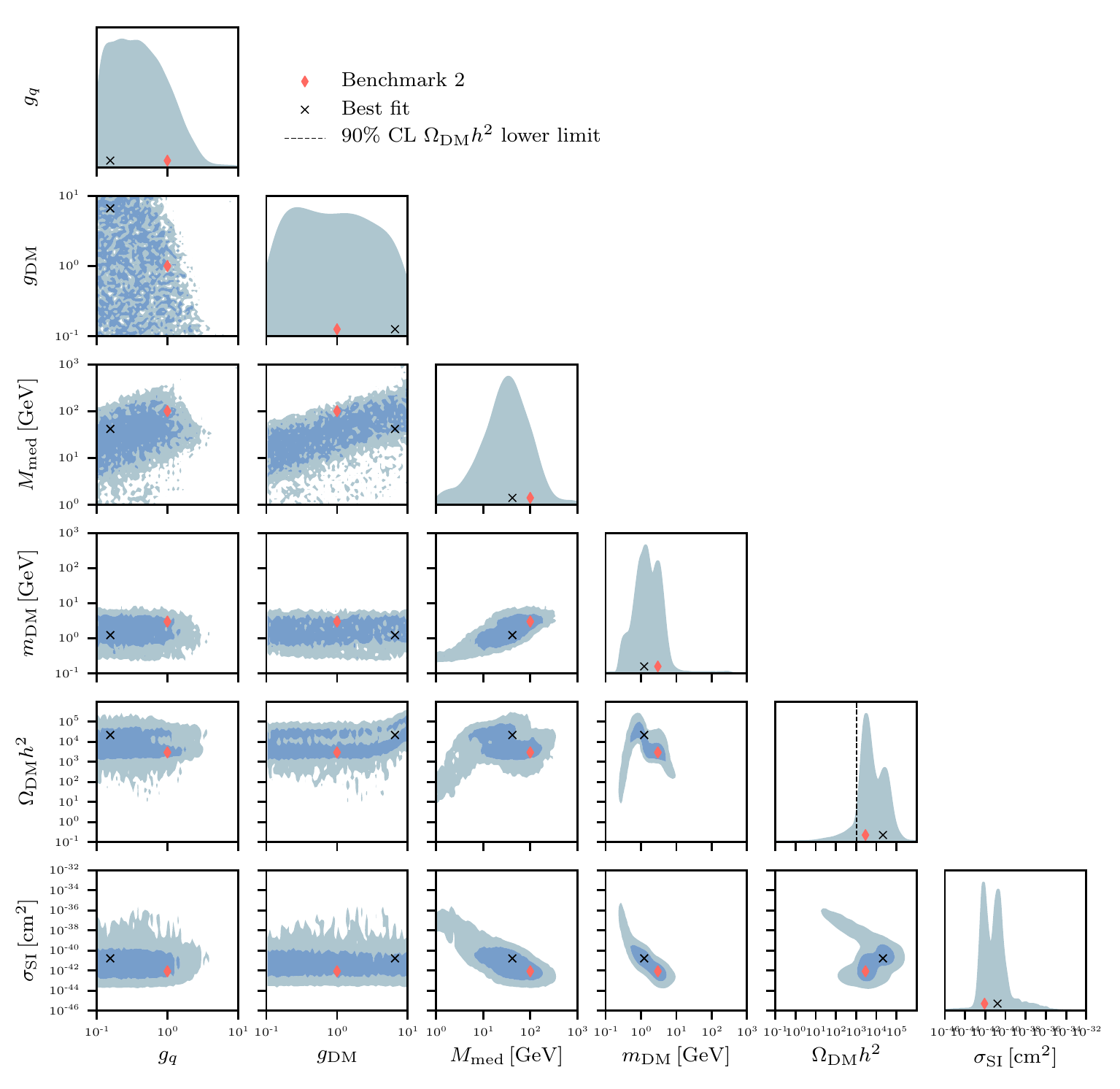}
\caption{
\textbf{Scalar mediator:}
Posteriors over the model parameters and the phenomenologically interesting quantities assuming future excesses in CRESST and the ATLAS monojet search, assuming a~3\% systematic error on the Standard Model rates.
The full one-dimensional posteriors are shown. In the two-dimensional case, the~68\% (dark blue) and~95\% (light blue) credibility regions are shown.
Also marked are the best fit point (black cross), and the benchmark model (red diamond). These results are similar to the posterior distributions in figures~\ref{fig:scalar_1pc_params} and~\ref{fig:scalar_1pc_pheno} so we are able to draw the same conclusions.
\label{fig:scalar_3pc}}
\end{figure}

In this appendix we include supplementary figures showing the results of our global scan when we increase the systematic error on the Standard Model background rates for the future monojet analysis from~1\% to~3\% (as discussed in section~\ref{subsec:futuremono}). Figures~\ref{fig:axialvector_3pc} and~\ref{fig:scalar_3pc} show the posteriors for a~3\% systematic error for the axial-vector and scalar simplified models, respectively. These should be compared with figures~\ref{fig:axialvector_1pc_params} to~\ref{fig:scalar_1pc_pheno}, the results with a~1\% systematic error.

In the axial-vector case, degrading the impact of the monojet excess by increasing the systematic error reduces our ability to reconstruct the model parameters.
The reconstruction of the couplings $\gSM$ and $\gDM$ especially suffers in comparison with the $1\%$ reconstruction, but the mass reconstruction also suffers to a lesser degree.
The reconstruction of the phenomenological observables is consequently also impacted, and the freeze-out density is much less constrained. However, it is still consistent with the observationally inferred value.

In the scalar mediator case, increasing to 3\% systematic error means our ability to reconstruct the mediator-quark coupling $\gSM$ deteriorates. However, the phenomenological consequences do not change. Namely, the freeze-out density is reconstructed at a much higher value than the observationally inferred value, while $\sigma_{\rm{SI}}$ is reconstructed around the true value.

% axial 3%
% lg oh2 : (-0.45977438074670246, 1.4435236738762498) (two-sided)
% lg g_MED : (-1.3136593058979977, -0.5335642951749685) (two-sided)
% lg m_MED : (2.5658726990012752, 2.862493381758829) (two-sided)
% lg sigmaSD : (-41.89092878367574, -41.44055017152269) (two-sided)
% lg m_DM : (1.6755977563626818, 2.2796574312182387) (two-sided)
% lg g_DM : (-0.6805618592663694, 0.18748370536973236) (two-sided)

% scalar 3%
% lg m_MED : (0.969462225546054, 1.9604680146321485) (two-sided)
% lg g_DM : (-0.6547934890823994, 0.5957965268065862) (two-sided)
% lg m_DM : (-0.17799512020527208, 0.5468706343449118) (two-sided)
% lg sigmaSI : (-42.25367507585997, -40.12515162941321) (two-sided)
% lg g_MED : 0.1258864971907298 (ul)
% lg oh2 : 3.0301321312879916 (ll)

%%%%%%%%%%%%%%%%%%%%%%%%%%%%%%%%%%%%%%%%%%%%%%%%%%%%%%%%%%%%%%%%%%%%%%%%%%%%
%%%%%%%%%%%%%%%%%%%%%%%%%%%%%%%%%%%%%%%%%%%%%%%%%%%%%%%%%%%%%%%%%%%%%%%%%%%%

%%%%%%%%%%%%%%%%%%%%%%%%%%%%%%%%%%%%%%%%%%%%%%%%%%%%%%%%%%%%%%%%%%%%%%%%%%%%
%%%%%%%%%%%%%%%%%%%%%%%%%%%%%%%%%%%%%%%%%%%%%%%%%%%%%%%%%%%%%%%%%%%%%%%%%%%%

%%%%%%%%%%%%%%%%%%%%%%%%%%%%%%%%%%%%%%%%%%%%%%%%%%%%%%%%%%%%%%%%%%%%%%%%%%%%
%%%%%%%%%%%%%%%%%%%%%%%%%%%%%%%%%%%%%%%%%%%%%%%%%%%%%%%%%%%%%%%%%%%%%%%%%%%%
\section{Profile likelihood results  \label{sec:addfigsprofile}}
%%%%%%%%%%%%%%%%%%%%%%%%%%%%%%%%%%%%%%%%%%%%%%%%%%%%%%%%%%%%%%%%%%%%%%%%%%%%
%%%%%%%%%%%%%%%%%%%%%%%%%%%%%%%%%%%%%%%%%%%%%%%%%%%%%%%%%%%%%%%%%%%%%%%%%%%%

\begin{figure}[t!]
\centering
\includegraphics{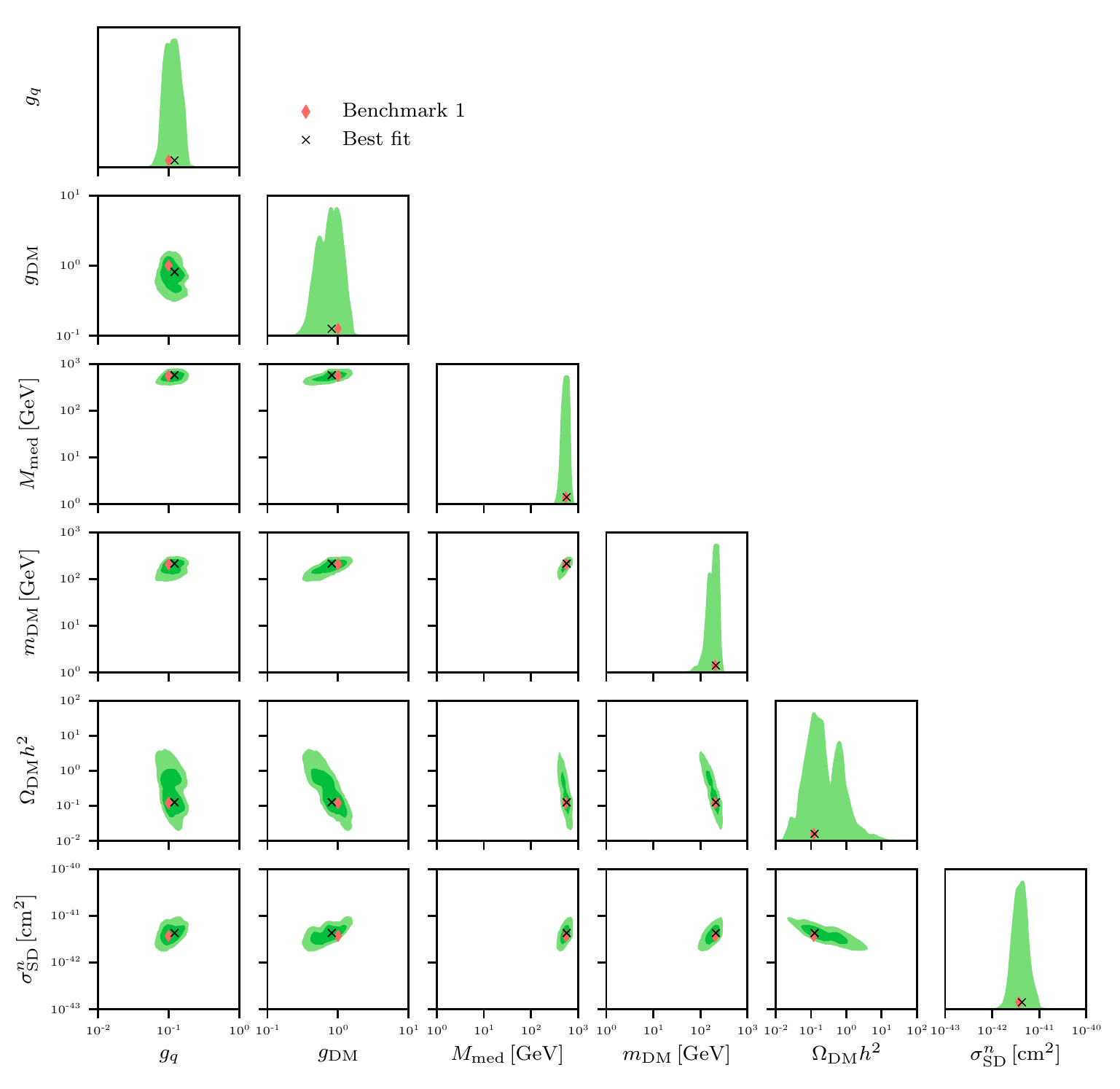}
\caption{
\textbf{Axial-vector mediator:}
Profile likelihoods over the model parameters and the phenomenologically interesting quantities assuming future excesses in an ATLAS monojet search (with 1\% syst.) and LZ.
The full one-dimensional profile likelihoods are shown. In the two-dimensional case, the 68\% (dark green) and 95\% (light green) confidence regions are shown.
Also marked are the best fit point (black cross), and the benchmark model (red diamond). These results are similar to the posterior distributions in figures~\ref{fig:axialvector_1pc_params} and~\ref{fig:axialvector_1pc_pheno} so we are able to draw the same conclusions.
\label{fig:axialvector_1pc_proflike}}
\end{figure}

\begin{figure}[t!]
\centering
\includegraphics{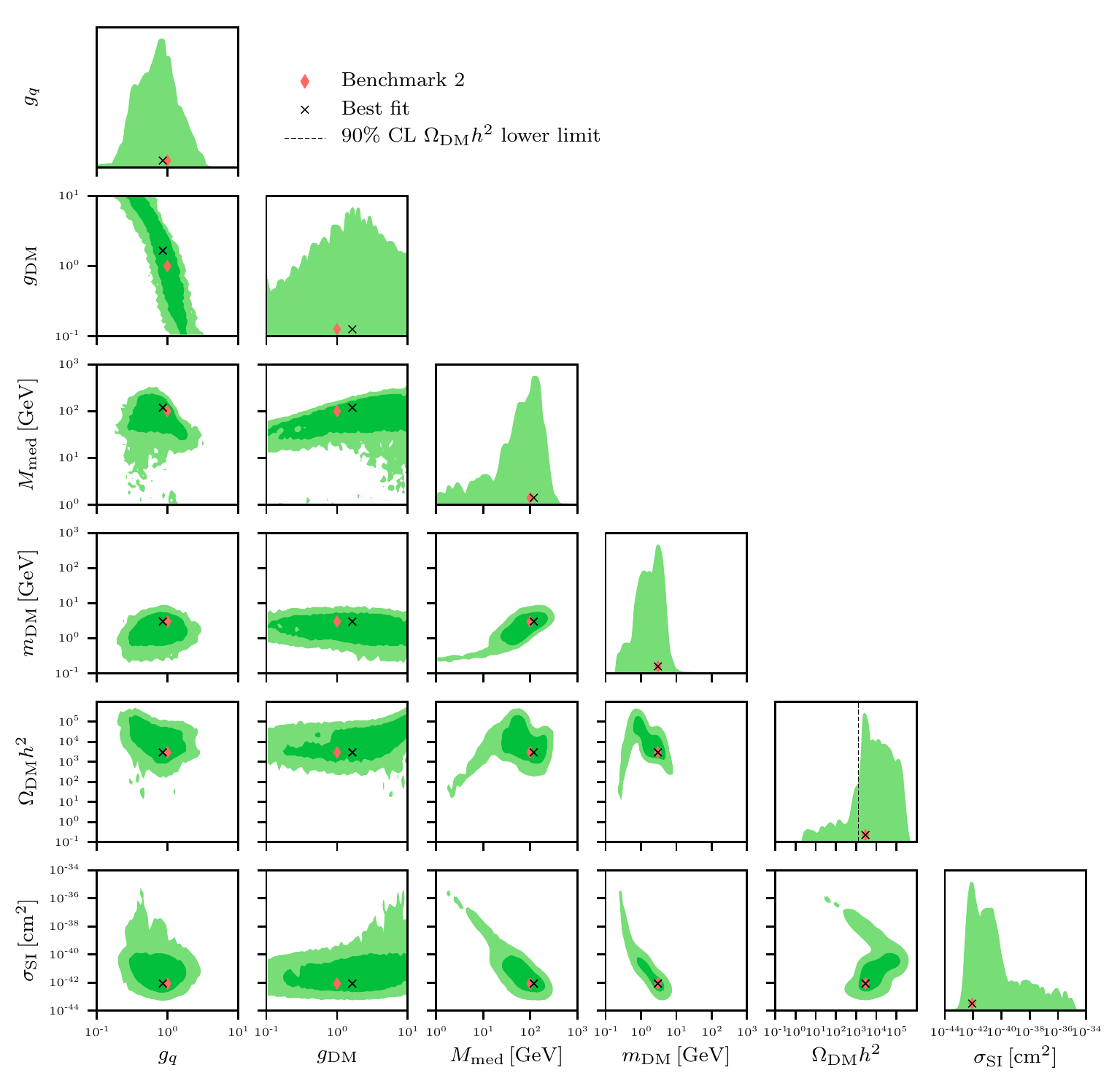}
\caption{
\textbf{Scalar mediator:}
Profile likelihoods over the model parameters and the phenomenologically interesting quantities assuming future excesses in an ATLAS monojet search (with 1\% syst.) and CRESST.
The full one-dimensional profile likelihoods are shown. In the two-dimensional case, the 68\% (dark green) and 95\% (light green) confidence regions are shown.
Also marked are the best fit point (black cross), and the benchmark model (red diamond). These results are similar to the posterior distributions in figures~\ref{fig:scalar_1pc_params} and~\ref{fig:scalar_1pc_pheno} so we are able to draw the same conclusions.
\label{fig:scalar_1pc_proflike}}
\end{figure}

In figures~\ref{fig:axialvector_1pc_proflike} and~\ref{fig:scalar_1pc_proflike}, we show the profile likelihoods for the axial-vector and scalar mediator cases, respectively.
These are the equivalents of the posteriors shown in figures~\ref{fig:axialvector_1pc_params} to~\ref{fig:scalar_1pc_pheno}, and assume a~1\% systematic error on the Standard Model rates for the future monojet analysis.
To visualise the full multidimensional likelihood we profile it with respect to some parameter(s) of interest~$\alpha$.
If we denote the parameters we suppress with $\beta$, we define the profile likelihood $\mathcal{L}(\mathbf{D}|\alpha)$ in terms of the full likelihood $\mathcal{L}(\mathbf{D}|\alpha,\beta)$ as
\begin{equation}
\mathcal{L}(\mathbf{D}|\alpha) = \max_{\beta} \mathcal{L}(\mathbf{D}|\alpha,\beta).
\end{equation}

The fundamental difference between profiling (maximising) and marginalising (integrating, see eq.~\eqref{eq:marg}) makes profile likelihoods and marginalised posteriors complementary; 
with both we can identify the regions that are highly tuned.
In these specific cases the posteriors and the profile likelihoods do not differ significantly so we are able to draw the same conclusions.

%%%%%%%%%%%%%%%%%%%%%%%%%%%%%%%%%%%%%%%%%%%%%%%%%%%%%%%%%%%%%%%%%%%%%%%%%%%%
%%%%%%%%%%%%%%%%%%%%%%%%%%%%%%%%%%%%%%%%%%%%%%%%%%%%%%%%%%%%%%%%%%%%%%%%%%%%

%%%%%%%%%%%%%%%%%%%%%%%%%%%%%%%%%%%%%%%%%%%%%%%%%%%%%%%%%%%%%%%%%%%%%%%%%%%%
%%%%%%%%%%%%%%%%%%%%%%%%%%%%%%%%%%%%%%%%%%%%%%%%%%%%%%%%%%%%%%%%%%%%%%%%%%%%
\FloatBarrier
\bibliographystyle{JHEP}
\bibliography{refs}

\providecommand{\href}[2]{#2}\begingroup\raggedright\begin{thebibliography}{100}

\bibitem{Bertone:2010zza}
G.~Bertone, ed., {\em {Particle Dark Matter: Observations, Models and
  Searches}}.
\newblock Cambridge Univ. Press, Cambridge, 2010.

\bibitem{Jungman:1995df}
G.~Jungman, M.~Kamionkowski, and K.~Griest, {\it {Supersymmetric dark matter}},
   {\em Phys.Rept.} {\bf 267} (1996) 195--373,
  [\href{http://arxiv.org/abs/hep-ph/9506380}{{\tt hep-ph/9506380}}].

\bibitem{Bergstrom00}
L.~{Bergstr\"om}, {\it {Non-baryonic dark matter: observational evidence and
  detection methods}},  {\em Rep. Prog. Phys.} {\bf 63} (2000) 793--841,
  [\href{http://arxiv.org/abs/hep-ph/0002126}{{\tt hep-ph/0002126}}].

\bibitem{Bertone05}
G.~Bertone, D.~Hooper, and J.~Silk, {\it {Particle dark matter: Evidence,
  candidates and constraints}},  {\em Phys. Rept.} {\bf 405} (2005) 279--390,
  [\href{http://arxiv.org/abs/hep-ph/0404175}{{\tt hep-ph/0404175}}].

\bibitem{Lee:1977ua}
B.~W. Lee and S.~Weinberg, {\it {Cosmological Lower Bound on Heavy Neutrino
  Masses}},  {\em Phys. Rev. Lett.} {\bf 39} (1977) 165--168.

\bibitem{Hut:1977zn}
P.~Hut, {\it {Limits on Masses and Number of Neutral Weakly Interacting
  Particles}},  {\em Phys. Lett.} {\bf 69B} (1977) 85.

\bibitem{Wolfram:1978gp}
S.~Wolfram, {\it {Abundances of Stable Particles Produced in the Early
  Universe}},  {\em Phys. Lett.} {\bf 82B} (1979) 65--68.

\bibitem{Steigman:1979kw}
G.~Steigman, {\it {Cosmology Confronts Particle Physics}},  {\em Ann. Rev.
  Nucl. Part. Sci.} {\bf 29} (1979) 313--338.

\bibitem{Bernstein:1985th}
J.~Bernstein, L.~S. Brown, and G.~Feinberg, {\it {The Cosmological Heavy
  Neutrino Problem Revisited}},  {\em Phys. Rev.} {\bf D32} (1985) 3261.

\bibitem{Scherrer:1985zt}
R.~J. Scherrer and M.~S. Turner, {\it {On the Relic, Cosmic Abundance of Stable
  Weakly Interacting Massive Particles}},  {\em Phys. Rev.} {\bf D33} (1986)
  1585. [Erratum: Phys. Rev.D34,3263(1986)].

\bibitem{Srednicki:1988ce}
M.~Srednicki, R.~Watkins, and K.~A. Olive, {\it {Calculations of Relic
  Densities in the Early Universe}},  {\em Nucl. Phys.} {\bf B310} (1988) 693.

\bibitem{Ade:2015xua}
{\bf Planck} Collaboration, P.~A.~R. Ade et~al., {\it {Planck 2015 results.
  XIII. Cosmological parameters}},  {\em Astron. Astrophys.} {\bf 594} (2016)
  A13, [\href{http://arxiv.org/abs/1502.01589}{{\tt arXiv:1502.01589}}].

\bibitem{Steigman:2012nb}
G.~Steigman, B.~Dasgupta, and J.~F. Beacom, {\it {Precise Relic WIMP Abundance
  and its Impact on Searches for Dark Matter Annihilation}},  {\em Phys. Rev.}
  {\bf D86} (2012) 023506, [\href{http://arxiv.org/abs/1204.3622}{{\tt
  arXiv:1204.3622}}].

\bibitem{Battaglieri:2017aum}
M.~Battaglieri et~al., {\it {US Cosmic Visions: New Ideas in Dark Matter 2017:
  Community Report}},  \href{http://arxiv.org/abs/1707.04591}{{\tt
  arXiv:1707.04591}}.

\bibitem{Bertone:2010at}
G.~Bertone, {\it {The moment of truth for WIMP Dark Matter}},  {\em Nature}
  {\bf 468} (2010) 389--393, [\href{http://arxiv.org/abs/1011.3532}{{\tt
  arXiv:1011.3532}}].

\bibitem{Arcadi:2017kky}
G.~Arcadi, M.~Dutra, P.~Ghosh, M.~Lindner, Y.~Mambrini, M.~Pierre, S.~Profumo,
  and F.~S. Queiroz, {\it {The Waning of the WIMP? A Review of Models,
  Searches, and Constraints}},  \href{http://arxiv.org/abs/1703.07364}{{\tt
  arXiv:1703.07364}}.

\bibitem{Baltz:2006fm}
E.~A. Baltz, M.~Battaglia, M.~E. Peskin, and T.~Wizansky, {\it {Determination
  of dark matter properties at high-energy colliders}},  {\em Phys. Rev.} {\bf
  D74} (2006) 103521, [\href{http://arxiv.org/abs/hep-ph/0602187}{{\tt
  hep-ph/0602187}}].

\bibitem{Bertone:2007xj}
G.~Bertone, D.~G. Cerdeno, J.~I. Collar, and B.~C. Odom, {\it {WIMP
  identification through a combined measurement of axial and scalar
  couplings}},  {\em Phys. Rev. Lett.} {\bf 99} (2007) 151301,
  [\href{http://arxiv.org/abs/0705.2502}{{\tt arXiv:0705.2502}}].

\bibitem{Roszkowski:2017dou}
L.~Roszkowski, S.~Trojanowski, and K.~Turzynski, {\it {Towards understanding
  thermal history of the Universe through direct and indirect detection of dark
  matter}},  {\em JCAP} {\bf 1710} (2017), no.~10 005,
  [\href{http://arxiv.org/abs/1703.00841}{{\tt arXiv:1703.00841}}].

\bibitem{Baum:2017kfa}
S.~Baum, R.~Catena, J.~Conrad, K.~Freese, and M.~B. Krauss, {\it {Determining
  Dark Matter properties with a XENONnT/LZ signal and LHC-Run3 mono-jet
  searches}},  \href{http://arxiv.org/abs/1709.06051}{{\tt arXiv:1709.06051}}.

\bibitem{Allanach:2005kz}
B.~C. Allanach and C.~G. Lester, {\it {Multi-dimensional mSUGRA likelihood
  maps}},  {\em Phys. Rev.} {\bf D73} (2006) 015013,
  [\href{http://arxiv.org/abs/hep-ph/0507283}{{\tt hep-ph/0507283}}].

\bibitem{Hooper:2006wv}
D.~Hooper and A.~M. Taylor, {\it {Determining Supersymmetric Parameters With
  Dark Matter Experiments}},  {\em JCAP} {\bf 0703} (2007) 017,
  [\href{http://arxiv.org/abs/hep-ph/0607086}{{\tt hep-ph/0607086}}].

\bibitem{Bernal:2008zk}
N.~Bernal, A.~Goudelis, Y.~Mambrini, and C.~Munoz, {\it {Determining the WIMP
  mass using the complementarity between direct and indirect searches and the
  ILC}},  {\em JCAP} {\bf 0901} (2009) 046,
  [\href{http://arxiv.org/abs/0804.1976}{{\tt arXiv:0804.1976}}].

\bibitem{Trotta:2008bp}
R.~Trotta, F.~Feroz, M.~P. Hobson, L.~Roszkowski, and R.~Ruiz~de Austri, {\it
  {The Impact of priors and observables on parameter inferences in the
  Constrained MSSM}},  {\em JHEP} {\bf 12} (2008) 024,
  [\href{http://arxiv.org/abs/0809.3792}{{\tt arXiv:0809.3792}}].

\bibitem{Bertone:2010rv}
G.~Bertone, D.~G. Cerdeno, M.~Fornasa, R.~Ruiz~de Austri, and R.~Trotta, {\it
  {Identification of Dark Matter particles with LHC and direct detection
  data}},  {\em Phys. Rev.} {\bf D82} (2010) 055008,
  [\href{http://arxiv.org/abs/1005.4280}{{\tt arXiv:1005.4280}}].

\bibitem{Buchmueller:2011ki}
O.~Buchmueller et~al., {\it {Supersymmetry and Dark Matter in Light of LHC 2010
  and Xenon100 Data}},  {\em Eur. Phys. J.} {\bf C71} (2011) 1722,
  [\href{http://arxiv.org/abs/1106.2529}{{\tt arXiv:1106.2529}}].

\bibitem{Strege:2011pk}
C.~Strege, G.~Bertone, D.~G. Cerdeno, M.~Fornasa, R.~Ruiz~de Austri, and
  R.~Trotta, {\it {Updated global fits of the cMSSM including the latest LHC
  SUSY and Higgs searches and XENON100 data}},  {\em JCAP} {\bf 1203} (2012)
  030, [\href{http://arxiv.org/abs/1112.4192}{{\tt arXiv:1112.4192}}].

\bibitem{Roszkowski:2012uf}
L.~Roszkowski, E.~M. Sessolo, and Y.-L.~S. Tsai, {\it {Bayesian Implications of
  Current LHC Supersymmetry and Dark Matter Detection Searches for the
  Constrained MSSM}},  {\em Phys. Rev.} {\bf D86} (2012) 095005,
  [\href{http://arxiv.org/abs/1202.1503}{{\tt arXiv:1202.1503}}].

\bibitem{Mambrini:2012ue}
Y.~Mambrini, M.~H.~G. Tytgat, G.~Zaharijas, and B.~Zaldivar, {\it
  {Complementarity of Galactic radio and collider data in constraining WIMP
  dark matter models}},  {\em JCAP} {\bf 1211} (2012) 038,
  [\href{http://arxiv.org/abs/1206.2352}{{\tt arXiv:1206.2352}}].

\bibitem{Arbey:2013iza}
A.~Arbey, M.~Battaglia, and F.~Mahmoudi, {\it {Combining monojet,
  supersymmetry, and dark matter searches}},  {\em Phys. Rev.} {\bf D89}
  (2014), no.~7 077701, [\href{http://arxiv.org/abs/1311.7641}{{\tt
  arXiv:1311.7641}}].

\bibitem{Strege:2014ija}
C.~Strege, G.~Bertone, G.~J. Besjes, S.~Caron, R.~Ruiz~de Austri, A.~Strubig,
  and R.~Trotta, {\it {Profile likelihood maps of a 15-dimensional MSSM}},
  {\em JHEP} {\bf 09} (2014) 081, [\href{http://arxiv.org/abs/1405.0622}{{\tt
  arXiv:1405.0622}}].

\bibitem{Demir:2014jqa}
D.~A. Demir and C.~S. Un, {\it {Stop on Top: SUSY Parameter Regions,
  Fine-Tuning Constraints}},  {\em Phys. Rev.} {\bf D90} (2014) 095015,
  [\href{http://arxiv.org/abs/1407.1481}{{\tt arXiv:1407.1481}}].

\bibitem{Cerdeno:2015ega}
D.~G. Cerdeno, M.~Peiro, and S.~Robles, {\it {Fits to the Fermi-LAT GeV excess
  with RH sneutrino dark matter: implications for direct and indirect dark
  matter searches and the LHC}},  {\em Phys. Rev.} {\bf D91} (2015), no.~12
  123530, [\href{http://arxiv.org/abs/1501.01296}{{\tt arXiv:1501.01296}}].

\bibitem{deVries:2015hva}
K.~J. de~Vries et~al., {\it {The pMSSM10 after LHC Run 1}},  {\em Eur. Phys.
  J.} {\bf C75} (2015), no.~9 422, [\href{http://arxiv.org/abs/1504.03260}{{\tt
  arXiv:1504.03260}}].

\bibitem{Buchmueller:2015uqa}
O.~Buchmueller, M.~Citron, J.~Ellis, S.~Guha, J.~Marrouche, K.~A. Olive,
  K.~de~Vries, and J.~Zheng, {\it {Collider Interplay for Supersymmetry, Higgs
  and Dark Matter}},  {\em Eur. Phys. J.} {\bf C75} (2015), no.~10 469,
  [\href{http://arxiv.org/abs/1505.04702}{{\tt arXiv:1505.04702}}]. [Erratum:
  Eur. Phys. J.C76,no.4,190(2016)].

\bibitem{Bertone:2015tza}
G.~Bertone, F.~Calore, S.~Caron, R.~Ruiz, J.~S. Kim, R.~Trotta, and C.~Weniger,
  {\it {Global analysis of the pMSSM in light of the Fermi GeV excess:
  prospects for the LHC Run-II and astroparticle experiments}},  {\em JCAP}
  {\bf 1604} (2016), no.~04 037, [\href{http://arxiv.org/abs/1507.07008}{{\tt
  arXiv:1507.07008}}].

\bibitem{Bagnaschi:2015eha}
E.~A. Bagnaschi et~al., {\it {Supersymmetric Dark Matter after LHC Run 1}},
  {\em Eur. Phys. J.} {\bf C75} (2015) 500,
  [\href{http://arxiv.org/abs/1508.01173}{{\tt arXiv:1508.01173}}].

\bibitem{Dutta:2015exw}
J.~Dutta, P.~Konar, S.~Mondal, B.~Mukhopadhyaya, and S.~K. Rai, {\it {A Revisit
  to a Compressed Supersymmetric Spectrum with 125 GeV Higgs}},  {\em JHEP}
  {\bf 01} (2016) 051, [\href{http://arxiv.org/abs/1511.09284}{{\tt
  arXiv:1511.09284}}].

\bibitem{Ross:2016pml}
G.~G. Ross, K.~Schmidt-Hoberg, and F.~Staub, {\it {On the MSSM Higgsino mass
  and fine tuning}},  {\em Phys. Lett.} {\bf B759} (2016) 110--114,
  [\href{http://arxiv.org/abs/1603.09347}{{\tt arXiv:1603.09347}}].

\bibitem{Liem:2016xpm}
S.~Liem, G.~Bertone, F.~Calore, R.~Ruiz~de Austri, T.~M.~P. Tait, R.~Trotta,
  and C.~Weniger, {\it {Effective field theory of dark matter: a global
  analysis}},  {\em JHEP} {\bf 09} (2016) 077,
  [\href{http://arxiv.org/abs/1603.05994}{{\tt arXiv:1603.05994}}].

\bibitem{Roszkowski:2016bhs}
L.~Roszkowski, E.~M. Sessolo, S.~Trojanowski, and A.~J. Williams, {\it
  {Reconstructing WIMP properties through an interplay of signal measurements
  in direct detection, Fermi-LAT, and CTA searches for dark matter}},  {\em
  JCAP} {\bf 1608} (2016), no.~08 033,
  [\href{http://arxiv.org/abs/1603.06519}{{\tt arXiv:1603.06519}}].

\bibitem{Caron:2016hib}
S.~Caron, J.~S. Kim, K.~Rolbiecki, R.~Ruiz~de Austri, and B.~Stienen, {\it {The
  BSM-AI project: SUSY-AI -- generalizing LHC limits on supersymmetry with
  machine learning}},  {\em Eur. Phys. J.} {\bf C77} (2017), no.~4 257,
  [\href{http://arxiv.org/abs/1605.02797}{{\tt arXiv:1605.02797}}].

\bibitem{Barr:2016sho}
A.~Barr and J.~Liu, {\it {Analysing parameter space correlations of recent 13
  TeV gluino and squark searches in the pMSSM}},  {\em Eur. Phys. J.} {\bf C77}
  (2017), no.~3 202, [\href{http://arxiv.org/abs/1608.05379}{{\tt
  arXiv:1608.05379}}].

\bibitem{Bagnaschi:2016afc}
E.~Bagnaschi et~al., {\it {Likelihood Analysis of Supersymmetric SU(5) GUTs}},
  {\em Eur. Phys. J.} {\bf C77} (2017), no.~2 104,
  [\href{http://arxiv.org/abs/1610.10084}{{\tt arXiv:1610.10084}}].

\bibitem{Bagnaschi:2016xfg}
E.~Bagnaschi et~al., {\it {Likelihood Analysis of the Minimal AMSB Model}},
  {\em Eur. Phys. J.} {\bf C77} (2017), no.~4 268,
  [\href{http://arxiv.org/abs/1612.05210}{{\tt arXiv:1612.05210}}].

\bibitem{Rogers:2016jrx}
H.~Rogers, D.~G. Cerdeno, P.~Cushman, F.~Livet, and V.~Mandic, {\it
  {Multidimensional effective field theory analysis for direct detection of
  dark matter}},  {\em Phys. Rev.} {\bf D95} (2017), no.~8 082003,
  [\href{http://arxiv.org/abs/1612.09038}{{\tt arXiv:1612.09038}}].

\bibitem{Athron:2017ard}
{\bf GAMBIT} Collaboration, P.~Athron et~al., {\it {GAMBIT: the global and
  modular beyond-the-standard-model inference tool}},  {\em Eur. Phys. J.} {\bf
  C77} (2017), no.~11 784, [\href{http://arxiv.org/abs/1705.07908}{{\tt
  arXiv:1705.07908}}].

\bibitem{Athron:2017yua}
{\bf GAMBIT} Collaboration, P.~Athron et~al., {\it {A global fit of the MSSM
  with GAMBIT}},  {\em Eur. Phys. J.} {\bf C77} (2017), no.~12 879,
  [\href{http://arxiv.org/abs/1705.07917}{{\tt arXiv:1705.07917}}].

\bibitem{Athron:2017kgt}
{\bf GAMBIT} Collaboration, P.~Athron et~al., {\it {Status of the scalar
  singlet dark matter model}},  {\em Eur. Phys. J.} {\bf C77} (2017), no.~8
  568, [\href{http://arxiv.org/abs/1705.07931}{{\tt arXiv:1705.07931}}].

\bibitem{Athron:2017qdc}
{\bf GAMBIT} Collaboration, P.~Athron et~al., {\it {Global fits of GUT-scale
  SUSY models with GAMBIT}},  {\em Eur. Phys. J.} {\bf C77} (2017), no.~12 824,
  [\href{http://arxiv.org/abs/1705.07935}{{\tt arXiv:1705.07935}}].

\bibitem{Bagnaschi:2017tru}
E.~Bagnaschi et~al., {\it {Likelihood Analysis of the pMSSM11 in Light of LHC
  13-TeV Data}},  \href{http://arxiv.org/abs/1710.11091}{{\tt
  arXiv:1710.11091}}.

\bibitem{Costa:2017gup}
J.~C. Costa et~al., {\it {Likelihood Analysis of the Sub-GUT MSSM in Light of
  LHC 13-TeV Data}},  \href{http://arxiv.org/abs/1711.00458}{{\tt
  arXiv:1711.00458}}.

\bibitem{Alves:2011wf}
{\bf LHC New Physics Working Group} Collaboration, D.~Alves, {\it {Simplified
  Models for LHC New Physics Searches}},  {\em J. Phys.} {\bf G39} (2012)
  105005, [\href{http://arxiv.org/abs/1105.2838}{{\tt arXiv:1105.2838}}].

\bibitem{Goodman:2011jq}
J.~Goodman and W.~Shepherd, {\it {LHC Bounds on UV-Complete Models of Dark
  Matter}},  \href{http://arxiv.org/abs/1111.2359}{{\tt arXiv:1111.2359}}.

\bibitem{An:2012va}
H.~An, X.~Ji, and L.-T. Wang, {\it {Light Dark Matter and $Z'$ Dark Force at
  Colliders}},  {\em JHEP} {\bf 07} (2012) 182,
  [\href{http://arxiv.org/abs/1202.2894}{{\tt arXiv:1202.2894}}].

\bibitem{Frandsen:2012rk}
M.~T. Frandsen, F.~Kahlhoefer, A.~Preston, S.~Sarkar, and K.~Schmidt-Hoberg,
  {\it {LHC and Tevatron Bounds on the Dark Matter Direct Detection
  Cross-Section for Vector Mediators}},  {\em JHEP} {\bf 07} (2012) 123,
  [\href{http://arxiv.org/abs/1204.3839}{{\tt arXiv:1204.3839}}].

\bibitem{Fox:2012ru}
P.~J. Fox and C.~Williams, {\it {Next-to-Leading Order Predictions for Dark
  Matter Production at Hadron Colliders}},  {\em Phys. Rev.} {\bf D87} (2013),
  no.~5 054030, [\href{http://arxiv.org/abs/1211.6390}{{\tt arXiv:1211.6390}}].

\bibitem{Dreiner:2013vla}
H.~Dreiner, D.~Schmeier, and J.~Tattersall, {\it {Contact Interactions Probe
  Effective Dark Matter Models at the LHC}},  {\em Europhys. Lett.} {\bf 102}
  (2013), no.~5 51001, [\href{http://arxiv.org/abs/1303.3348}{{\tt
  arXiv:1303.3348}}].

\bibitem{Cotta:2013jna}
R.~C. Cotta, A.~Rajaraman, T.~M.~P. Tait, and A.~M. Wijangco, {\it {Particle
  Physics Implications and Constraints on Dark Matter Interpretations of the
  CDMS Signal}},  {\em Phys. Rev.} {\bf D90} (2014), no.~1 013020,
  [\href{http://arxiv.org/abs/1305.6609}{{\tt arXiv:1305.6609}}].

\bibitem{Buchmueller:2013dya}
O.~Buchmueller, M.~J. Dolan, and C.~McCabe, {\it {Beyond Effective Field Theory
  for Dark Matter Searches at the LHC}},  {\em JHEP} {\bf 01} (2014) 025,
  [\href{http://arxiv.org/abs/1308.6799}{{\tt arXiv:1308.6799}}].

\bibitem{Papucci:2014iwa}
M.~Papucci, A.~Vichi, and K.~M. Zurek, {\it {Monojet versus the rest of the
  world I: t-channel models}},  {\em JHEP} {\bf 11} (2014) 024,
  [\href{http://arxiv.org/abs/1402.2285}{{\tt arXiv:1402.2285}}].

\bibitem{Buchmueller:2014yoa}
O.~Buchmueller, M.~J. Dolan, S.~A. Malik, and C.~McCabe, {\it {Characterising
  dark matter searches at colliders and direct detection experiments: Vector
  mediators}},  {\em JHEP} {\bf 01} (2015) 037,
  [\href{http://arxiv.org/abs/1407.8257}{{\tt arXiv:1407.8257}}].

\bibitem{Abdallah:2014hon}
J.~Abdallah et~al., {\it {Simplified Models for Dark Matter and Missing Energy
  Searches at the LHC}},  \href{http://arxiv.org/abs/1409.2893}{{\tt
  arXiv:1409.2893}}.

\bibitem{Malik:2014ggr}
S.~A. Malik et~al., {\it {Interplay and Characterization of Dark Matter
  Searches at Colliders and in Direct Detection Experiments}},  {\em Phys. Dark
  Univ.} {\bf 9-10} (2015) 51--58, [\href{http://arxiv.org/abs/1409.4075}{{\tt
  arXiv:1409.4075}}].

\bibitem{Buckley:2014fba}
M.~R. Buckley, D.~Feld, and D.~Goncalves, {\it {Scalar Simplified Models for
  Dark Matter}},  {\em Phys. Rev.} {\bf D91} (2015) 015017,
  [\href{http://arxiv.org/abs/1410.6497}{{\tt arXiv:1410.6497}}].

\bibitem{Harris:2014hga}
P.~Harris, V.~V. Khoze, M.~Spannowsky, and C.~Williams, {\it {Constraining Dark
  Sectors at Colliders: Beyond the Effective Theory Approach}},  {\em Phys.
  Rev.} {\bf D91} (2015) 055009, [\href{http://arxiv.org/abs/1411.0535}{{\tt
  arXiv:1411.0535}}].

\bibitem{Jacques:2015zha}
T.~Jacques and K.~NordstrÃÂ¶m, {\it {Mapping monojet constraints onto
  Simplified Dark Matter Models}},  {\em JHEP} {\bf 06} (2015) 142,
  [\href{http://arxiv.org/abs/1502.05721}{{\tt arXiv:1502.05721}}].

\bibitem{Haisch:2015ioa}
U.~Haisch and E.~Re, {\it {Simplified dark matter top-quark interactions at the
  LHC}},  {\em JHEP} {\bf 06} (2015) 078,
  [\href{http://arxiv.org/abs/1503.00691}{{\tt arXiv:1503.00691}}].

\bibitem{Buchmueller:2015eea}
O.~Buchmueller, S.~A. Malik, C.~McCabe, and B.~Penning, {\it {Constraining Dark
  Matter Interactions with Pseudoscalar and Scalar Mediators Using Collider
  Searches for Multijets plus Missing Transverse Energy}},  {\em Phys. Rev.
  Lett.} {\bf 115} (2015), no.~18 181802,
  [\href{http://arxiv.org/abs/1505.07826}{{\tt arXiv:1505.07826}}].

\bibitem{Abdallah:2015ter}
J.~Abdallah et~al., {\it {Simplified Models for Dark Matter Searches at the
  LHC}},  {\em Phys. Dark Univ.} {\bf 9-10} (2015) 8--23,
  [\href{http://arxiv.org/abs/1506.03116}{{\tt arXiv:1506.03116}}].

\bibitem{Abercrombie:2015wmb}
D.~Abercrombie et~al., {\it {Dark Matter Benchmark Models for Early LHC Run-2
  Searches: Report of the ATLAS/CMS Dark Matter Forum}},
  \href{http://arxiv.org/abs/1507.00966}{{\tt arXiv:1507.00966}}.

\bibitem{Brennan:2016xjh}
A.~J. Brennan, M.~F. McDonald, J.~Gramling, and T.~D. Jacques, {\it {Collide
  and Conquer: Constraints on Simplified Dark Matter Models using Mono-X
  Collider Searches}},  {\em JHEP} {\bf 05} (2016) 112,
  [\href{http://arxiv.org/abs/1603.01366}{{\tt arXiv:1603.01366}}].

\bibitem{Boveia:2016mrp}
G.~Busoni et~al., {\it {Recommendations on presenting LHC searches for missing
  transverse energy signals using simplified $s$-channel models of dark
  matter}},  \href{http://arxiv.org/abs/1603.04156}{{\tt arXiv:1603.04156}}.

\bibitem{Bertone:2016mdy}
G.~Bertone, M.~P. Deisenroth, J.~S. Kim, S.~Liem, R.~Ruiz~de Austri, and
  M.~Welling, {\it {Accelerating the BSM interpretation of LHC data with
  machine learning}},  \href{http://arxiv.org/abs/1611.02704}{{\tt
  arXiv:1611.02704}}.

\bibitem{Frate:2017mai}
M.~Frate, K.~Cranmer, S.~Kalia, A.~Vandenberg-Rodes, and D.~Whiteson, {\it
  {Modeling Smooth Backgrounds and Generic Localized Signals with Gaussian
  Processes}},  \href{http://arxiv.org/abs/1709.05681}{{\tt arXiv:1709.05681}}.

\bibitem{Chala:2015ama}
M.~Chala, F.~Kahlhoefer, M.~McCullough, G.~Nardini, and K.~Schmidt-Hoberg, {\it
  {Constraining Dark Sectors with Monojets and Dijets}},  {\em JHEP} {\bf 07}
  (2015) 089, [\href{http://arxiv.org/abs/1503.05916}{{\tt arXiv:1503.05916}}].

\bibitem{Khachatryan:2016ecr}
{\bf CMS} Collaboration, V.~Khachatryan et~al., {\it {Search for narrow
  resonances in dijet final states at $\sqrt(s)=$ 8 TeV with the novel CMS
  technique of data scouting}},  {\em Phys. Rev. Lett.} {\bf 117} (2016), no.~3
  031802, [\href{http://arxiv.org/abs/1604.08907}{{\tt arXiv:1604.08907}}].

\bibitem{Sirunyan:2016iap}
{\bf CMS} Collaboration, A.~M. Sirunyan et~al., {\it {Search for dijet
  resonances in protonÃ¢ÂÂproton collisions at $\sqrt{s}$ = 13 TeV and
  constraints on dark matter and other models}},  {\em Phys. Lett.} {\bf B769}
  (2017) 520--542, [\href{http://arxiv.org/abs/1611.03568}{{\tt
  arXiv:1611.03568}}].

\bibitem{ATLAS:2016bvn}
{\bf ATLAS} Collaboration, T.~A. collaboration, {\it {Search for new light
  resonances decaying to jet pairs and produced in association with a photon or
  a jet in proton-proton collisions at $\sqrt{s}=13$~TeV with the ATLAS
  detector}},  {\em ATLAS-CONF-2016-070} (2016).

\bibitem{ATLAS:2016xiv}
{\bf ATLAS} Collaboration, T.~A. collaboration, {\it {Search for light dijet
  resonances with the ATLAS detector using a Trigger-Level Analysis in LHC pp
  collisions at $\sqrt{s}=13$~TeV}},  {\em ATLAS-CONF-2016-030} (2016).

\bibitem{Aaltonen:2008dn}
{\bf CDF} Collaboration, T.~Aaltonen et~al., {\it {Search for new particles
  decaying into dijets in proton-antiproton collisions at $\sqrt{s}$ = 1.96
  TeV}},  {\em Phys. Rev.} {\bf D79} (2009) 112002,
  [\href{http://arxiv.org/abs/0812.4036}{{\tt arXiv:0812.4036}}].

\bibitem{Fairbairn:2016iuf}
M.~Fairbairn, J.~Heal, F.~Kahlhoefer, and P.~Tunney, {\it {Constraints on Z'
  models from LHC dijet searches and implications for dark matter}},  {\em
  JHEP} {\bf 09} (2016) 018, [\href{http://arxiv.org/abs/1605.07940}{{\tt
  arXiv:1605.07940}}].

\bibitem{ATLASsummary}
ATLAS DM Simplified Model Exclusions summary plot, July 2017,
  \url{https://atlas.web.cern.ch/Atlas/GROUPS/PHYSICS/CombinedSummaryPlots/EXOTICS/ATLAS_DarkMatter_Summary_ModifiedCoupling/ATLAS_DarkMatter_Summary_ModifiedCoupling.png}
  (Accessed 1 Sep.\ 2017).

\bibitem{Akerib:2017kat}
{\bf LUX} Collaboration, D.~S. Akerib et~al., {\it {Limits on spin-dependent
  WIMP-nucleon cross section obtained from the complete LUX exposure}},  {\em
  Phys. Rev. Lett.} {\bf 118} (2017), no.~25 251302,
  [\href{http://arxiv.org/abs/1705.03380}{{\tt arXiv:1705.03380}}].

\bibitem{Akerib:2016lao}
{\bf LUX} Collaboration, D.~S. Akerib et~al., {\it {Results on the
  Spin-Dependent Scattering of Weakly Interacting Massive Particles on Nucleons
  from the Run 3 Data of the LUX Experiment}},  {\em Phys. Rev. Lett.} {\bf
  116} (2016), no.~16 161302, [\href{http://arxiv.org/abs/1602.03489}{{\tt
  arXiv:1602.03489}}].

\bibitem{ATLAS:2017dnw}
{\bf ATLAS} Collaboration, T.~A. collaboration, {\it {Search for dark matter
  and other new phenomena in events with an energetic jet and large missing
  transverse momentum using the ATLAS detector}},  {\em ATLAS-CONF-2017-060}
  (2017).

\bibitem{Aaboud:2017phn}
{\bf ATLAS} Collaboration, M.~Aaboud et~al., {\it {Search for dark matter and
  other new phenomena in events with an energetic jet and large missing
  transverse momentum using the ATLAS detector}},
  \href{http://arxiv.org/abs/1711.03301}{{\tt arXiv:1711.03301}}.

\bibitem{CMS:2017tbk}
{\bf CMS} Collaboration, C.~Collaboration, {\it {Search for new physics in
  final states with an energetic jet or a hadronically decaying W or Z boson
  using $35.9~\mathrm{fb}^{-1}$ of data at $\sqrt{s} = 13~\mathrm{TeV}$}},
  {\em CMS-PAS-EXO-16-048} (2017).

\bibitem{Albert:2017onk}
A.~Albert et~al., {\it {Recommendations of the LHC Dark Matter Working Group:
  Comparing LHC searches for heavy mediators of dark matter production in
  visible and invisible decay channels}},
  \href{http://arxiv.org/abs/1703.05703}{{\tt arXiv:1703.05703}}.

\bibitem{Aaboud:2017dor}
{\bf ATLAS} Collaboration, M.~Aaboud et~al., {\it {Search for dark matter at
  $\sqrt{s}=13$ TeV in final states containing an energetic photon and large
  missing transverse momentum with the ATLAS detector}},  {\em Eur. Phys. J.}
  {\bf C77} (2017), no.~6 393, [\href{http://arxiv.org/abs/1704.03848}{{\tt
  arXiv:1704.03848}}].

\bibitem{Barducci:2016pcb}
D.~Barducci, G.~Belanger, J.~Bernon, F.~Boudjema, J.~Da~Silva, S.~Kraml,
  U.~Laa, and A.~Pukhov, {\it {Collider limits on new physics within
  micrOMEGAs4.3}},  \href{http://arxiv.org/abs/1606.03834}{{\tt
  arXiv:1606.03834}}.

\bibitem{DMsimp_feynrules}
A.~Martini and K.~Mawatari, ``Spin-0 and spin-1 simplified model feynrules
  implementation.'' \url{http://feynrules.irmp.ucl.ac.be/wiki/DMsimp}.

\bibitem{Busoni:2014gta}
G.~Busoni, A.~De~Simone, T.~Jacques, E.~Morgante, and A.~Riotto, {\it {Making
  the Most of the Relic Density for Dark Matter Searches at the LHC 14 TeV
  Run}},  {\em JCAP} {\bf 1503} (2015), no.~03 022,
  [\href{http://arxiv.org/abs/1410.7409}{{\tt arXiv:1410.7409}}].

\bibitem{Gondolo:1990dk}
P.~Gondolo and G.~Gelmini, {\it {Cosmic abundances of stable particles:
  Improved analysis}},  {\em Nucl. Phys.} {\bf B360} (1991) 145--179.

\bibitem{Ackermann:2015zua}
{\bf Fermi-LAT} Collaboration, M.~Ackermann et~al., {\it {Searching for Dark
  Matter Annihilation from Milky Way Dwarf Spheroidal Galaxies with Six Years
  of Fermi Large Area Telescope Data}},  {\em Phys. Rev. Lett.} {\bf 115}
  (2015), no.~23 231301, [\href{http://arxiv.org/abs/1503.02641}{{\tt
  arXiv:1503.02641}}].

\bibitem{Charles:2016pgz}
{\bf Fermi-LAT} Collaboration, E.~Charles et~al., {\it {Sensitivity Projections
  for Dark Matter Searches with the Fermi Large Area Telescope}},  {\em Phys.
  Rept.} {\bf 636} (2016) 1--46, [\href{http://arxiv.org/abs/1605.02016}{{\tt
  arXiv:1605.02016}}].

\bibitem{Silverwood:2014yza}
H.~Silverwood, C.~Weniger, P.~Scott, and G.~Bertone, {\it {A realistic
  assessment of the CTA sensitivity to dark matter annihilation}},  {\em JCAP}
  {\bf 1503} (2015), no.~03 055, [\href{http://arxiv.org/abs/1408.4131}{{\tt
  arXiv:1408.4131}}].

\bibitem{Lefranc:2015pza}
V.~Lefranc, E.~Moulin, P.~Panci, and J.~Silk, {\it {Prospects for Annihilating
  Dark Matter in the inner Galactic halo by the Cherenkov Telescope Array}},
  {\em Phys. Rev.} {\bf D91} (2015), no.~12 122003,
  [\href{http://arxiv.org/abs/1502.05064}{{\tt arXiv:1502.05064}}].

\bibitem{Agashe:2014kda}
{\bf Particle Data Group} Collaboration, K.~A. Olive et~al., {\it {Review of
  Particle Physics}},  {\em Chin. Phys.} {\bf C38} (2014) 090001.

\bibitem{Crivellin:2014qxa}
A.~Crivellin, F.~D'Eramo, and M.~Procura, {\it {New Constraints on Dark Matter
  Effective Theories from Standard Model Loops}},  {\em Phys. Rev. Lett.} {\bf
  112} (2014) 191304, [\href{http://arxiv.org/abs/1402.1173}{{\tt
  arXiv:1402.1173}}].

\bibitem{DEramo:2014nmf}
F.~D'Eramo and M.~Procura, {\it {Connecting Dark Matter UV Complete Models to
  Direct Detection Rates via Effective Field Theory}},  {\em JHEP} {\bf 04}
  (2015) 054, [\href{http://arxiv.org/abs/1411.3342}{{\tt arXiv:1411.3342}}].

\bibitem{DEramo:2016gos}
F.~D'Eramo, B.~J. Kavanagh, and P.~Panci, {\it {You can hide but you have to
  run: direct detection with vector mediators}},  {\em JHEP} {\bf 08} (2016)
  111, [\href{http://arxiv.org/abs/1605.04917}{{\tt arXiv:1605.04917}}].

\bibitem{Hoferichter:2015ipa}
M.~Hoferichter, P.~Klos, and A.~Schwenk, {\it {Chiral power counting of one-
  and two-body currents in direct detection of dark matter}},  {\em Phys.
  Lett.} {\bf B746} (2015) 410--416,
  [\href{http://arxiv.org/abs/1503.04811}{{\tt arXiv:1503.04811}}].

\bibitem{Bishara:2016hek}
F.~Bishara, J.~Brod, B.~Grinstein, and J.~Zupan, {\it {Chiral Effective Theory
  of Dark Matter Direct Detection}},  {\em JCAP} {\bf 1702} (2017), no.~02 009,
  [\href{http://arxiv.org/abs/1611.00368}{{\tt arXiv:1611.00368}}].

\bibitem{Bishara:2017pfq}
F.~Bishara, J.~Brod, B.~Grinstein, and J.~Zupan, {\it {From quarks to nucleons
  in dark matter direct detection}},  {\em JHEP} {\bf 11} (2017) 059,
  [\href{http://arxiv.org/abs/1707.06998}{{\tt arXiv:1707.06998}}].

\bibitem{Bishara:2017nnn}
F.~Bishara, J.~Brod, B.~Grinstein, and J.~Zupan, {\it {DirectDM: a tool for
  dark matter direct detection}},  \href{http://arxiv.org/abs/1708.02678}{{\tt
  arXiv:1708.02678}}.

\bibitem{Fu:2016ega}
{\bf PandaX-II} Collaboration, C.~Fu et~al., {\it {Spin-Dependent Weakly
  Interacting Massive Particle Nucleon Cross Section Limits from First Data of
  PandaX-II Experiment}},  {\em Phys. Rev. Lett.} {\bf 118} (2017), no.~7
  071301, [\href{http://arxiv.org/abs/1611.06553}{{\tt arXiv:1611.06553}}].

\bibitem{Amole:2017dex}
{\bf PICO} Collaboration, C.~Amole et~al., {\it {Dark Matter Search Results
  from the PICO-60 C$_3$F$_8$ Bubble Chamber}},  {\em Phys. Rev. Lett.} {\bf
  118} (2017), no.~25 251301, [\href{http://arxiv.org/abs/1702.07666}{{\tt
  arXiv:1702.07666}}].

\bibitem{Mount:2017qzi}
B.~J. Mount et~al., {\it {LUX-ZEPLIN (LZ) Technical Design Report}},
  \href{http://arxiv.org/abs/1703.09144}{{\tt arXiv:1703.09144}}.

\bibitem{Cao:2014jsa}
{\bf PandaX} Collaboration, X.~Cao et~al., {\it {PandaX: A Liquid Xenon Dark
  Matter Experiment at CJPL}},  {\em Sci. China Phys. Mech. Astron.} {\bf 57}
  (2014) 1476--1494, [\href{http://arxiv.org/abs/1405.2882}{{\tt
  arXiv:1405.2882}}].

\bibitem{Aprile:2015uzo}
{\bf XENON} Collaboration, E.~Aprile et~al., {\it {Physics reach of the XENON1T
  dark matter experiment}},  {\em JCAP} {\bf 1604} (2016), no.~04 027,
  [\href{http://arxiv.org/abs/1512.07501}{{\tt arXiv:1512.07501}}].

\bibitem{Aalbers:2016jon}
{\bf DARWIN} Collaboration, J.~Aalbers et~al., {\it {DARWIN: towards the
  ultimate dark matter detector}},  {\em JCAP} {\bf 1611} (2016) 017,
  [\href{http://arxiv.org/abs/1606.07001}{{\tt arXiv:1606.07001}}].

\bibitem{Baudis:2013bba}
L.~Baudis, G.~Kessler, P.~Klos, R.~F. Lang, J.~Menendez, S.~Reichard, and
  A.~Schwenk, {\it {Signatures of Dark Matter Scattering Inelastically Off
  Nuclei}},  {\em Phys. Rev.} {\bf D88} (2013), no.~11 115014,
  [\href{http://arxiv.org/abs/1309.0825}{{\tt arXiv:1309.0825}}].

\bibitem{McCabe:2015eia}
C.~McCabe, {\it {Prospects for dark matter detection with inelastic transitions
  of xenon}},  {\em JCAP} {\bf 1605} (2016), no.~05 033,
  [\href{http://arxiv.org/abs/1512.00460}{{\tt arXiv:1512.00460}}].

\bibitem{Aartsen:2016zhm}
{\bf IceCube} Collaboration, M.~G. Aartsen et~al., {\it {Search for
  annihilating dark matter in the Sun with 3 years of IceCube data}},  {\em
  Eur. Phys. J.} {\bf C77} (2017), no.~3 146,
  [\href{http://arxiv.org/abs/1612.05949}{{\tt arXiv:1612.05949}}].

\bibitem{PICO500}
S.~Fallows, TeVPA Conference 2017 presentation,
  \url{https://atlas.web.cern.ch/Atlas/GROUPS/PHYSICS/CombinedSummaryPlots/EXOTICS/ATLAS_DarkMatter_Summary_ModifiedCoupling/ATLAS_DarkMatter_Summary_ModifiedCoupling.png}
  (Accessed 17 Aug.\ 2017).

\bibitem{Gelmini:2006pq}
G.~Gelmini, P.~Gondolo, A.~Soldatenko, and C.~E. Yaguna, {\it {The Effect of a
  late decaying scalar on the neutralino relic density}},  {\em Phys. Rev.}
  {\bf D74} (2006) 083514, [\href{http://arxiv.org/abs/hep-ph/0605016}{{\tt
  hep-ph/0605016}}].

\bibitem{Rehagen:2015zma}
T.~Rehagen and G.~B. Gelmini, {\it {Low reheating temperatures in monomial and
  binomial inflationary potentials}},  {\em JCAP} {\bf 1506} (2015), no.~06
  039, [\href{http://arxiv.org/abs/1504.03768}{{\tt arXiv:1504.03768}}].

\bibitem{Akerib:2016vxi}
{\bf LUX} Collaboration, D.~S. Akerib et~al., {\it {Results from a search for
  dark matter in the complete LUX exposure}},  {\em Phys. Rev. Lett.} {\bf 118}
  (2017), no.~2 021303, [\href{http://arxiv.org/abs/1608.07648}{{\tt
  arXiv:1608.07648}}].

\bibitem{Agnese:2017jvy}
{\bf SuperCDMS} Collaboration, R.~Agnese et~al., {\it {Low-Mass Dark Matter
  Search with CDMSlite}},  {\em Submitted to: Phys. Rev. D} (2017)
  [\href{http://arxiv.org/abs/1707.01632}{{\tt arXiv:1707.01632}}].

\bibitem{Angloher:2015ewa}
{\bf CRESST} Collaboration, G.~Angloher et~al., {\it {Results on light dark
  matter particles with a low-threshold CRESST-II detector}},  {\em Eur. Phys.
  J.} {\bf C76} (2016), no.~1 25, [\href{http://arxiv.org/abs/1509.01515}{{\tt
  arXiv:1509.01515}}].

\bibitem{Angloher:2015eza}
{\bf CRESST} Collaboration, G.~Angloher et~al., {\it {Probing low WIMP masses
  with the next generation of CRESST detector}},
  \href{http://arxiv.org/abs/1503.08065}{{\tt arXiv:1503.08065}}.

\bibitem{Haisch:2012kf}
U.~Haisch, F.~Kahlhoefer, and J.~Unwin, {\it {The impact of heavy-quark loops
  on LHC dark matter searches}},  {\em JHEP} {\bf 07} (2013) 125,
  [\href{http://arxiv.org/abs/1208.4605}{{\tt arXiv:1208.4605}}].

\bibitem{Aaboud:2017aeu}
{\bf ATLAS} Collaboration, M.~Aaboud et~al., {\it {Search for top-squark pair
  production in final states with one lepton, jets, and missing transverse
  momentum using 36 fb$^{-1}$ of $\sqrt{s}=13$ TeV pp collision data with the
  ATLAS detector}},  \href{http://arxiv.org/abs/1711.11520}{{\tt
  arXiv:1711.11520}}.

\bibitem{Sirunyan:2017hci}
{\bf CMS} Collaboration, A.~M. Sirunyan et~al., {\it {Search for dark matter
  produced with an energetic jet or a hadronically decaying W or Z boson at $
  \sqrt{s}=13 $ TeV}},  {\em JHEP} {\bf 07} (2017) 014,
  [\href{http://arxiv.org/abs/1703.01651}{{\tt arXiv:1703.01651}}].

\bibitem{Sirunyan:2017xgm}
{\bf CMS} Collaboration, A.~M. Sirunyan et~al., {\it {Search for dark matter
  produced in association with heavy-flavor quarks in proton-proton collisions
  at sqrt(s)=13 TeV}},  \href{http://arxiv.org/abs/1706.02581}{{\tt
  arXiv:1706.02581}}.

\bibitem{Hoferichter:2017olk}
M.~Hoferichter, P.~Klos, J.~Menendez, and A.~Schwenk, {\it {Improved limits for
  Higgs-portal dark matter from LHC searches}},  {\em Phys. Rev. Lett.} {\bf
  119} (2017), no.~18 181803, [\href{http://arxiv.org/abs/1708.02245}{{\tt
  arXiv:1708.02245}}].

\bibitem{Aprile:2017iyp}
{\bf XENON} Collaboration, E.~Aprile et~al., {\it {First Dark Matter Search
  Results from the XENON1T Experiment}},  {\em Phys. Rev. Lett.} {\bf 119}
  (2017), no.~18 181301, [\href{http://arxiv.org/abs/1705.06655}{{\tt
  arXiv:1705.06655}}].

\bibitem{Agnese:2016cpb}
{\bf SuperCDMS} Collaboration, R.~Agnese et~al., {\it {Projected Sensitivity of
  the SuperCDMS SNOLAB experiment}},  {\em Phys. Rev.} {\bf D95} (2017), no.~8
  082002, [\href{http://arxiv.org/abs/1610.00006}{{\tt arXiv:1610.00006}}].

\bibitem{Bell:2016ekl}
N.~F. Bell, G.~Busoni, and I.~W. Sanderson, {\it {Self-consistent Dark Matter
  Simplified Models with an s-channel scalar mediator}},  {\em JCAP} {\bf 1703}
  (2017), no.~03 015, [\href{http://arxiv.org/abs/1612.03475}{{\tt
  arXiv:1612.03475}}].

\bibitem{Kahlhoefer:2017dnp}
F.~Kahlhoefer, {\it {Review of LHC Dark Matter Searches}},  {\em Int. J. Mod.
  Phys.} {\bf A32} (2017), no.~13 1730006,
  [\href{http://arxiv.org/abs/1702.02430}{{\tt arXiv:1702.02430}}].

\bibitem{Cui:2017juz}
Y.~Cui and F.~D'Eramo, {\it {Surprises from complete vector portal theories:
  New insights into the dark sector and its interplay with Higgs physics}},
  {\em Phys. Rev.} {\bf D96} (2017), no.~9 095006,
  [\href{http://arxiv.org/abs/1705.03897}{{\tt arXiv:1705.03897}}].

\bibitem{Kahlhoefer:2015bea}
F.~Kahlhoefer, K.~Schmidt-Hoberg, T.~Schwetz, and S.~Vogl, {\it {Implications
  of unitarity and gauge invariance for simplified dark matter models}},  {\em
  JHEP} {\bf 02} (2016) 016, [\href{http://arxiv.org/abs/1510.02110}{{\tt
  arXiv:1510.02110}}].

\bibitem{Duerr:2016tmh}
M.~Duerr, F.~Kahlhoefer, K.~Schmidt-Hoberg, T.~Schwetz, and S.~Vogl, {\it {How
  to save the WIMP: global analysis of a dark matter model with two s-channel
  mediators}},  {\em JHEP} {\bf 09} (2016) 042,
  [\href{http://arxiv.org/abs/1606.07609}{{\tt arXiv:1606.07609}}].

\bibitem{Englert:2016joy}
C.~Englert, M.~McCullough, and M.~Spannowsky, {\it {S-Channel Dark Matter
  Simplified Models and Unitarity}},  {\em Phys. Dark Univ.} {\bf 14} (2016)
  48--56, [\href{http://arxiv.org/abs/1604.07975}{{\tt arXiv:1604.07975}}].

\bibitem{DAmbrosio:2002vsn}
G.~D'Ambrosio, G.~F. Giudice, G.~Isidori, and A.~Strumia, {\it {Minimal flavor
  violation: An Effective field theory approach}},  {\em Nucl. Phys.} {\bf
  B645} (2002) 155--187, [\href{http://arxiv.org/abs/hep-ph/0207036}{{\tt
  hep-ph/0207036}}].

\bibitem{Liu:2011dh}
J.-Y. Liu, Y.~Tang, and Y.-L. Wu, {\it {Searching for $Z^{'}$ Gauge Boson in an
  Anomaly-Free U(1)$'$ Gauge Family Model}},  {\em J. Phys.} {\bf G39} (2012)
  055003, [\href{http://arxiv.org/abs/1108.5012}{{\tt arXiv:1108.5012}}].

\bibitem{Duerr:2013dza}
M.~Duerr, P.~Fileviez~Perez, and M.~B. Wise, {\it {Gauge Theory for Baryon and
  Lepton Numbers with Leptoquarks}},  {\em Phys. Rev. Lett.} {\bf 110} (2013)
  231801, [\href{http://arxiv.org/abs/1304.0576}{{\tt arXiv:1304.0576}}].

\bibitem{Perez:2014qfa}
P.~Fileviez~Perez, S.~Ohmer, and H.~H. Patel, {\it {Minimal Theory for
  Lepto-Baryons}},  {\em Phys. Lett.} {\bf B735} (2014) 283--287,
  [\href{http://arxiv.org/abs/1403.8029}{{\tt arXiv:1403.8029}}].

\bibitem{Ekstedt:2016wyi}
A.~Ekstedt, R.~Enberg, G.~Ingelman, J.~Lofgren, and T.~Mandal, {\it
  {Constraining minimal anomaly free $\mathrm{U}(1)$ extensions of the Standard
  Model}},  {\em JHEP} {\bf 11} (2016) 071,
  [\href{http://arxiv.org/abs/1605.04855}{{\tt arXiv:1605.04855}}].

\bibitem{Jacques:2016dqz}
T.~Jacques, A.~Katz, E.~Morgante, D.~Racco, M.~Rameez, and A.~Riotto, {\it
  {Complementarity of DM searches in a consistent simplified model: the case of
  $ZÃ¢ÂÂ²$}},  {\em JHEP} {\bf 10} (2016) 071,
  [\href{http://arxiv.org/abs/1605.06513}{{\tt arXiv:1605.06513}}].

\bibitem{Ismail:2016tod}
A.~Ismail, W.-Y. Keung, K.-H. Tsao, and J.~Unwin, {\it {Axial vector
  $ZÃ¢ÂÂ²$ and anomaly cancellation}},  {\em Nucl. Phys.} {\bf B918}
  (2017) 220--244, [\href{http://arxiv.org/abs/1609.02188}{{\tt
  arXiv:1609.02188}}].

\bibitem{Ellis:2017tkh}
J.~Ellis, M.~Fairbairn, and P.~Tunney, {\it {Anomaly-Free Dark Matter Models
  are not so Simple}},  {\em JHEP} {\bf 08} (2017) 053,
  [\href{http://arxiv.org/abs/1704.03850}{{\tt arXiv:1704.03850}}].

\bibitem{Ismail:2017ulg}
A.~Ismail, A.~Katz, and D.~Racco, {\it {On dark matter interactions with the
  Standard Model through an anomalous $Z′$}},  {\em JHEP} {\bf 10} (2017)
  165, [\href{http://arxiv.org/abs/1707.00709}{{\tt arXiv:1707.00709}}].

\bibitem{Dror:2017nsg}
J.~A. Dror, R.~Lasenby, and M.~Pospelov, {\it {Dark forces coupled to
  nonconserved currents}},  {\em Phys. Rev.} {\bf D96} (2017), no.~7 075036,
  [\href{http://arxiv.org/abs/1707.01503}{{\tt arXiv:1707.01503}}].

\bibitem{Bell:2016uhg}
N.~F. Bell, Y.~Cai, and R.~K. Leane, {\it {Impact of mass generation for spin-1
  mediator simplified models}},  {\em JCAP} {\bf 1701} (2017), no.~01 039,
  [\href{http://arxiv.org/abs/1610.03063}{{\tt arXiv:1610.03063}}].

\bibitem{Duerr:2017uap}
M.~Duerr, A.~Grohsjean, F.~Kahlhoefer, B.~Penning, K.~Schmidt-Hoberg, and
  C.~Schwanenberger, {\it {Hunting the dark Higgs}},  {\em JHEP} {\bf 04}
  (2017) 143, [\href{http://arxiv.org/abs/1701.08780}{{\tt arXiv:1701.08780}}].

\bibitem{Aaboud:2016tnv}
{\bf ATLAS} Collaboration, M.~Aaboud et~al., {\it {Search for new phenomena in
  final states with an energetic jet and large missing transverse momentum in
  $pp$ collisions at $\sqrt{s}=13$Ã¢ÂÂÃ¢ÂÂTeV using the ATLAS
  detector}},  {\em Phys. Rev.} {\bf D94} (2016), no.~3 032005,
  [\href{http://arxiv.org/abs/1604.07773}{{\tt arXiv:1604.07773}}].

\bibitem{Nason:2004rx}
P.~Nason, {\it {A New method for combining NLO QCD with shower Monte Carlo
  algorithms}},  {\em JHEP} {\bf 11} (2004) 040,
  [\href{http://arxiv.org/abs/hep-ph/0409146}{{\tt hep-ph/0409146}}].

\bibitem{Frixione:2007vw}
S.~Frixione, P.~Nason, and C.~Oleari, {\it {Matching NLO QCD computations with
  Parton Shower simulations: the POWHEG method}},  {\em JHEP} {\bf 11} (2007)
  070, [\href{http://arxiv.org/abs/0709.2092}{{\tt arXiv:0709.2092}}].

\bibitem{Alioli:2010xd}
S.~Alioli, P.~Nason, C.~Oleari, and E.~Re, {\it {A general framework for
  implementing NLO calculations in shower Monte Carlo programs: the POWHEG
  BOX}},  {\em JHEP} {\bf 06} (2010) 043,
  [\href{http://arxiv.org/abs/1002.2581}{{\tt arXiv:1002.2581}}].

\bibitem{Haisch:2013ata}
U.~Haisch, F.~Kahlhoefer, and E.~Re, {\it {QCD effects in mono-jet searches for
  dark matter}},  {\em JHEP} {\bf 12} (2013) 007,
  [\href{http://arxiv.org/abs/1310.4491}{{\tt arXiv:1310.4491}}].

\bibitem{Sjostrand:2007gs}
T.~Sjostrand, S.~Mrenna, and P.~Z. Skands, {\it {A Brief Introduction to PYTHIA
  8.1}},  {\em Comput. Phys. Commun.} {\bf 178} (2008) 852--867,
  [\href{http://arxiv.org/abs/0710.3820}{{\tt arXiv:0710.3820}}].

\bibitem{Ball:2014uwa}
{\bf NNPDF} Collaboration, R.~D. Ball et~al., {\it {Parton distributions for
  the LHC Run II}},  {\em JHEP} {\bf 04} (2015) 040,
  [\href{http://arxiv.org/abs/1410.8849}{{\tt arXiv:1410.8849}}].

\bibitem{Drees:2013wra}
M.~Drees, H.~Dreiner, D.~Schmeier, J.~Tattersall, and J.~S. Kim, {\it
  {CheckMATE: Confronting your Favourite New Physics Model with LHC Data}},
  {\em Comput. Phys. Commun.} {\bf 187} (2015) 227--265,
  [\href{http://arxiv.org/abs/1312.2591}{{\tt arXiv:1312.2591}}].

\bibitem{Kim:2015wza}
J.~S. Kim, D.~Schmeier, J.~Tattersall, and K.~Rolbiecki, {\it {A framework to
  create customised LHC analyses within CheckMATE}},  {\em Comput. Phys.
  Commun.} {\bf 196} (2015) 535--562,
  [\href{http://arxiv.org/abs/1503.01123}{{\tt arXiv:1503.01123}}].

\bibitem{Dercks:2016npn}
D.~Dercks, N.~Desai, J.~S. Kim, K.~Rolbiecki, J.~Tattersall, and T.~Weber, {\it
  {CheckMATE 2: From the model to the limit}},
  \href{http://arxiv.org/abs/1611.09856}{{\tt arXiv:1611.09856}}.

\bibitem{deFavereau:2013fsa}
{\bf DELPHES 3} Collaboration, J.~de~Favereau, C.~Delaere, P.~Demin,
  A.~Giammanco, V.~LemaÃÂ®tre, A.~Mertens, and M.~Selvaggi, {\it {DELPHES
  3, A modular framework for fast simulation of a generic collider
  experiment}},  {\em JHEP} {\bf 02} (2014) 057,
  [\href{http://arxiv.org/abs/1307.6346}{{\tt arXiv:1307.6346}}].

\bibitem{Cacciari:2011ma}
M.~Cacciari, G.~P. Salam, and G.~Soyez, {\it {FastJet User Manual}},  {\em Eur.
  Phys. J.} {\bf C72} (2012) 1896, [\href{http://arxiv.org/abs/1111.6097}{{\tt
  arXiv:1111.6097}}].

\bibitem{checkmatewebpage}
{\tt CheckMATE} webpage, \url{https://checkmate.hepforge.org/}.

\bibitem{Alwall:2014hca}
J.~Alwall, R.~Frederix, S.~Frixione, V.~Hirschi, F.~Maltoni, O.~Mattelaer,
  H.~S. Shao, T.~Stelzer, P.~Torrielli, and M.~Zaro, {\it {The automated
  computation of tree-level and next-to-leading order differential cross
  sections, and their matching to parton shower simulations}},  {\em JHEP} {\bf
  07} (2014) 079, [\href{http://arxiv.org/abs/1405.0301}{{\tt
  arXiv:1405.0301}}].

\bibitem{Alloul:2013bka}
A.~Alloul, N.~D. Christensen, C.~Degrande, C.~Duhr, and B.~Fuks, {\it
  {FeynRules 2.0 - A complete toolbox for tree-level phenomenology}},  {\em
  Comput. Phys. Commun.} {\bf 185} (2014) 2250--2300,
  [\href{http://arxiv.org/abs/1310.1921}{{\tt arXiv:1310.1921}}].

\bibitem{McCabe:2010zh}
C.~McCabe, {\it {The Astrophysical Uncertainties Of Dark Matter Direct
  Detection Experiments}},  {\em Phys. Rev.} {\bf D82} (2010) 023530,
  [\href{http://arxiv.org/abs/1005.0579}{{\tt arXiv:1005.0579}}].

\bibitem{Bozorgnia:2016ogo}
N.~Bozorgnia, F.~Calore, M.~Schaller, M.~Lovell, G.~Bertone, C.~S. Frenk, R.~A.
  Crain, J.~F. Navarro, J.~Schaye, and T.~Theuns, {\it {Simulated Milky Way
  analogues: implications for dark matter direct searches}},  {\em JCAP} {\bf
  1605} (2016), no.~05 024, [\href{http://arxiv.org/abs/1601.04707}{{\tt
  arXiv:1601.04707}}].

\bibitem{Akerib:2015cja}
{\bf LZ} Collaboration, D.~S. Akerib et~al., {\it {LUX-ZEPLIN (LZ) Conceptual
  Design Report}},  \href{http://arxiv.org/abs/1509.02910}{{\tt
  arXiv:1509.02910}}.

\bibitem{Schumann:2015cpa}
M.~Schumann, L.~Baudis, L.~BÃÅtikofer, A.~Kish, and M.~Selvi, {\it {Dark
  matter sensitivity of multi-ton liquid xenon detectors}},  {\em JCAP} {\bf
  1510} (2015), no.~10 016, [\href{http://arxiv.org/abs/1506.08309}{{\tt
  arXiv:1506.08309}}].

\bibitem{McCabe:2017rln}
C.~McCabe, {\it {New constraints and discovery potential of sub-GeV dark matter
  with xenon detectors}},  {\em Phys. Rev.} {\bf D96} (2017), no.~4 043010,
  [\href{http://arxiv.org/abs/1702.04730}{{\tt arXiv:1702.04730}}].

\bibitem{Klos:2013rwa}
P.~Klos, J.~MenÃÂ©ndez, D.~Gazit, and A.~Schwenk, {\it {Large-scale
  nuclear structure calculations for spin-dependent WIMP scattering with chiral
  effective field theory currents}},  {\em Phys. Rev.} {\bf D88} (2013), no.~8
  083516, [\href{http://arxiv.org/abs/1304.7684}{{\tt arXiv:1304.7684}}].
  [Erratum: Phys. Rev.D89,no.2,029901(2014)].

\bibitem{Angloher:2014myn}
{\bf CRESST-II} Collaboration, G.~Angloher et~al., {\it {Results on low mass
  WIMPs using an upgraded CRESST-II detector}},  {\em Eur. Phys. J.} {\bf C74}
  (2014), no.~12 3184, [\href{http://arxiv.org/abs/1407.3146}{{\tt
  arXiv:1407.3146}}].

\bibitem{Petricca:2017zdp}
{\bf CRESST} Collaboration, F.~Petricca et~al., {\it {First results on low-mass
  dark matter from the CRESST-III experiment}},  in {\em {15th International
  Conference on Topics in Astroparticle and Underground Physics (TAUP 2017)
  Sudbury, Ontario, Canada, July 24-28, 2017}}, 2017.
\newblock \href{http://arxiv.org/abs/1711.07692}{{\tt arXiv:1711.07692}}.

\bibitem{Dolan:2017xbu}
M.~J. Dolan, F.~Kahlhoefer, and C.~McCabe, {\it {Direct Detection of sub-GeV
  Dark Matter with Electrons from Nuclear Scattering}},
  \href{http://arxiv.org/abs/1711.09906}{{\tt arXiv:1711.09906}}.

\bibitem{cresst-pv}
CRESST, 2016.
\newblock Private communication.

\bibitem{Feroz:2008xx}
F.~Feroz, M.~P. Hobson, and M.~Bridges, {\it {MultiNest: an efficient and
  robust Bayesian inference tool for cosmology and particle physics}},  {\em
  Mon. Not. Roy. Astron. Soc.} {\bf 398} (2009) 1601--1614,
  [\href{http://arxiv.org/abs/0809.3437}{{\tt arXiv:0809.3437}}].

\bibitem{Skilling04}
J.~{Skilling}, {\it {Nested Sampling}},  in {\em American Institute of Physics
  Conference Series} (R.~{Fischer}, R.~{Preuss}, and U.~V. {Toussaint}, eds.),
  pp.~395--405, Nov., 2004.

\bibitem{Workgroup:2017lvb}
{\bf GAMBIT Dark Matter Workgroup} Collaboration, T.~Bringmann et~al., {\it
  {DarkBit: A GAMBIT module for computing dark matter observables and
  likelihoods}},  \href{http://arxiv.org/abs/1705.07920}{{\tt
  arXiv:1705.07920}}.

\bibitem{Ho:1995:RDF:844379.844681}
T.~K. Ho, {\it Random decision forests},  in {\em Proceedings of the Third
  International Conference on Document Analysis and Recognition (Volume 1) -
  Volume 1}, ICDAR '95, (Washington, DC, USA), pp.~278--, IEEE Computer
  Society, 1995.

\bibitem{Breiman2001}
L.~Breiman, {\it Random forests},  {\em Machine Learning} {\bf 45} (Oct, 2001)
  5--32.

\bibitem{2015arXiv150202843D}
M.~P. {Deisenroth} and J.~W. {Ng}, {\it {Distributed Gaussian Processes}},
  {\em ArXiv e-prints} (Feb., 2015)
  [\href{http://arxiv.org/abs/1502.02843}{{\tt arXiv:1502.02843}}].

\bibitem{Green:2008rd}
A.~M. Green, {\it {Determining the WIMP mass from a single direct detection
  experiment, a more detailed study}},  {\em JCAP} {\bf 0807} (2008) 005,
  [\href{http://arxiv.org/abs/0805.1704}{{\tt arXiv:0805.1704}}].

\bibitem{Angloher:2017zkf}
{\bf CRESST} Collaboration, G.~Angloher et~al., {\it {Description of CRESST-II
  data}},  \href{http://arxiv.org/abs/1701.08157}{{\tt arXiv:1701.08157}}.

\bibitem{chollet2015keras}
F.~Chollet et~al., ``Keras.'' \url{https://github.com/fchollet/keras}, 2015.

\bibitem{Jin:2015aa}
X.~{Jin}, C.~{Xu}, J.~{Feng}, Y.~{Wei}, J.~{Xiong}, and S.~{Yan}, {\it {Deep
  Learning with S-shaped Rectified Linear Activation Units}},  {\em ArXiv
  e-prints} (Dec., 2015) [\href{http://arxiv.org/abs/1512.07030}{{\tt
  arXiv:1512.07030}}].

\bibitem{Kingma:2014aa}
D.~P. {Kingma} and J.~{Ba}, {\it {Adam: A Method for Stochastic Optimization}},
   {\em ArXiv e-prints} (Dec., 2014)
  [\href{http://arxiv.org/abs/1412.6980}{{\tt arXiv:1412.6980}}].

\bibitem{Rasmussen:2005:GPM:1162254}
C.~E. Rasmussen and C.~K.~I. Williams, {\em Gaussian Processes for Machine
  Learning (Adaptive Computation and Machine Learning)}.
\newblock The MIT Press, 2005.

\bibitem{scikit-learn}
F.~Pedregosa, G.~Varoquaux, A.~Gramfort, V.~Michel, B.~Thirion, O.~Grisel,
  M.~Blondel, P.~Prettenhofer, R.~Weiss, V.~Dubourg, J.~Vanderplas, A.~Passos,
  D.~Cournapeau, M.~Brucher, M.~Perrot, and E.~Duchesnay, {\it Scikit-learn:
  Machine learning in {P}ython},  {\em Journal of Machine Learning Research}
  {\bf 12} (2011) 2825--2830.

\end{thebibliography}\endgroup

%\clearpage
\FloatBarrier
%%%%%%%%%%%%%%%%%%%%%%%%%%%%%%%%%%%%%%%%%%%%%%%%%%%%%%%%%%

\end{document}